\documentclass[aps,pre,10pt,showpacs,floatfix,onecolumn,nofootinbib,a4paper]{revtex4-2}
\usepackage{amssymb, amsmath, amsthm, mathtools}
\usepackage{bm}
\usepackage{mathrsfs}
\usepackage{hyperref,bookmark}
\usepackage{graphicx}
\usepackage{epsfig,color}
\usepackage{mathrsfs}
\usepackage{verbatim}
\usepackage{url}
\usepackage{subcaption}
\usepackage{float}
\usepackage[greek,english]{babel}
\usepackage{dcolumn}

\newcommand{\tr}{\operatorname{tr}}
\newcommand{\dd}{\operatorname{d}\!}

\newcommand{\diver}{\operatorname{div}}
\newcommand{\curl}{\operatorname{curl}}
\newcommand{\arccot}{\operatorname{arccot}}
\newcommand{\rad}{\mathrm{rad}}
\newcommand{\n}{\bm{n}}
\newcommand{\e}{\bm{e}}
\newcommand{\nper}{\bm{n}_\perp}
\newcommand{\normal}{\bm{\nu}}
\newcommand{\body}{\mathscr{B}}
\newcommand{\free}{\mathscr{F}}
\newcommand{\boundary}{\partial\mathscr{B}}
\newcommand{\Req}{R_\mathrm{e}}
\newcommand{\ga}{\gamma_\mathrm{a}}
\newcommand{\vt}{\vartheta}
\newcommand{\p}{\bm{p}}
\newcommand{\framen}{(\n,\nper,\e_\vt)}
\newcommand{\framer}{(\e_r,\e_\vt,\n)}
\newcommand{\azero}{\alpha_\mathrm{c}}
\newcommand{\aone}{\alpha_\mathrm{c}'}
\newcommand{\atwo}{\alpha_\mathrm{c}''}
\newcommand{\conf}{\mathsf{S}}

\newcommand{\Fa}{F_\alpha}
\newcommand{\vae}{\varepsilon}

\begin{document}
\latintext

\title{Nematic Tactoid Population}
\author{Silvia Paparini}
\email{paparinisilvia@gmail.com}
\author{Epifanio G. Virga}
\email{eg.virga@unipv.it}
\affiliation{Dipartimento di Matematica, Universit\`a di Pavia, Via Ferrata 5, 27100 Pavia, Italy}

\date{\today}

\begin{abstract}
Tactoids are pointed, spindle-like droplets of nematic liquid crystal in an isotropic fluid.  They have long been observed in inorganic and organic nematics, in thermotropic phases as well as lyotropic colloidal aggregates. The variational problem of determining the optimal shape of a nematic droplet is formidable and has only been attacked in selected classes of shapes and director fields. Here, by considering a novel class of admissible solutions for a bipolar droplet, we study the prevalence in the population of all equilibrium shapes of each of the three that may be optimal (tactoids primarily among them). We show how the prevalence of a shape is affected by the drop's volume $V_0$ and the saddle-splay constant $K_{24}$ of the material. Tactoids, in particular, prevail for small $V_0$ and small $K_{24}$ (appropriately scaled). Our class of shapes (and director fields) is sufficiently different from those employed so far to unveil a rather different role of $K_{24}$.  
\end{abstract}

\maketitle

\section{Introduction}\label{sec:intro}
Tactoids have a long and intriguing history. The name \emph{tactoid} (in German, \emph{taktoid}) comes from the Greek \greektext takt'os\latintext, meaning \emph{ordered}; it was coined by Zocher and Jacobsohn \cite{zocher:taktosole} to designate spindle-like aggregates of elongated colloidal particles dispersed in sols (typically aqueous). Originally, such particles were composed of monocrystals of vanadium pentoxide ($\mathrm{V}_2\mathrm{O}_5$) grown by aging, first\footnote{We learn, however, in the historical review \cite{sonin:inorganic} (which is highly recommended to the reader) that an earlier experiment performed in 1904 by Q. Majorana had already found magnetically induced birefringence in a sol of inorganic particles (FeOOH).} studied in \cite{zocher:freiwillige} and further characterized in \cite{watson:comparative}.

Later, once Stanley~\cite{stanley:isolation} had succeeded in extracting tobacco mosaic virus (TMV) from infected plants, tactoids made again their appearance in aqueous sols where TMV had been dispersed with a concentration higher than 2\% by weight \cite{bawden:liquid}.\footnote{A more recent study attempting to characterize this special system can be found in \cite{fraden:magnetic-field}.} Remarkable is the evidence of tactoids in TMV sols collected in \cite{bernal:x-ray}, whose diagrams and pictures of pointed shapes we found inspirational.\footnote{The original aim of \cite{bernal:x-ray} was to measure inter-tactoid distances as a function of pH and ionic strength.} Onsager himself says that explaining the formation of TMV is one motivation for his seminal paper \cite{onsager:effects} on the coexistence of nematic and isotropic phases as sole consequence of steric interactions.\footnote{Although the connection between colloidal aggregates and liquid crystals was already clear to Onsager, it took Zocher a longer time to include what he had called nematic (and smectic) \emph{superphases} \cite{zocher:taktosole_meso,zocher:neuere} ``into the realm of liquid crystals, though their physico-chemical nature is very different from that of relatively low molecular organic substances exhibiting mesophases'' \cite[p.\,178]{zocher:nematic}.} Chromonic liquid crystals,\footnote{Disparate materials can be classified as chromonic liquid crystals; they include dyes, drugs \cite{dickinson:aggregate,tam-chang:chromonic}, nucleotides \cite{mariani:small}, and DNA oligomers \cite{zanchetta:phase,nakata:end-to-end}. See also the review \cite{lydon:chromonic} and the thesis \cite{zhou:lyotropic}.} which are constituted by molecular aggregates whose length distribution is affected by both temperature and concentration, have shown hosts of tactoids  \cite{nastishin:optical,tortora:chiral,kim:morphogenesis,jeong:chiral_2014,peng:chirality}.\footnote{Examples of tactoids in other materials can also be found in \cite{oakes:growth,verhoeff:tactoids}.}

Now, we know that tactoids are ubiquitous in liquid crystals, irrespective of whether the latter are thermotropic (when the ordered phase is induced by a change in temperature) or lyotropic (when the ordered phase is induced by a change in concentration). A classical phenomenological theory first proposed by Oseen~\cite{oseen:theory} and then formalized by Frank~\cite{frank:theory} has proven valid in describing the elastic cost associated with static distortions of the nematic director $\n$, the mesoscopic unit vector field designating the local average orientation of the elementary constituents of the phase (be they molecules or super-molecular constructs). Since tactoids are droplets surrounded by an isotropic fluid,\footnote{This could either be the isotropic melt (or vapor) of the same substance (in thermotropic materials) or the isotropic component in phase-coexistence (in lyotropic materials).}  the Oseen-Frank elastic energy, which accounts for distortions in bulk, does not suffice to describe the whole energetic landscape. A surface energy at the interface separating the droplet from the surrounding fluid must also be included. 

A heuristic argument has often been sketched, which builds on the purely entropic model put forward by Onsager \cite{onsager:effects} that only accounts for steric, excluded-volume interactions between the particles constituting the phase. It holds that at the interface particles would tend to lie parallel to the boundary of the droplet, as this would enhance their mutual sliding and so increase the entropy of the interface (a proper statistical model arriving at the same conclusion was offered in \cite{parsons:molecular}). Although there is experimental evidence showing that spherical droplets of (mostly thermotropic) liquid crystals may have normal as well as tangential anchoring at the interface \cite{candau:magnetic,volovik:topological,kurik:negative-positive}, since the earliest works \cite{chandrasekhar:surface,dubois-violette:emulsions} tactoids have been studied under the assumption that the nematic director $\n$ is tangent to the boundary. Actually, most studies have assumed an axisymmetric shape for tactoids with $\n$ along the meridians on their boundaries. In such \emph{bipolar} configurations, the poles are doubly singular, because both the surface normal and the nematic director have there \emph{defects}.

Williams~\cite{williams:nematic} made the first systematic attempt to find both the equilibrium shape of droplets subject to tangential surface anchoring and the equilibrium director field inside them.\footnote{In the special case where all elastic constants in the Oseen-Frank theory are equal.} In its most general formulation, the problem soon appeared formidable. Nonetheless, analytic estimates and numerical computations suggested that tactoids ``are difficult to observe, since very small drops and very low surface tension interface are required'' \cite[p.\,12]{williams:nematic}.\footnote{It is perhaps for this reason that in a subsequent paper Williams~\cite{williams:transitions} considered only bipolar spherical droplets.}

Such a disheartening conclusion did not deter further studies. Tactoids and their mathematical description have recently witnessed a surge of interest in a series of papers by several authors \cite{kaznacheev:nature,prinsen:shape,kaznacheev:influence,prinsen:continuous,prinsen:parity}. Despite a number of differences, they have one feature in common: being directly or indirectly influenced by the work of Williams~\cite{williams:nematic}, they adopt a special representation for both the droplet's shape and the nematic director that makes the saddle-splay constant of the Oseen-Frank energy, the most elusive to experimental detection, to feature as mere renormalization of the splay constant, thus playing a marginal role in the occurrence of tactoids.

What makes the saddle-splay constant elusive is its being related to a surface (elastic) energy. On this basis, one would expect it to be more effective in deciding how large the population of tactoids can be compared to other equilibrium shapes. This is precisely what this paper is about: to widen the class of admissible droplet's shapes and directors to identify the agents responsible for the growth and decay of tactoids' population. We shall how a  change in the class of shapes may alter considerably the whole scene.

Section~\ref{sec:shape_class} is devoted to the illustration of the class of shapes (and director fields) adopted in this paper. In particular, we realize that in our class there are both \emph{genuine} tactoids (those with pointed tips) and shapes that, although perfectly smooth, look very much like sharply pointed spindles. We introduce a full shape taxonomy that helps us navigate the configuration space. Not all admissible shapes are convex, but it is shown in Sec.~\ref{sec:optimal_shapes} that all optimal shapes are so. They can be tactoids or spheroids (or something in between), depending on the values of two dimensionless parameters, one related to the droplet's size and the other related to the saddle-splay constant. In Sec.~\ref{sec:polupations}, we collect the main results of this paper by examining the circumstances that determine the prevalence of one shape over the others. In Sec.~\ref{sec:conclusions}, we summarize our findings and see how Williams' pessimistic conclusion can be mellowed. The paper is closed by three mathematical appendices, where we illustrate the details of our development that in the main text could easily hamper the reader.

\section{Class of Shapes}\label{sec:shape_class}
We describe nematic liquid crystals within the classical theory, which features a \emph{director} $\n$ as the only mesoscopic descriptor of local molecular order. The spatial distortions of a director field $\n$ is measured by its gradient $\nabla\n$. The elastic energy-density $f_\mathrm{OF}$ associated with a director distortion is given by the celebrated Oseen-Frank formula
(see, e.g., \cite[Ch.\,3]{degennes:physics} or \cite[Ch.\,3]{virga:variational}),
\begin{equation}\label{eq:frank_energy}
f_\mathrm{OF} := \frac{1}{2}K_{11}(\diver\n)^{2} + \frac{1}{2}K_{22}(\n\cdot\curl\n)^{2} + \frac{1}{2}K_{33}|\n\times\curl\n|^{2} + K_{24}\big(\tr(\nabla\n)^{2}-(\diver\n)^{2}\big),
\end{equation}
where $K_{11}$, $K_{22}$, $K_{33}$, and $K_{24}$ are the \emph{splay}, \emph{twist}, \emph{bend}, and \emph{saddle-splay} elastic constants, respectively, each corresponding to a particular  elastic mode.\footnote{Recently, a different modal decomposition has been put forward for $f_\mathrm{OF}$ \cite{selinger:interpretation}, which has also been given a graphical representation in terms of an \emph{octupolar} tensor \cite{pedrini:liquid}. Such a novel decomposition, however, is not particularly germane to the topic at hand, and so here we shall stick to tradition.}

The saddle-splay term is a null Lagrangian \cite{ericksen:nilpotent} and an integration over the bulk reduces it to a surface energy. We shall see in the following that this is indeed a key feature in the role that $K_{24}$ plays in determining the population of different droplet shapes.

In the original works of Oseen~\cite{oseen:theory} and Frank~\cite{frank:theory}, $f_\mathrm{OF}$ in \eqref{eq:frank_energy} stems from requiring the elastic energy-density, which estimates the local distortional cost, to be a quadratic form of the measure of distortion that complies with frame-indifference and is invariant under the nematic symmetry (demanding $\n$ to be equivalent to $-\n$). To ensure that such a cost is never negative, the elastic constants in \eqref{eq:frank_energy} must satisfy Ericknsen's inequalities \cite{ericksen:inequalities},
\begin{equation}
\label{eq:ericksen_inequalities}
K_{11}\geqq K_{24}\geqq0,\quad K_{22}\geqq K_{24}\geqq0,\quad K_{33}\geqq0,
\end{equation} 
which will be taken as valid throughout this paper.

Here we study a free-boundary problem, where a given quantity of nematic liquid crystal occupying the volume $V_0$ (we treat liquid crystals as incompressible fluids) is surrounded by an isotropic fluid (which could well be its own melt) and can take on any desired shape. We shall call $\body$ the region in space occupied by the material at equilibrium; its shape will be the primary unknown of our problem.

The bulk elastic energy distributed over $\body$ must be supplemented by the interfacial energy concentrated on the boundary $\boundary$, where the liquid crystal comes in contact with the isotropic environment that surrounds it. There, an anisotropic surface tension $\ga$,depending on the orientation of $\n$ relative to the outer unit normal $\normal$ to $\boundary$ comes into play. Following \cite{prinsen:shape},we represent $\ga$ as
\begin{equation}
\label{eq:gamma_a}
\ga:=\gamma\big(1+\omega(\n\cdot\normal)^2\big),
\end{equation}
where $\gamma>0$ is the \emph{isotropic} surface tension and $\omega$ is a dimensionless \emph{anchoring strength}, which we take to satisfy $\omega\geqq0$. Actually, \eqref{eq:gamma_a} is far more general than needed, as here we shall assume that $\n$ obeys the \emph{degenerate} boundary condition on $\boundary$, which requires
\begin{equation}
\label{eq:degenerate_boundary_condition}
\n\cdot\normal\equiv0.
\end{equation}
As also shown in \cite{prinsen:shape}, for sufficiently small droplets, this assumption is untenable, as $\n$ tends to be uniform throughout $\body$, making \eqref{eq:degenerate_boundary_condition} impossible (see also \cite[Ch.\,5]{virga:variational}). Our development  will be based on assumption \eqref{eq:degenerate_boundary_condition}; although $\omega$ will never feature explicitly below, it must be taken to be sufficiently large so as to make \eqref{eq:degenerate_boundary_condition} valid.\footnote{A more precise estimate that also involves the droplet's size will be presented in Sec.~\ref{sec:admissible}.}

Equilibrium is attained whenever the total free energy is minimized, that is, whenever $\body$ minimizes the shape functional
\begin{equation}
\label{eq:free_energy_functional}
\free[\body]:=\int_{\body} f_\mathrm{OF}\dd V+\gamma A(\boundary),
\end{equation}
subject to the constraint
\begin{equation}
\label{eq:volume_constraint}
V(\body)=V_0,
\end{equation}
where $f_\mathrm{OF}$ is as in \eqref{eq:frank_energy}, $A$ is the area measure, and $V$ is the volume measure. Minimizers of $\free$ will be sought for in a special class of shapes and director fields, which we now describe in detail.

\subsection{Shape representation and director retraction}\label{sec:retraction}
\begin{figure}[h!] 
	\includegraphics[width=.25\linewidth]{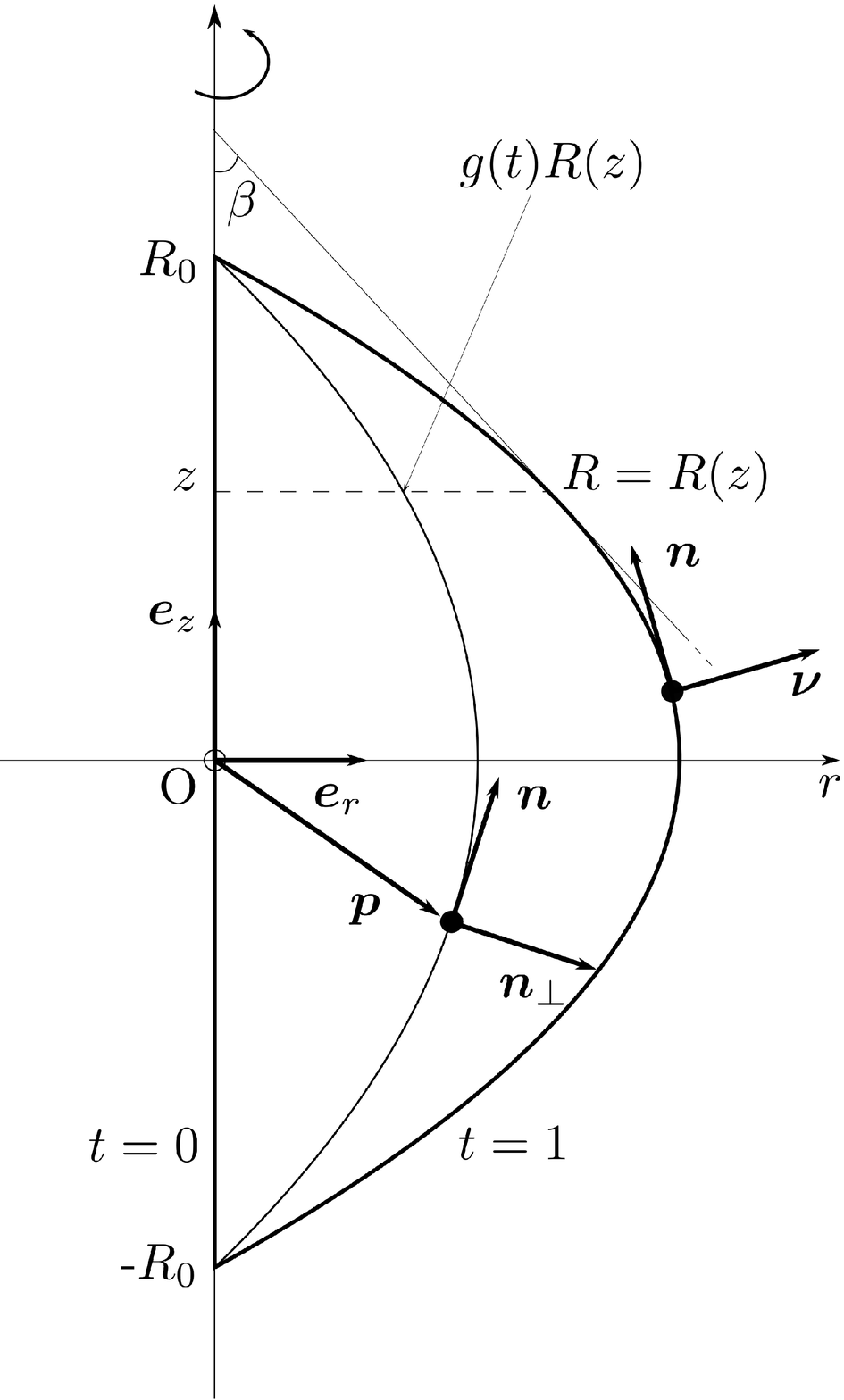}
	\caption{Cross-section of the drop with a meridian plane. The function $R(z)$ represents the boundary $\boundary$, while the function $R_t(z)=g(t)R(z)$ is the retraction of $R(z)$ described in the text. The director field $\n$ is everywhere tangent to the retracted curves; $\nper=\e_\vt\times\n$ is the orthogonal field that agrees with the outer unit normal $\normal$ on $\boundary$. The tangent to $R(z)$ for $z\geqq0$ makes the angle $\beta$ with the $z$-axis; it is instrumental to the definition of the tactoidal measure $\tau$ illustrated in Appendix~\ref{sec:tau}.}
	\label{fig:shape_cross_section}
\end{figure}

We shall represent $\body$ as a region in three-dimensional space axisymmetric  about the $z$-axis of a standard cylindrical frame $(\e_r, \e_\vartheta, \e_z)$ and mirror-symmetric relative to the equatorial plane $(\e_r,\e_\vt)$. As shown in Fig.~\ref{fig:shape_cross_section}, the boundary $\boundary$ is obtained by rotating  the graph of a function of class $C^1$, 
$R=R(z)$, which  describes the radius of the drop's cross-section at height $z\in\left[-R_0,R_0\right]$. $R$ is taken to an even function, 
\begin{equation}
\label{eq:R_even}
R(z)=R(-z),\quad z\in[-R_0,R_0]. 
\end{equation}
The points on the $z$-axis at  $z=\pm R_0$, where $R$ vanishes, are the \emph{poles} of the drop. On the equator, which falls at $z=0$,  smoothness and symmetry require that $R'(0)=0$, where a prime denotes differentiation.

Whenever $R'(R_0)$ is finite, the shape $\body$ has pointed poles, it is a \emph{tactoid}, which we shall call \emph{genuine} to distinguish it from similar elongated shapes with a smooth boundary. On the other hand, whenever $R'$ is unbounded at the end-points of the interval $[-R_0,R_0]$, $\body$ has a smooth boundary; as shown below, its shape may appear in different forms and guises. Figure~\ref{fig:shape_cross_section} depicts the cross-section  with a meridian plane (say, at $\vt=0$) of a region $\body$ in the class we are considering; the full shape is generated by a $2\pi$-rotation around the $z$-axis. The director field $\n$ on $\boundary$ is taken to be oriented along the meridians. This is an additional hypothesis, also made in \cite{prinsen:shape,prinsen:continuous}, which is justified by assuming the twist constant $K_{22}$ sufficiently larger than the others: an axisymmetric  director field with no component out of the meridian plane is indeed twist-free.
For any given $\vt$,  the boundary curve is represented by the position vector (issued from the center of symmetry of $\body$)
\begin{equation}
\label{eq:curve_p_1}
\bm{p}_1(\vt,z):=R(z)\e_r+z\e_z,\quad-R_0\leqq z\leqq R_0,
\end{equation}
which  is \emph{retracted} inside $\body$ as the  curve
\begin{equation}
\label{eq:curve_p_t}
\bm{p}(t,\vt,z):=g(t)R(z)\e_r+z\e_z, \quad-R_0\leqq z\leqq R_0,
\end{equation}
where $t$ is the \emph{retraction parameter} ranging in $[0,1]$ and $g$ is any function of class $C^1$ strictly increasing on $[0,1]$ and such that $g(0)=0$ and $g(1)=1$ (an example would be $g(t)=t$). Clearly, for $t=1$, $\bm{p}(t,\vt,z)$ reduces to $\bm{p}_1(\vt,z)$ in \eqref{eq:curve_p_1}, whereas for $t=0$ it describes the polar axis $\bm{p}_0(z)=z\e_z$ (see Fig.~\ref{fig:shape_cross_section}). The family of retracted curves fill the whole of $\body$ by letting $\vt$ vary in $[0,2\pi)$. Thus, $(t,\vt,z)\in[0,1]\times[0,2\pi)\times[-R_0,R_0]$ are a new set of \emph{retracted} coordinates for the region $\body$. 

We not only retract the boundary $\boundary$ by letting $t$ vary in $[0,1]$, we also retract the meridian director field on $\boundary$, which is thus defined in the whole of $\body$ as the unit vector field tangent to the lines with given $(t,\vt)$ and varying $z$ (see Fig.~\ref{fig:shape_cross_section}). The director field produced with such a geometric construction possesses two point defects at the poles; they are two \emph{boojums} with equal topological  charge $m=+1$. By differentiating $\p$ in \eqref{eq:curve_p_t} with respect to $z$, keeping $(t,\vt)$ fixed, we easily obtain
\begin{eqnarray}
\label{eq:n}
\n=\frac{gR'\e_r+\e_z}{\sqrt{1+(gR')^2}}.
\end{eqnarray}
Letting $\nper$ be the unit vector orthogonal to $\n$ in the meridian plane, oriented so as to coincide with the outer unit normal $\normal$ on $\boundary$, we see form \eqref{eq:n} that 
\begin{equation}
\label{eq:n_perp}
\nper=\frac{\e_r-gR'\e_z}{\sqrt{1+(gR')^2}}.
\end{equation}
A positively oriented orthonormal frame $\framen$ is then obtained by appending the unit vector $\e_\vt=\n\times\nper$, everywhere orthogonal to the local meridian plane. This frame, however, is not the frame associated with the new coordinates $(t,\vt,z)$, as we now proceed to show.

Imagine a smooth curve in $\body$ parameterized as $\xi\mapsto(t(\xi),\vt(\xi),z(\xi))$. It follows from \eqref{eq:curve_p_t} that 
\begin{equation}
\label{eq:dot_curve_t}
\dot{\p}=(g'R\dot{t}+gR'\dot{z})\e_r+gR\dot{\vt}\e_\vt+\dot{z}\e_z,
\end{equation}
where a superimposed dot denotes differentiation with respect to the parameter $\xi$.\footnote{With a slight abuse of language, here a prime denotes differentiation both with respect to $t$ (in $g'$) and with respect to $z$ (in $R'$). No confusion should arise since, apart from $\p$, no other instance will occur of a function depending on both $(t,z)$.} A glance at \eqref{eq:n} suffices to show that \eqref{eq:dot_curve_t} can also be rewritten as
\begin{equation}\label{eq:dot_curve_t_rewritten}
\dot{\bm{p}}=g'R\dot{t}\e_r+gR\dot{\vt} \e_\vt+\sqrt{1+(gR')^2}\dot{z}\n,
\end{equation}
showing that $\framer$ is the (non-orthogonal) frame associated with the retracted coordinates $(t,\vt,z)$.

Equation \eqref{eq:dot_curve_t_rewritten} is especially expedient to derive the elementary volume $\dd V$ of $\body$ and the elementary area $\dd A_t$ for the retracted boundary $\boundary_t$ in $\body$. For the former we have that 
\begin{equation}
\label{eq:dV}
\dd V=\dd t\dd\vt\dd z(g'R)(gR)\sqrt{1+(gR')^2}\,\e_r\cdot\e_\vt\times\n =gg'R^2\dd t\dd \vt\dd z,
\end{equation}
where use has also been made of \eqref{eq:n}, while for the latter, 
\begin{equation}
\label{eq:dA}
\dd A_t=\dd\vt \dd z (gR)\sqrt{1+(gR')^2}\,\e_\vt\times\n\cdot\nper=gR\sqrt{1+(gR')^2}\dd\vt\dd z.
\end{equation}
Thus, the volume of a droplet $\body$ will be expressed in terms of the function $R$ as
\begin{equation}
\label{eq:volume}
V(\body)=\int_{0}^{1}gg'\dd t \int_0^{2\pi}\dd\vt\int_{-R_0}^{R_0}R^2\dd z   =\pi\int_{-R_0}^{R_0}R^2\dd z=V_0,
\end{equation}
and the area of the boundary $\boundary$ as
\begin{equation}
\label{eq:area}
A(\boundary)=2\pi\int_{-R_0}^{R_0} R\sqrt{1+R'^2}\dd z,
\end{equation}
which follows from \eqref{eq:dA} for $t=1$.

It is shown in Appendix~\ref{sec:retracted_coordinates}
how to derive from \eqref{eq:curve_p_t} the form taken by $\nabla\n$ in the orthonormal frame $\framen$; this reads as
\begin{equation}
\label{eq:nabla_n}
\nabla\n=\frac{gR''}{\big(1+(gR')^2\big)^{3/2}}\nper\otimes\n+
+\left(\frac{R'}{R}\frac{1}{\sqrt{1+(gR')^2}}-\frac{g^2R'R''}{\big(1+(gR')^2\big)^{3/2}}\right)\nper\otimes\nper+
\frac{R'}{R}\dfrac{1}{\sqrt{1+(gR')^2}}\e_\vt\otimes\e_\vt.
\end{equation}
The following expressions for the traditional measures of distortion are easy consequences of \eqref{eq:nabla_n}:
\begin{subequations}\label{eq:distortion_measures}
\begin{align}
\diver\n&=\frac{R'}{\sqrt{1+(gR')^2}}\left(\frac{2}{R}-\frac{g^2R''}{1+(gR')^2}\right),\label{eq:div}\\
\curl\n&=\frac{gR''}{\big(1+(gR')^2\big)^{3/2}}\e_\vt,\label{eq:curl}\\
\n\cdot\curl\n&=0,\label{eq:twist}\\
\n \times \curl\n&=-\frac{gR''}{\big(1+(gR')^2\big)^{3/2}}\nper,\label{eq:bend}\\
\tr(\nabla\n)^2-(\diver\n)^2&=-\frac{2R'^2}{R\big(1+(gR')^2\big)}\left(\frac{1}{R}-\frac{g^2R''}{1+(gR')^2}\right).\label{eq:saddle_splay}
\end{align}	
\end{subequations} 
In particular, \eqref{eq:twist} shows that, as expected, no twist is associated with the class of retracted meridian fields that we are considering. 

Inserting \eqref{eq:distortion_measures} in \eqref{eq:frank_energy}, we arrive at 
\begin{equation}
\label{eq:frank_energy_reduced}
f_\mathrm{OF}=\frac{1}{2}K_{11}\frac{g^4R'^2R''^2}{\big(1+(gR')^2\big)^3}
+(K_{11}-K_{24})\frac{2R'^2}{R\big(1+(gR')^2\big)}\left(\frac{1}{R}-\frac{g^2R''}{1+(gR')^2}\right)
+\frac{1}{2}K_{33}\dfrac{g^2R''^2}{\big(1+(gR')^2\big)^3},
\end{equation}
which shows how in our setting, at variance with \cite{prinsen:shape,prinsen:parity,prinsen:continuous,kaznacheev:nature,kaznacheev:influence}, the saddle-splay constant does not merely renormalize the  splay constant. For given $g$, the function $R$ represents here both the shape $\body$ of a droplet and the nematic director field inside it.

Before building upon \eqref{eq:free_energy_functional} and \eqref{eq:frank_energy_reduced} the free-energy functional that we shall study in the following, we find it useful to rescale all lengths to the one dictated by the volume constraint. We call $\Req$ the radius of the \emph{equivalent} sphere, which has volume $V_0$, and we rescale to $\Req$ both $z$ and $R(z)$, keeping their names unaltered.\footnote{An abuse of notation that we hope  the reader will tolerate.}  Letting 
\begin{equation}\label{eq:mu_definition}
\mu:=\frac{R_0}{\Req}, 
\end{equation}
by use of \eqref{eq:free_energy_functional}, \eqref{eq:frank_energy_reduced}, \eqref{eq:dV}, and \eqref{eq:area}, we arrive at the following reduced functional, $F[\mu;R]$, which is an appropriate dimensionless form of $\free$,
\begin{align}
\label{eq:F_energy_functional} 
F[\mu;R]:=\frac{\free[\body]}{2\pi K_{11}\Req}&=\int_0^1\dd t \int_{-\mu}^{\mu}gg' R^2\left[\frac{1}{2}\frac{g^4R'^2R''^2}{\big(1+(gR')^2\big)^3}+
\frac{2(1-k_{24})R'^2}{R\big(1+(gR')^2\big)}\left(\frac{1}{R}-\frac{g^2R''}{1+(gR')^2}\right)+
\frac{k_3}{2}\frac{g^2R''^2}{\big(1+(gR')^2\big)^3}\right]\dd z \nonumber\\
&+\alpha\int_{-\mu}^{\mu}R\sqrt{1+R'^2}\dd z\nonumber\\
&=\int_{-\mu}^{\mu}  \bigg\{\left[\frac{R^2R''^2}{4R'^2}\left(\frac{\ln(1+R'^2)}{R'^2}-\frac{1}{(1+R'^2)^2}\right)+(1-k_{24})\left[\left(1-\frac{RR''}{R'^2}\right)\right]\ln(1+R'^2) 
+\frac{RR''}{1+R'^2}\right]\nonumber\\
&\qquad\qquad+(k_3-3)\frac{R^2R''^2}{8(1+R'^2)^2}\bigg\}\dd z+\alpha\int_{-\mu}^{\mu}R\sqrt{1+R'^2} \dd z,
\end{align}
where the integration in $t$ is shown to be independent of the specific function $g$, provided it is monotonic and obeys the prescribed boundary conditions.
The following scaled elastic constants have been introduced in \eqref{eq:F_energy_functional},
\begin{equation}
\label{eq:elastic_constants_rescaled}
k_3:=\frac{K_{33}}{K_{11}},\quad k_{24}:=\dfrac{K_{24}}{K_{11}},
\end{equation}
the former is non-negative and the latter is subject to $0\leqq k_{24}\leqq1$, as a consequence of \eqref{eq:ericksen_inequalities}. Moreover,
\begin{equation}
\label{eq:alpha}
\alpha:=\frac{\gamma\Req}{K_{11}}
\end{equation}
is a reduced (dimensionless) volume.\footnote{When we say that a drop is either small or large, we mean precisely that either $\alpha\ll1$ or $\alpha\gg1$, respectively.\label{foot:alpha}} 

The variational problem that we thus face can be phrased as follows: Find a positive $\mu$ and a smooth, even function $R$ that obeys
\begin{equation}
\label{eq:R_end_points}
R(-\mu)=R(\mu)=0
\end{equation}
so as to minimize $F$ subject to the isoperimetric constraint \eqref{eq:volume}, which in the scaled variables reads simply as
\begin{equation}
\label{eq:volume_rescaled}
\int_{-\mu}^{\mu} R^2(z) \dd z=\dfrac{4}{3}.
\end{equation}

\subsection{Special Family of Shapes}\label{sec:special}
The variational problem just stated is rather challenging, especially if we wish to discuss the role played by the constitutive parameters $k_3$ and $k_{24}$ and by the reduced volume $\alpha$ in the population of minimizing shapes. Instead of embarking in a thorough numerical minimization of $F$, we rather resort to a special family of shapes described by a small number of parameters. We shall take the function $R$ in the special form\footnote{For $b=0$, $R$ in \eqref{eq:profile} reduces to the parabolic profile considered in \cite{williams:nematic}.}  
\begin{equation}
\label{eq:profile}
R(z)=a(\mu^2-z^2)+b\sqrt{\mu^2-z^2}.
\end{equation}

We now illustrates and classify the relatively large variety of shapes that can be represented through \eqref{eq:profile}. We begin by considering the constraints that the parameters $(a,b,\mu)$ are subject to.

First, $R(z)$ must be non-negative for all $-\mu\leqq z\leqq\mu$. It a simple matter to check that this requirement is equivalent to the inequalities
\begin{equation}
\label{eq:R_positivity}
b\geqq-\mu a\quad\text{and}\quad b^2\geqq(\mu a)^2\quad\text{or}\quad ab>0,
\end{equation}
which can be represented by letting
\begin{equation}
\label{eq:R_positivity_representation}
\mu a=\rho\cos\phi\quad\text{and}\quad b=\rho\sin\phi,\quad\text{with}\quad\rho>0\quad\text{and}\quad0\leqq\phi\leqq\frac{3\pi}{4}.
\end{equation}
Second, the isoperimetric constraint \eqref{eq:volume_rescaled} requires that 
\begin{equation}
\label{eq:volume_constraint_rescaled}
\mu^3\left(b^2+\frac{9\pi}{16}(\mu a)b+\frac45(\mu a)^2\right)=1.
\end{equation}

Inserting \eqref{eq:R_positivity_representation} into \eqref{eq:volume_constraint_rescaled}, we obtain $\rho$ and conclude that all admissible values of $a$ and $b$ are represented by
\begin{equation}
\label{eq:a_b_representation}
a=\frac{1}{\mu^{5/2}}\frac{\cos\phi}{\sqrt{h(\phi)}},\quad b=\frac{1}{\mu^{3/2}}\frac{\sin\phi}{\sqrt{h(\phi)}}\quad\text{with}\quad h(\phi):=\sin^2\phi+\frac{9\pi}{16}\sin\phi\cos\phi+\frac45\cos^2\phi>0,\quad 0\leqq\phi\leqq\frac{3\pi}{4}.
\end{equation}
Thus, $(\phi,\mu)$ are the only independent parameters that describe all admissible shapes in the special class \eqref{eq:profile}. We now explore the qualitative features of these shapes, corresponding to different regions in configuration space $\mathsf{S}:=\{(\phi,\mu): 0\leqq\phi\leqq\frac{3\pi}{4},\ \mu>0\}$.

\begin{figure}[h] 
	\includegraphics[width=.3\linewidth]{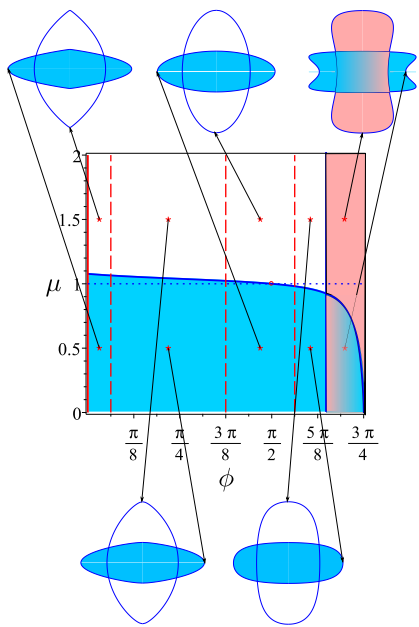}
	\caption{Configuration space with all admissible shapes described by \eqref{eq:profile}. The blue region below the graph of the function $\varpi(\phi)$ in \eqref{eq:prolate_curve} represents all prolate shapes. The pink vertical strip for $\phi_\mathrm{c}\leqq\phi\leqq\frac{3\pi}{4}$ represents the concave shapes that we have called \emph{dumbbells}; all shapes falling on the left of this strip are convex. The sphere is represented by the point $(\frac\pi2,1)$, marked by a red circle. According to the taxonomy introduced in Sec.~\ref{sec:taxonomy}, we also call \emph{tactoids} the shapes for $0\leqq\phi\leqq\frac{\pi}{16}$ (\emph{genuine} tactoids, only those for $\phi=0$, marked by a red line), \emph{bumped spheroids} those for $\frac{\pi}{16}\leqq\phi\leqq\frac{6\pi}{16}$, simply \emph{spheroids} those for $\frac{6\pi}{16}\leqq\phi\leqq\frac{9\pi}{16}$, and \emph{barrels} those for $\frac{9\pi}{16}\leqq\phi\leqq\phi_\mathrm{c}$, see also Table~\ref{tab:taxonomy}, and  Fig.~\ref{fig:gallery} for a fuller gallery of shapes. The barriers marking transitions from one family of shapes to another are represented by vertical dashed lines.}
	\label{fig:configuration_space}
\end{figure}

We first distinguish \emph{prolate} from \emph{oblate} shapes, the former are characterized by having height larger than width, that is, by the inequality $\mu\geqq R(0)$, which by \eqref{eq:profile} and \eqref{eq:a_b_representation} becomes
\begin{equation}
\label{eq:prolate_curve}
\mu\geqq\varpi(\phi):=\sqrt[3]{\frac{(\cos\phi+\sin\phi)^2}{h(\phi)}}.
\end{equation}  
The graph of $\varpi(\phi)$ is shown in Fig.~\ref{fig:configuration_space}: all shapes above it are prolate, all shapes below it are oblate. The round sphere, corresponding to the point to $(\frac\pi2,1)$, falls on the graph of $\varpi$ (it is denoted by a circle in Fig.~\ref{fig:configuration_space}).

We distinguish convex from concave shapes. The latter arise whenever $R'$ has an extra root in $-\mu\leqq z\leqq\mu$, besides $z=0$. It is easily seen that such an extra root requires that 
\begin{equation}
\label{eq:concavity_condition}
a<0\quad\text{and}\quad b<2\mu|a|.
\end{equation}
By \eqref{eq:a_b_representation}, these inequalities reduce to $\phi>\phi_\mathrm{c}:=\arccot\left(-\frac12\right)\doteq2.03\,\rad$. Thus the pink strip in $\mathsf{S}$ depicted in Fig.~\ref{fig:configuration_space} is where we find all concave shapes represented by \eqref{eq:profile}. The corresponding three-dimensional droplets $\body$ are axisymmetric \emph{dumbbells}, with a neck that narrows as $\phi$ approaches the boundary of $\mathsf{S}$ at $\frac{3\pi}{4}$, where it vanishes altogether and the droplet is severed. The strip of dumbbells is also traversed by the graph of $\varpi(\phi)$ (see Fig.~\ref{fig:configuration_space}), which means that some dumbbells are prolate (if they fall above the graph of $\varpi$), while others are oblate (if they fall below the graph of $\varpi$), although here this simply means that their height is either larger or smaller than their neck.

\subsection{Droplet Taxonomy}\label{sec:taxonomy}
Strictly speaking, as already remarked above, a three-dimensional shape $\body$ represented by \eqref{eq:profile} has pointed tips at the poles only for $\phi=0$, which according to \eqref{eq:a_b_representation} is the only value of $\phi$ that makes $b$ vanish. But, are we sure that for small enough values of $\phi$ the shape represented by \eqref{eq:profile} via \eqref{eq:a_b_representation} can be visually distinguished from a tactoid (in accord with the etymology of the word recalled in the Introduction)? The answer to this question is vital to our ``demographic'' quest. If we want to know how tactoids feature in the whole droplet population, we need to have a clear criterion to classify as tactoids also  those shapes which may not have pointed tips, but look like they have.

In Appendix~\ref{sec:tau}, we build a  quantitative criterion on a certain qualitative observation. There, we arrive at a \emph{tactoidal} measure, which here translates into a conventional classification rule. We propose to   call simply tactoids all shapes represented by the strip $0\leqq\phi\leqq\frac{\pi}{16}$ in  configuration space $\mathsf{S}$. Other strips are conventionally identified in $\mathsf{S}$, which describe other shape variants. Our full taxonomy is summarized in Table~\ref{tab:taxonomy} below.
\begin{table}[h]
	\caption{We identify five strips in configuration space $\mathsf{S}$, which correspond to five qualitatively different shapes for a droplet $\body$ represented by \eqref{eq:profile} via \eqref{eq:a_b_representation}. The names given below are somewhat self-explanatory; a visual illustration is provided by the gallery of shapes drawn  in Fig.~\ref{fig:gallery} for $\mu=1$. The transition shapes, which somehow belong to two adjacent classes, are characterized by the following values of $\phi$: $\frac{\pi}{16}$, $\frac{6\pi}{16}$, $\frac{9\pi}{16}$, and $\phi_\mathrm{c}=\arccot\left(-\frac12\right)\doteq2.03\,\rad$.}
\begin{ruledtabular}
	\begin{tabular}{ccccc}
		Tactoids&Bumped Spheroids&Spheroids&Barrels&Dumbbells\\
		\hline\\
		$0\leqq\phi\leqq\dfrac{\pi}{16}$&$\dfrac{\pi}{16}\leqq\phi\leqq\dfrac{6\pi}{16}$&$\dfrac{6\pi}{16}\leqq\phi\leqq\dfrac{9\pi}{16}$&$\dfrac{9\pi}{16}\leqq\phi\leqq\phi_\mathrm{c}$&$\phi_\mathrm{c}\leqq\phi\leqq\dfrac{3\pi}{4}$\\
		\\
	\end{tabular}
\end{ruledtabular}
\label{tab:taxonomy}
\end{table}

In this section, we shall be contented with illustrating our taxonomic criterion by drawing shapes for which $\mu=1$. We have two good reasons to do so.\footnote{One would actually suffice.} First, we have drawn a number of shapes for very different values of $\mu$ and always found our criterion qualitatively accurate. Second, as will be clear in Sec.~\ref{sec:optimal_shapes} below, the equilibrium shapes that minimize the free-energy functional never fall too far away from $\mu=1$.

\begin{figure}[h]
	\centering
	\begin{subfigure}[c]{0.19\linewidth}
		\centering
		\includegraphics[width=\linewidth]{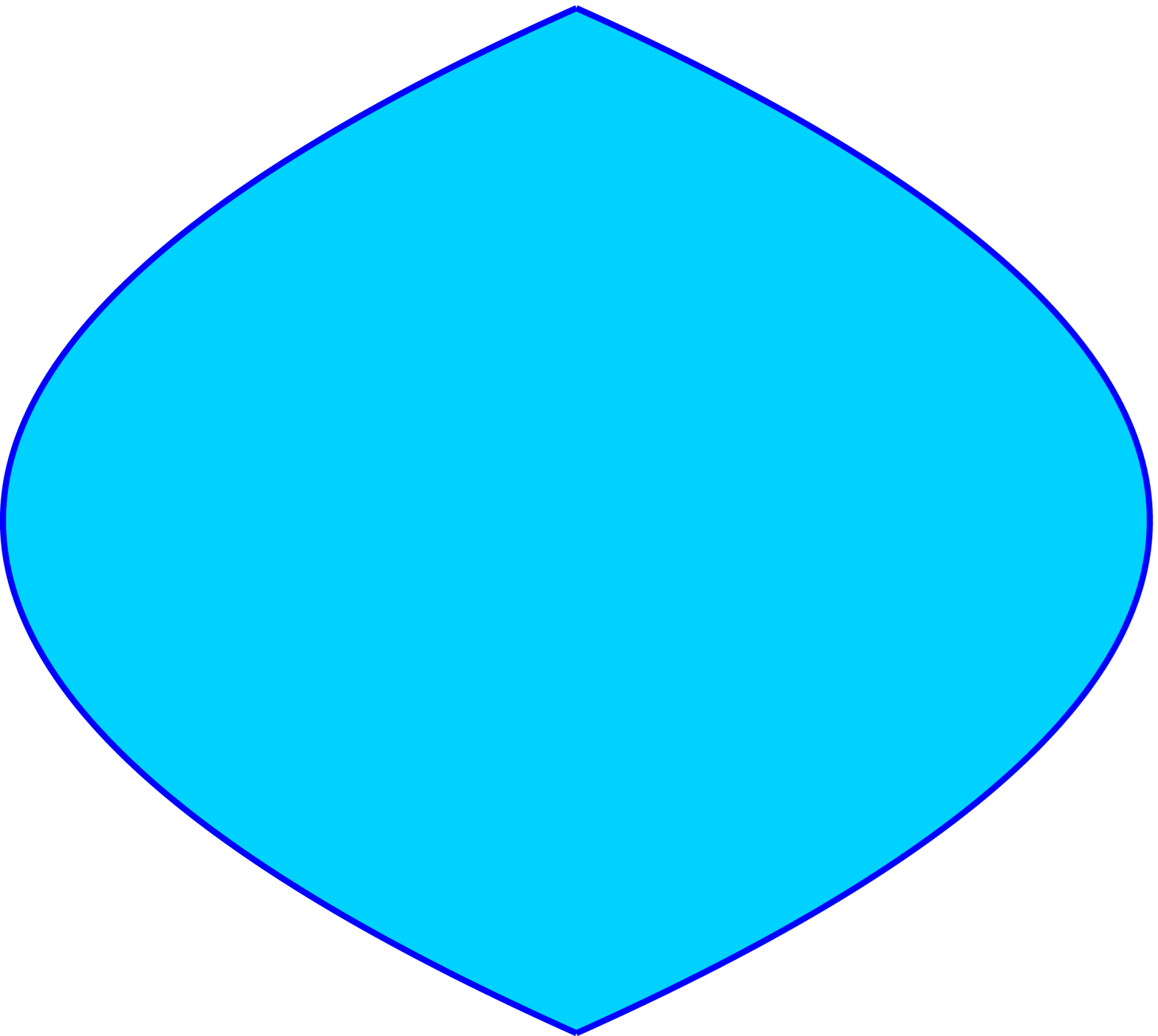}
		\caption{$\phi=0$\\genuine tactoid}
		\label{fig:gallery_0Pio16mu1}
	\end{subfigure}
	\begin{subfigure}[c]{0.19\linewidth}
		\centering
		\includegraphics[width=\linewidth]{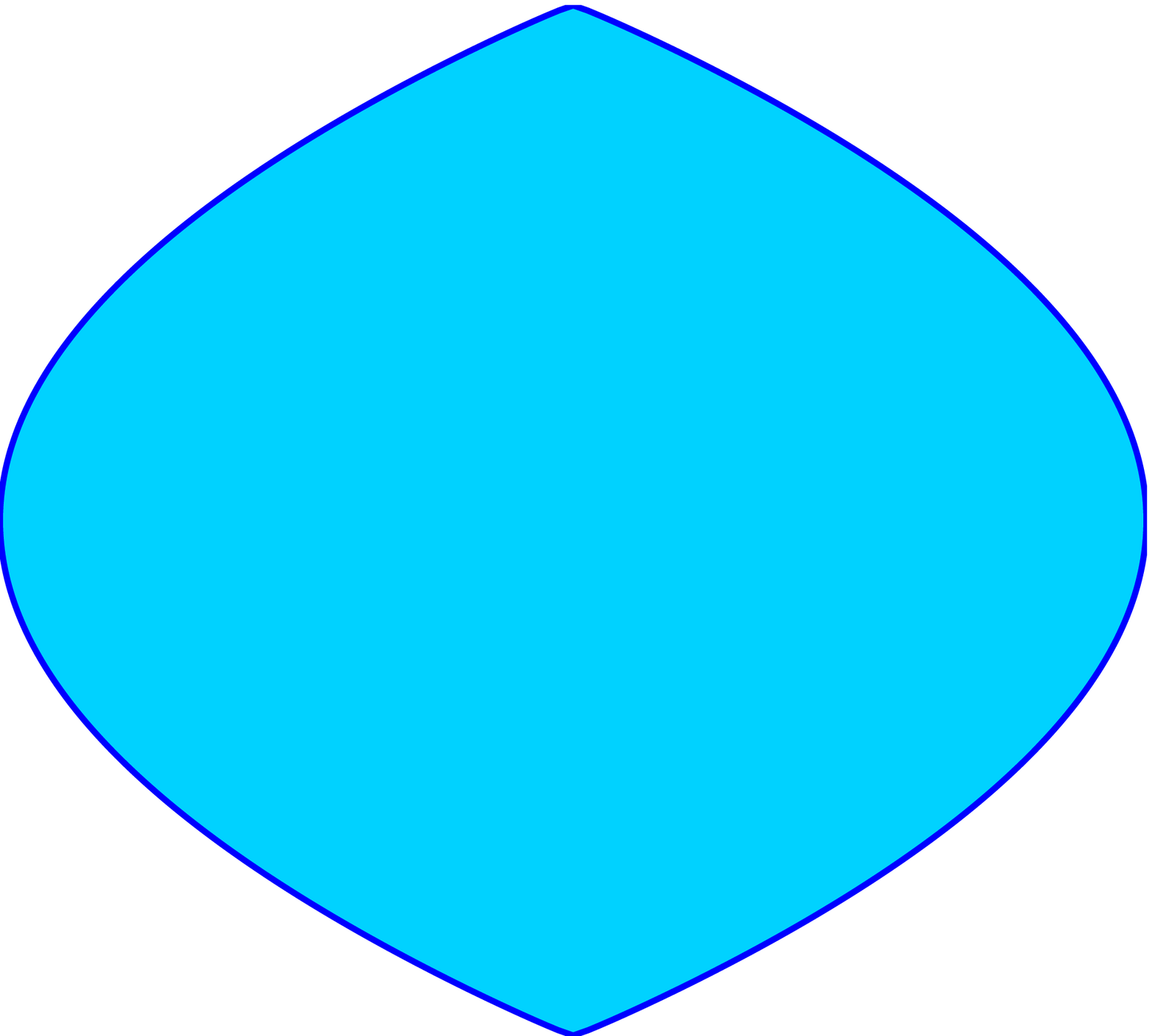}
		\caption{$\phi=\frac{\pi}{64}$\\tactoid}
		\label{fig:gallery_Pio64mu1}
	\end{subfigure}
	\begin{subfigure}[c]{0.19\linewidth}
	\centering
	\includegraphics[width=\linewidth]{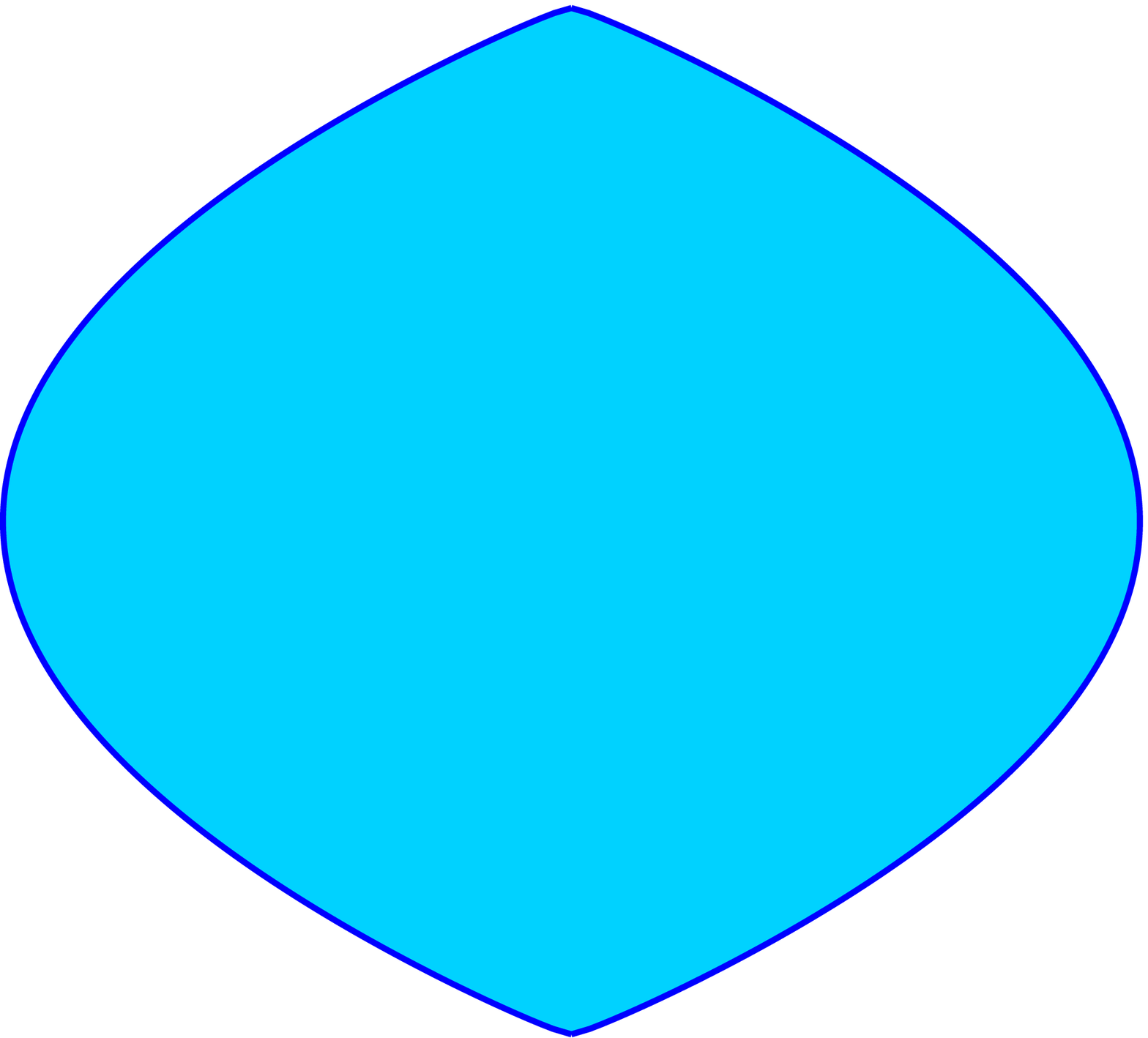}
	\caption{$\phi=\frac{\pi}{32}$\\tactoid}
	\label{fig:gallery_Pio32mu1}
\end{subfigure}
	\begin{subfigure}[c]{0.19\linewidth}
		\centering
		\includegraphics[width=\linewidth]{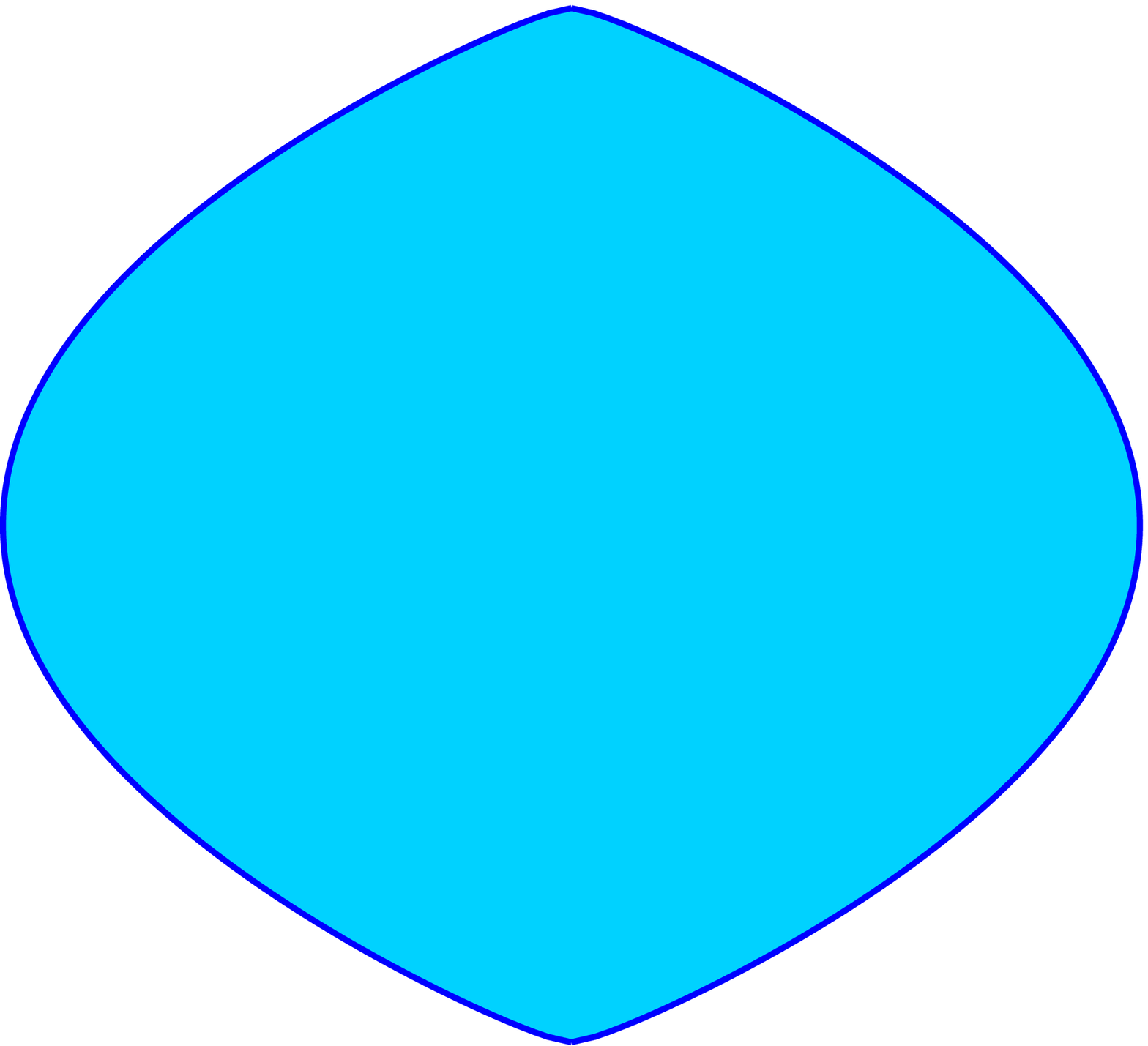}
		\caption{$\phi=\frac{\pi}{16}$\\transition shape}
		\label{fig:gallery_Pio16mu1}
	\end{subfigure}
	\begin{subfigure}[c]{0.19\linewidth}
		\centering
		\includegraphics[width=\linewidth]{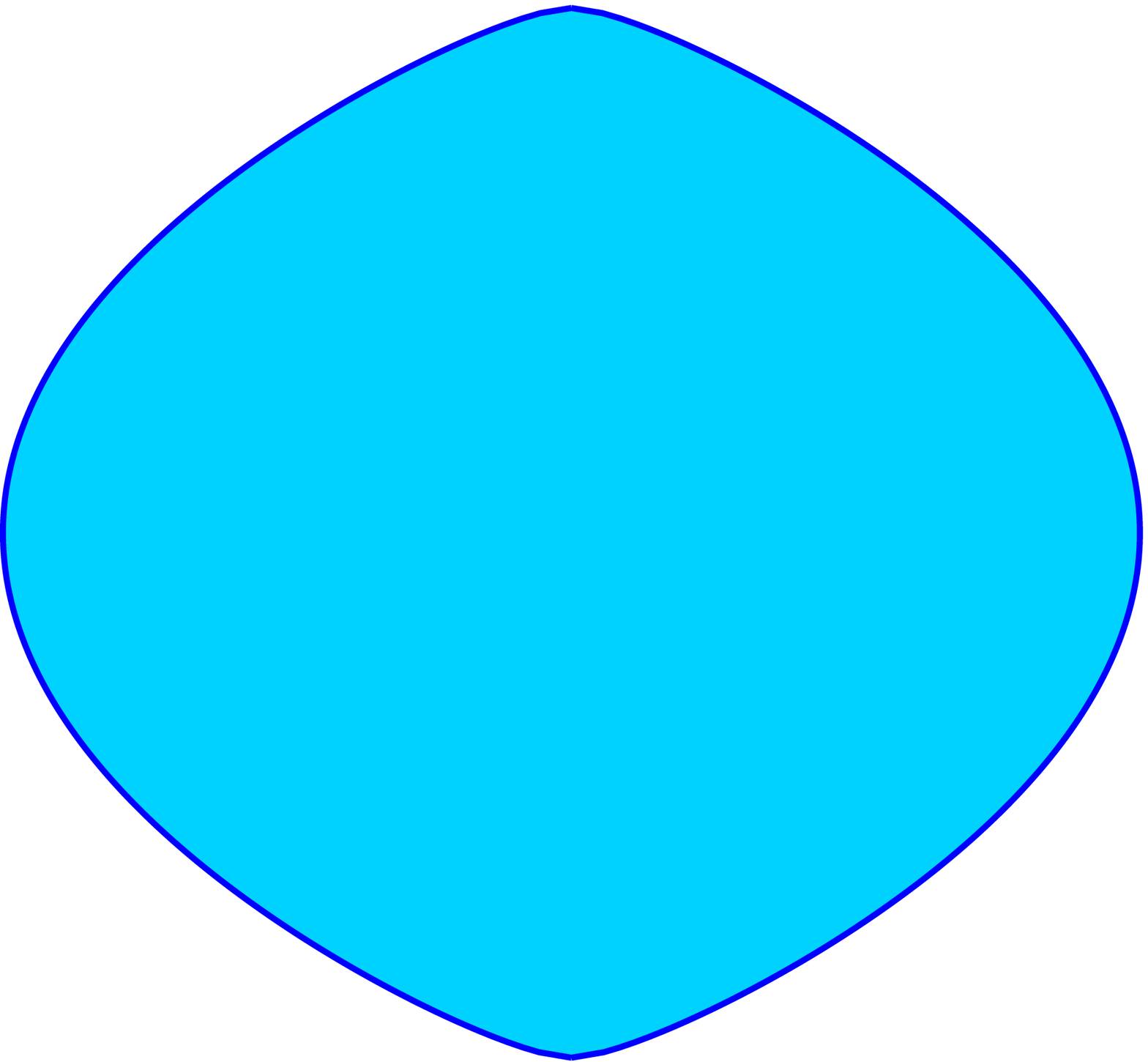}
		\caption{$\phi=\frac{2\pi}{16}$\\bumped spheroid}
		\label{fig:gallery_2Pio16mu1}
	\end{subfigure}
	\begin{subfigure}[c]{0.19\linewidth}
		\centering
		\includegraphics[width=\linewidth]{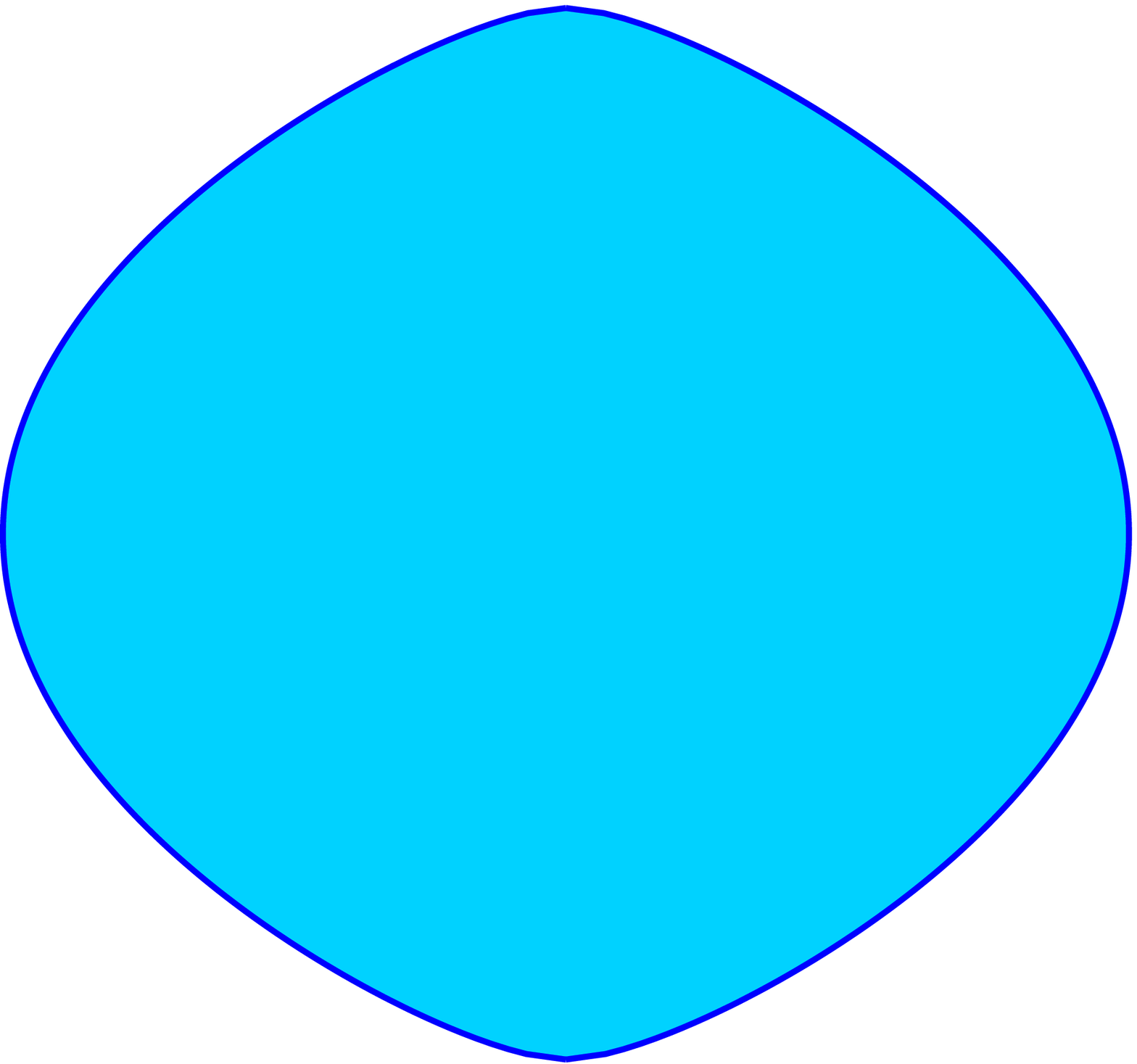}
		\caption{$\phi=\frac{3\pi}{16}$\\bumped spheroid}
		\label{fig:gallery_3Pio16mu1}
	\end{subfigure}
	\begin{subfigure}[c]{0.19\linewidth}
		\centering
		\includegraphics[width=\linewidth]{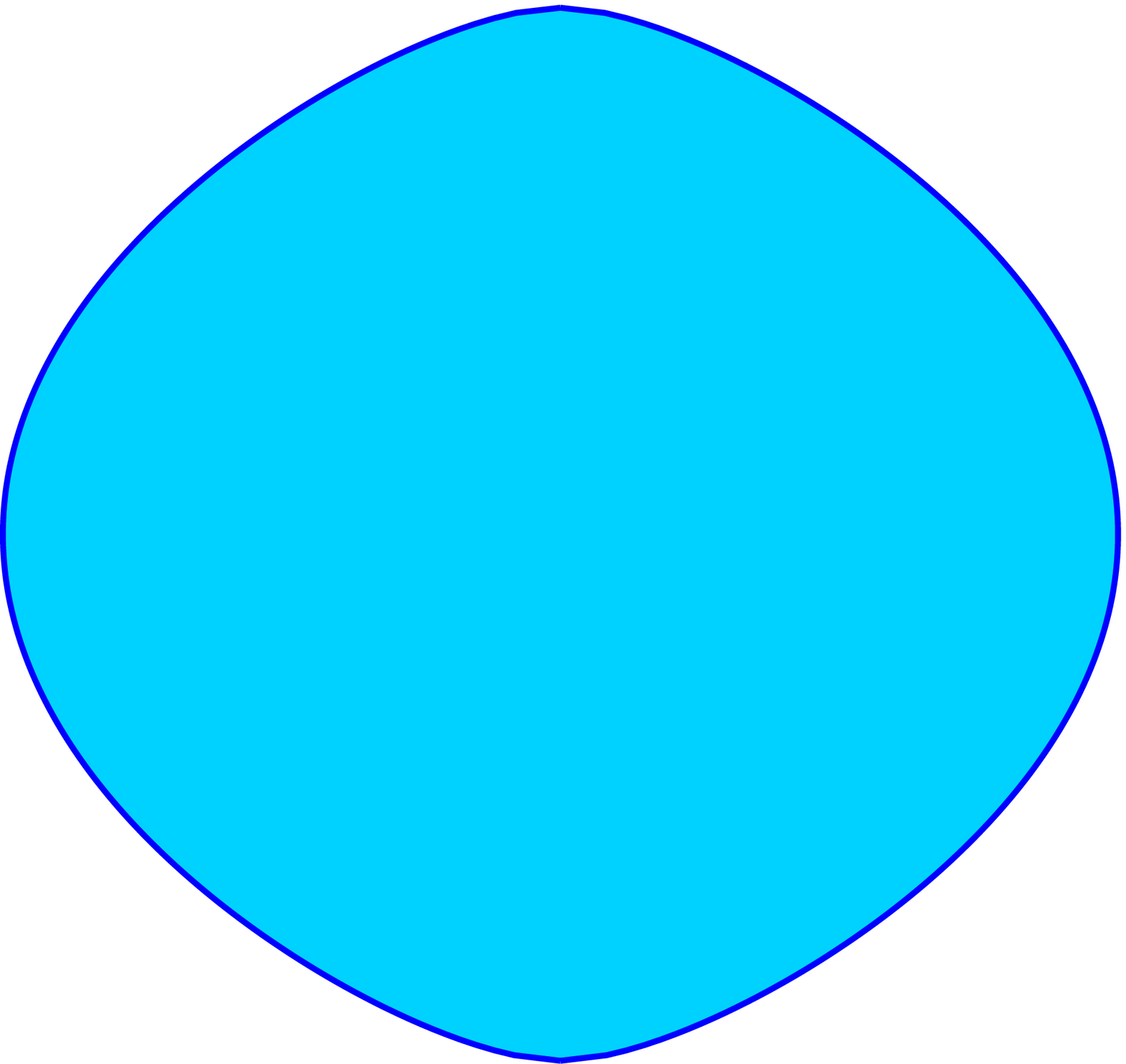}
		\caption{$\phi=\frac{4\pi}{16}$\\bumped spheroid}
		\label{fig:gallery_4Pio16mu1}
	\end{subfigure}
	\begin{subfigure}[c]{0.19\linewidth}
		\centering
		\includegraphics[width=\linewidth]{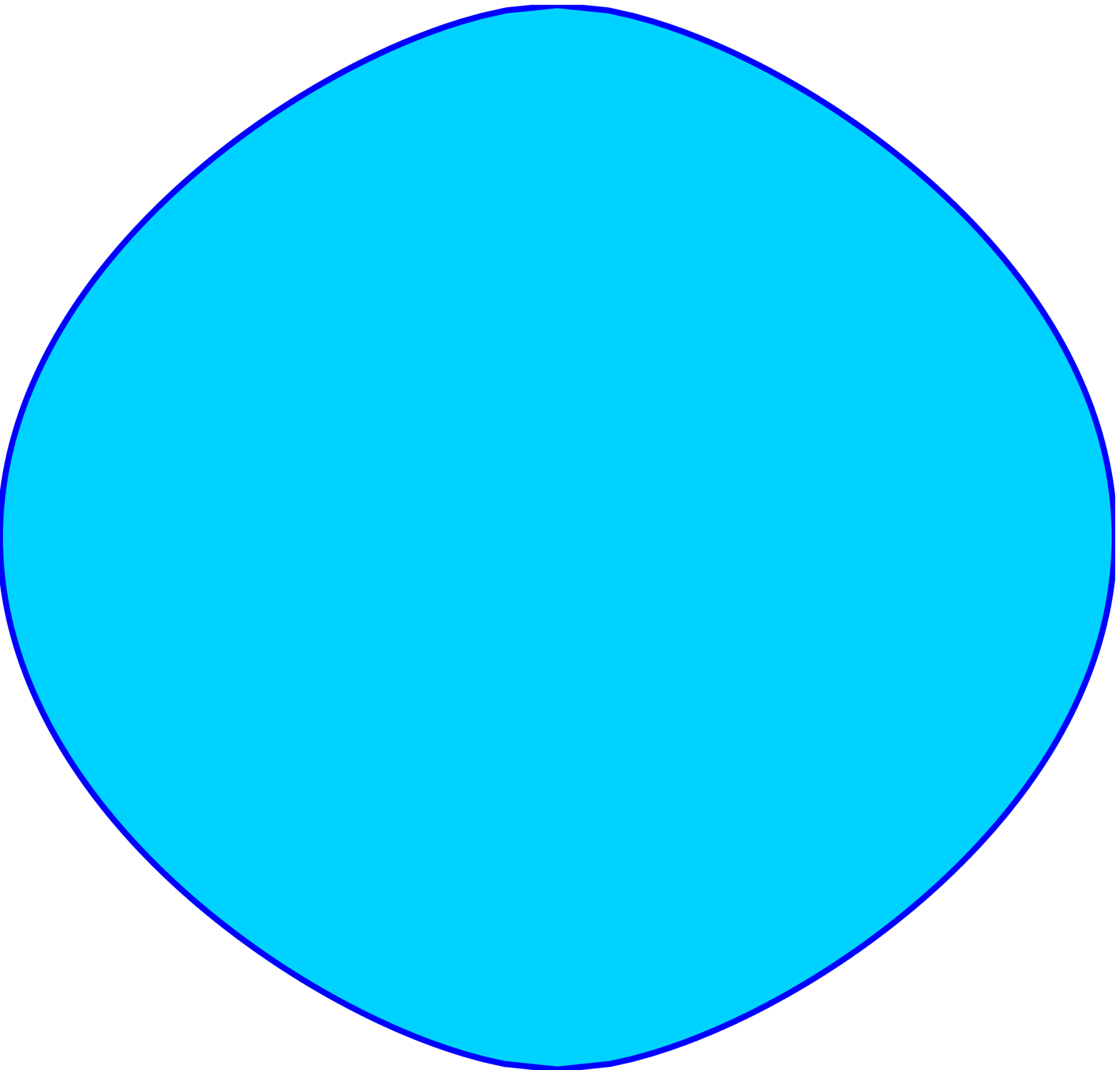}
		\caption{$\phi=\frac{5\pi}{16}$\\bumped spheroid}
		\label{fig:gallery_5Pio16mu1}
	\end{subfigure}
	\begin{subfigure}[c]{0.19\linewidth}
		\centering
		\includegraphics[width=\linewidth]{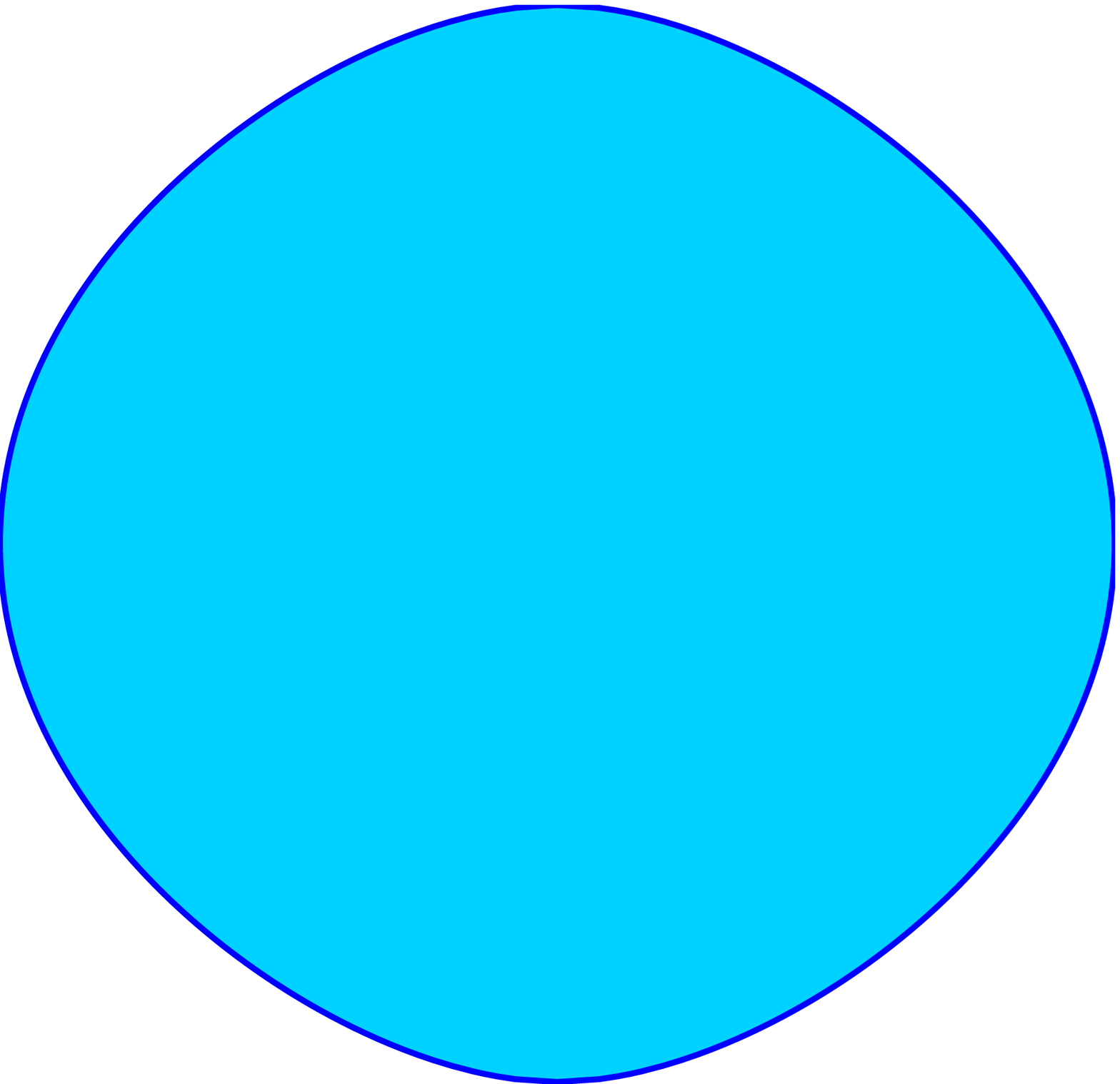}
		\caption{$\phi=\frac{6\pi}{16}$\\transition shape}
		\label{fig:gallery_6Pio16mu1}
	\end{subfigure}
	\begin{subfigure}[c]{0.19\linewidth}
		\centering
		\includegraphics[width=\linewidth]{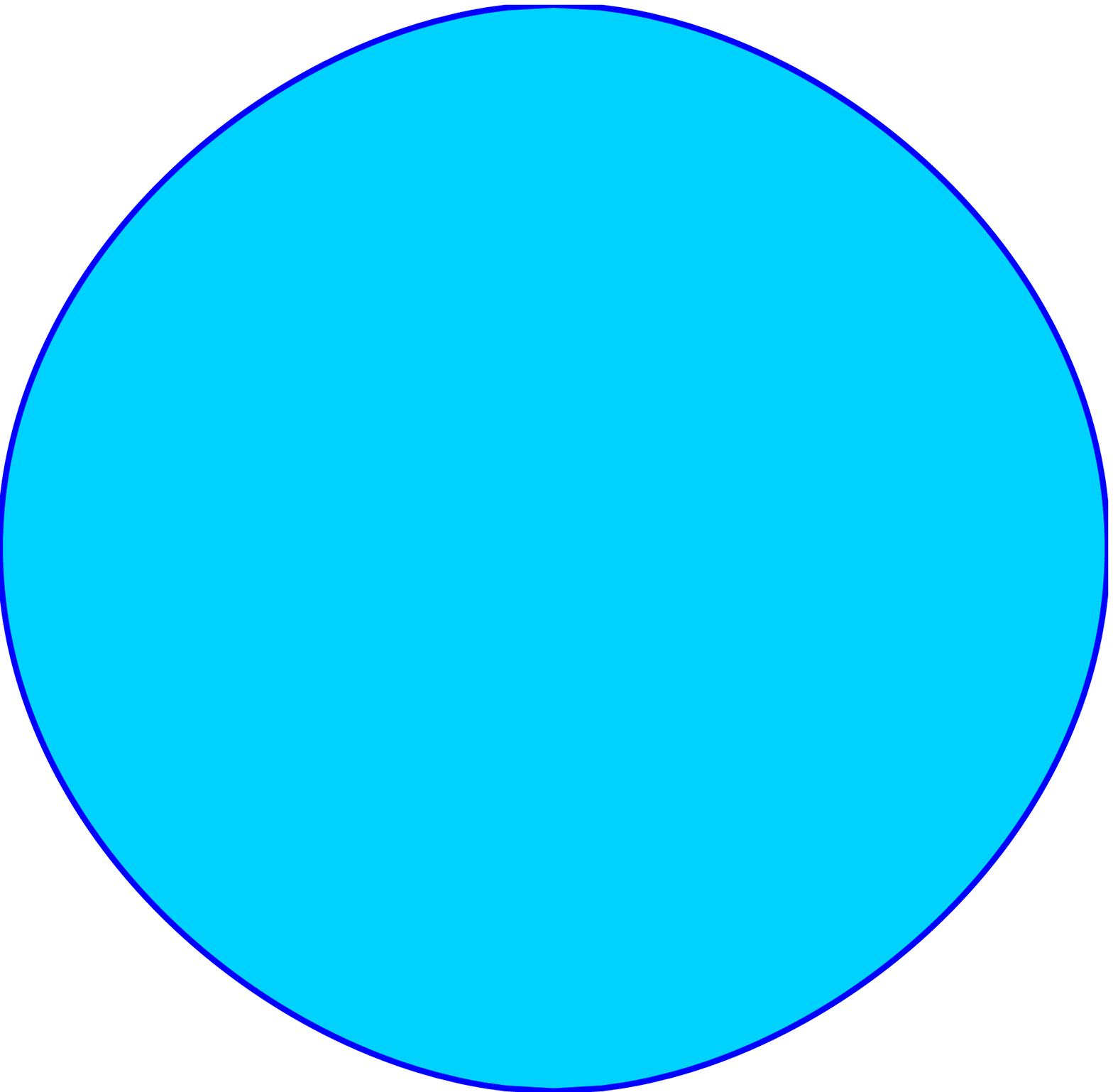}
		\caption{$\phi=\frac{7\pi}{16}$\\spheroid}
		\label{fig:gallery_7Pio16mu1}
	\end{subfigure}
	\begin{subfigure}[c]{0.19\linewidth}
		\centering
		\includegraphics[width=\linewidth]{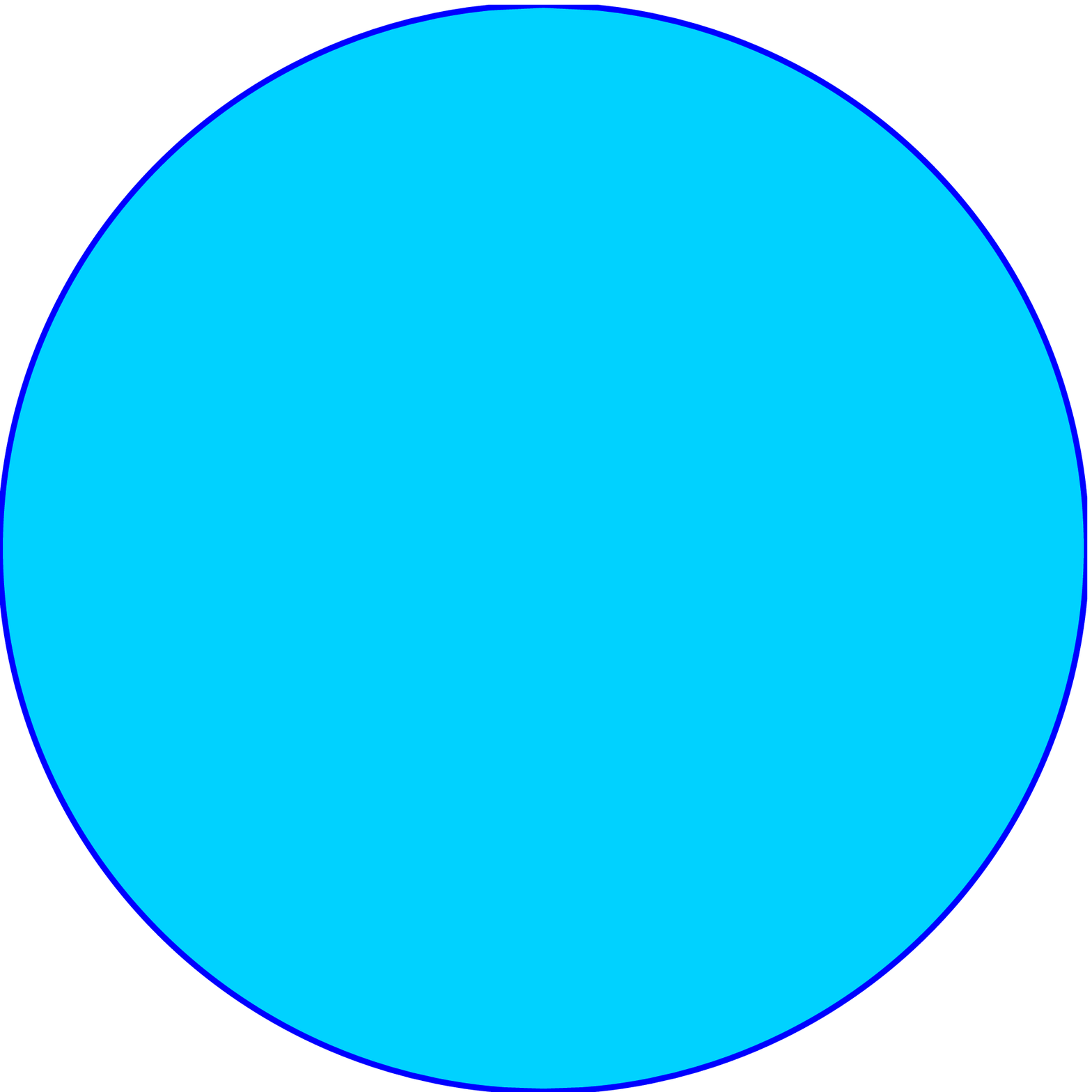}
		\caption{$\phi=\frac{8\pi}{16}$\\sphere}
		\label{fig:gallery_8Pio16mu1}
	\end{subfigure}
	\begin{subfigure}[c]{0.19\linewidth}
	\centering
	\includegraphics[width=\linewidth]{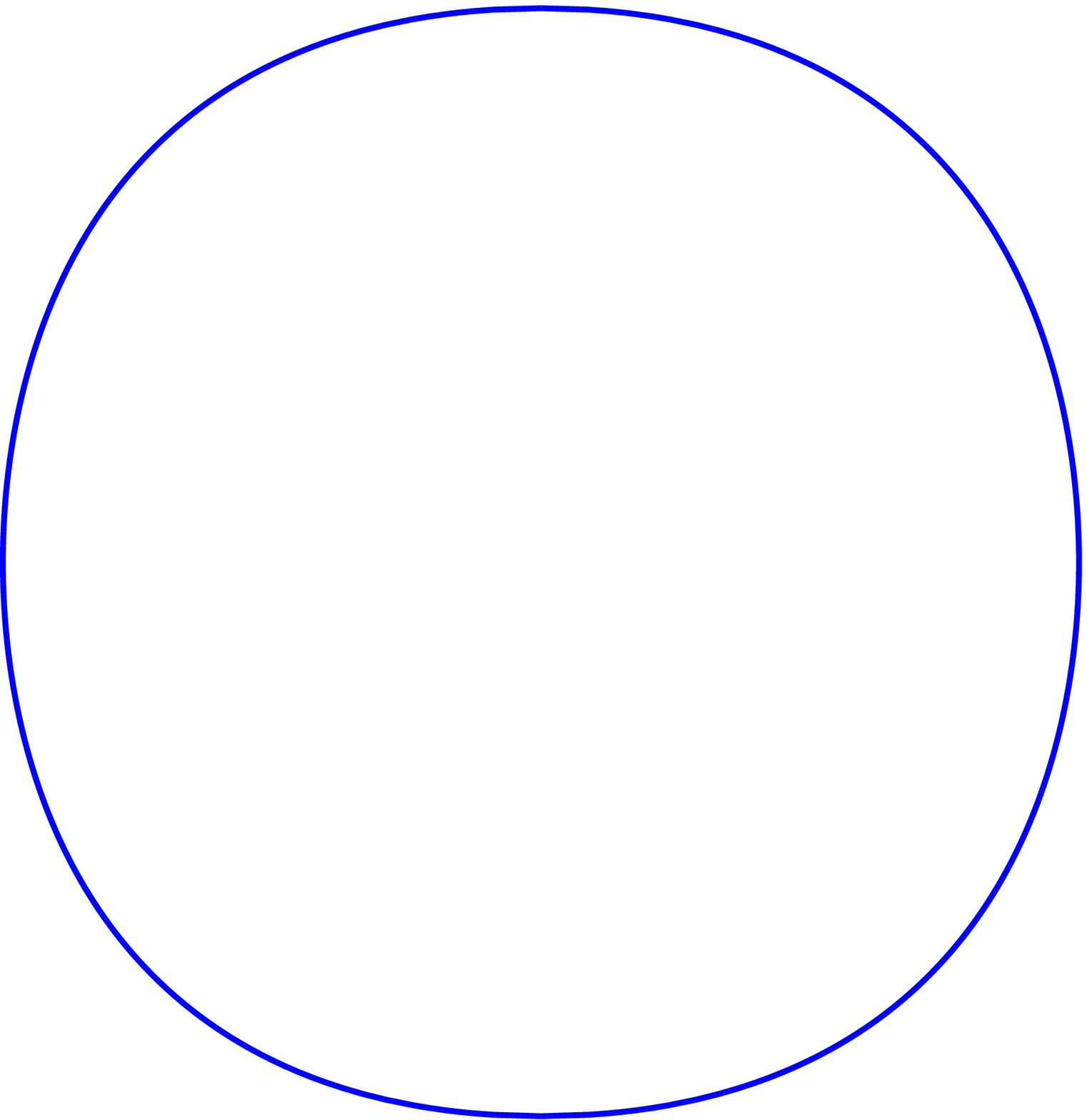}
	\caption{$\phi=\frac{9\pi}{16}$\\transition shape}
	\label{fig:gallery_9Pio16mu1}
\end{subfigure}
	\begin{subfigure}[c]{0.19\linewidth}
	\centering
	\includegraphics[width=\linewidth]{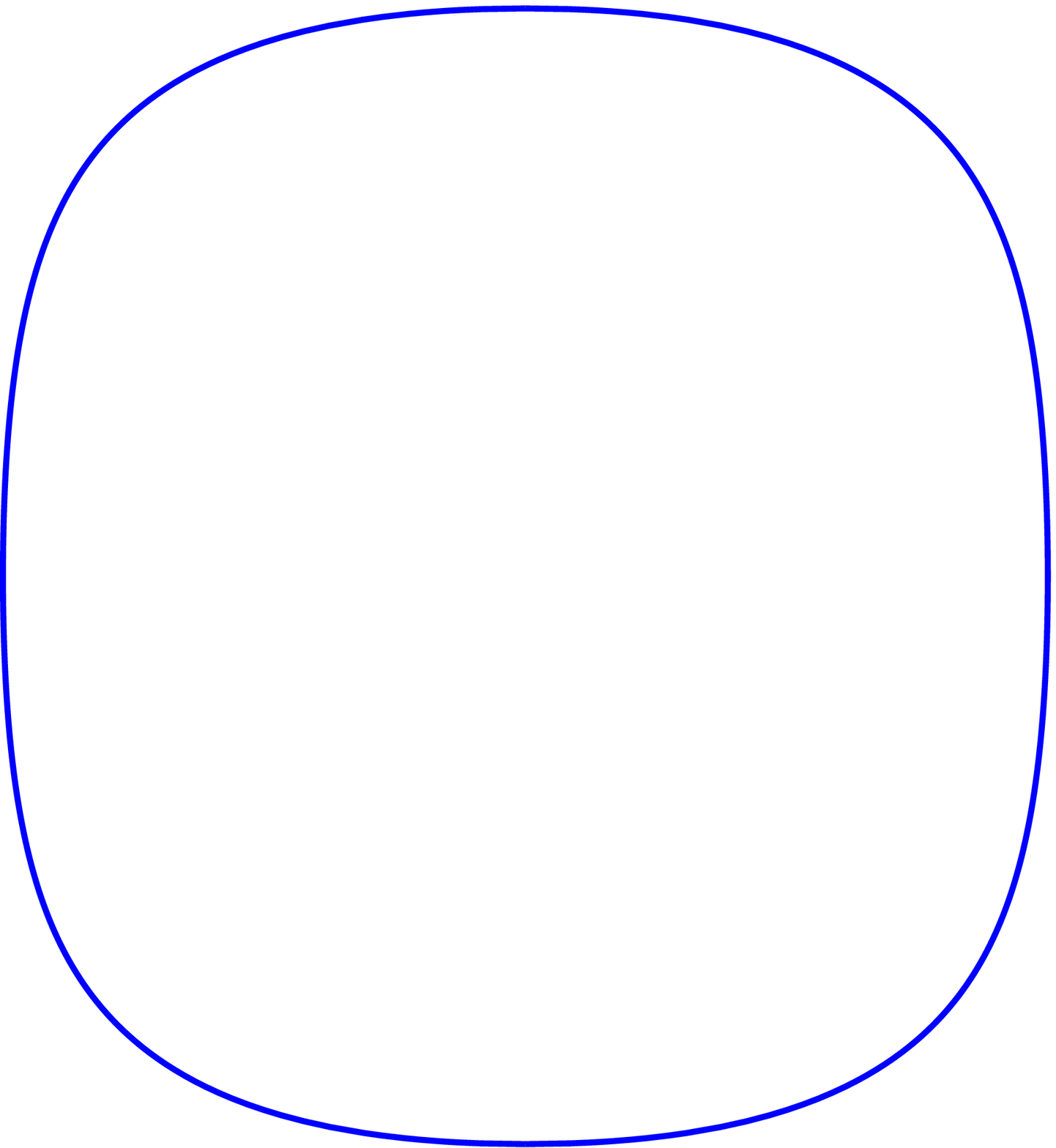}
	\caption{$\phi=\frac{10\pi}{16}$\\barrel}
	\label{fig:gallery_10Pio16mu1}
\end{subfigure}
\begin{subfigure}[c]{0.19\linewidth}
	\centering
	\includegraphics[width=\linewidth]{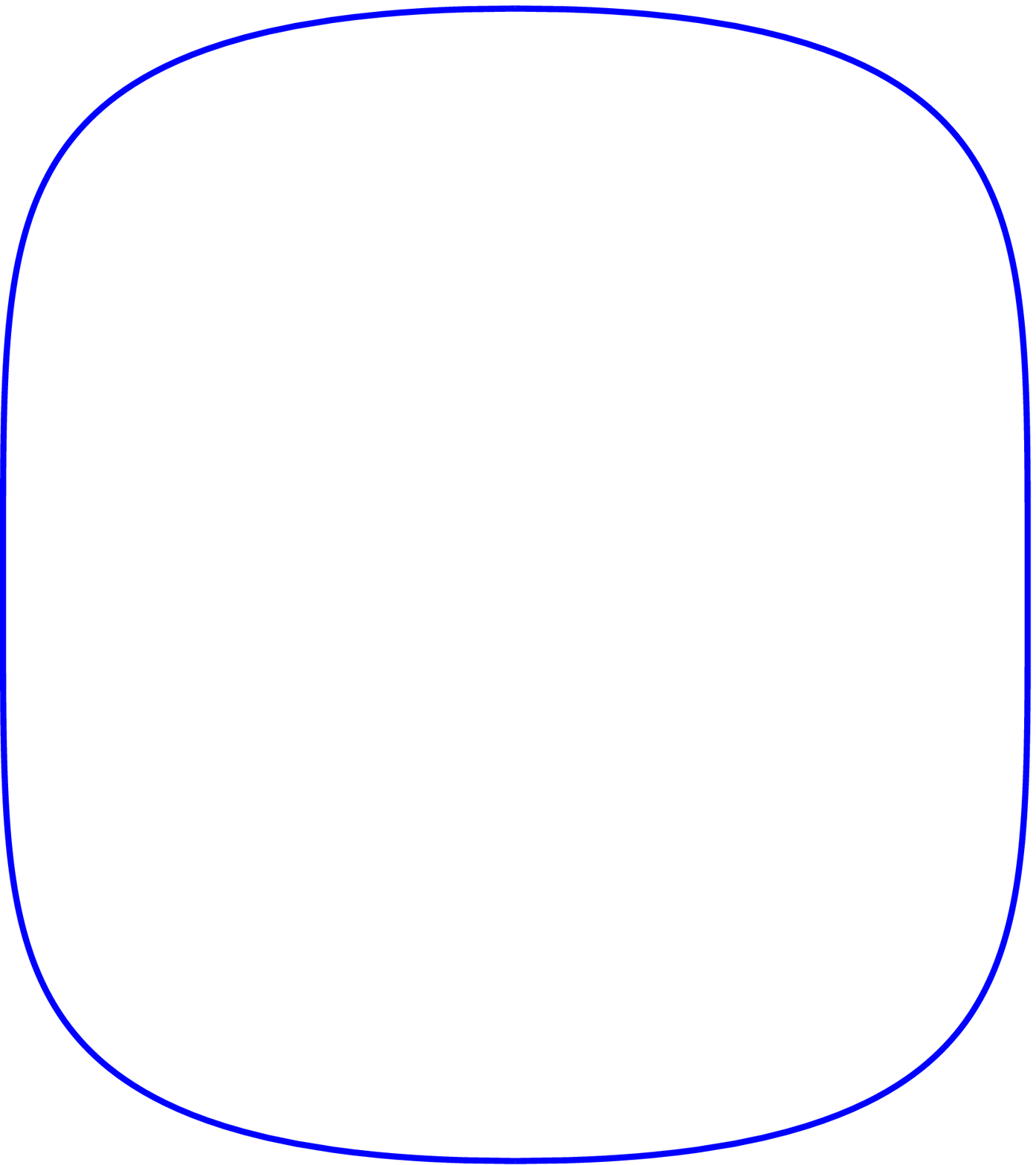}
	\caption{$\phi=\phi_\mathrm{c}$\\transition shape}
	\label{fig:gallery_phi_c_mu1}
\end{subfigure}
	\begin{subfigure}[c]{0.19\linewidth}
	\centering
	\includegraphics[width=\linewidth]{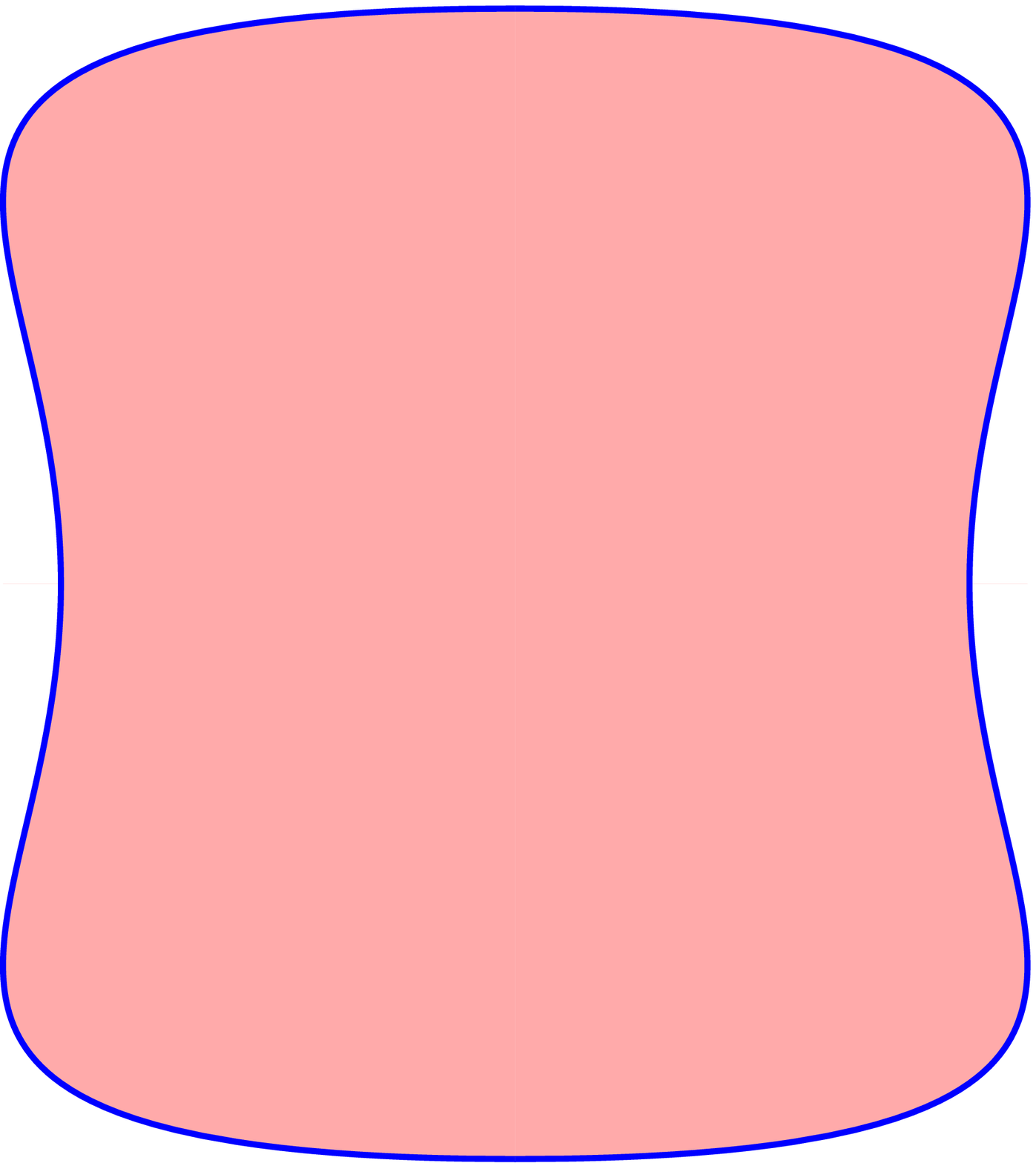}
	\caption{$\phi=\frac{11\pi}{16}$\\dumbbell}
	\label{fig:gallery_11Pio16mu1}
\end{subfigure}
	\caption{Gallery of shapes illustrating for $\mu=1$ the  taxonomy introduced in Table~\ref{tab:taxonomy}. In particular, the four transition shapes that somehow share features of two adjacent categories are also shown. The color coding of the shapes is the same used in Fig.~\ref{fig:configuration_space}.}
	\label{fig:gallery}
\end{figure}

Figure~\ref{fig:gallery} presents a gallery of meridian cross-sections of a droplet obtained from \eqref{eq:profile} and \eqref{eq:a_b_representation} for $\mu=1$  and a number of values of $\phi$ falling in the different categories listed in Table~\ref{tab:taxonomy}, including the transition shapes.
Clearly, the shapes in Figs.~\ref{fig:gallery_0Pio16mu1}-\ref{fig:gallery_Pio32mu1} are tactoids. On the other hand, the shapes shown in Figs.~\ref{fig:gallery_2Pio16mu1}-\ref{fig:gallery_5Pio16mu1} are definitely \emph{not} tactoidal, but they are not completely spherical either. We call them \emph{bumped spheroids} to highlight the fact that they exhibit a smooth bump where a tactoid would have a tapered tip. The difference between tactoids and bumped spheroids is a just a matter of how polar protrusions look like: pointed in the former, smoother in the latter. Keeping increasing $\phi$ from $\frac{6\pi}{16}$ to $\frac{9\pi}{16}$, the representative shapes gradually lose their bumps, justifying calling them simply \emph{spheroids}, see Figs.~\ref{fig:gallery_7Pio16mu1},\ref{fig:gallery_8Pio16mu1}. At $\phi=\frac{9\pi}{16}$, however, spheroids evolve into something else: they start resembling cylinders; we call them \emph{barrels}, see Fig.~\ref{fig:gallery_10Pio16mu1}. Beyond the transition shape at $\phi=\phi_ \mathrm{c}$ shown in Fig.~\ref{fig:gallery_phi_c_mu1}, the gallery of shapes is closed by a dumbbell falling in the  pink region of the configuration space  in Fig.~\ref{fig:configuration_space}, see Fig.~\ref{fig:gallery_11Pio16mu1}.

We shall see in the following sections where in configuration space $\mathsf{S}$ the free-energy functional $F$ in \eqref{eq:frank_energy_reduced}  attains its minimum, for given values of $k_3$ and $k_{24}$, and variable  $\alpha$. We shall learn which among the shapes illuminated in Fig.~\ref{fig:gallery} will be privileged as energy minimizers. In preparation for that, here we have set the language to describe a variety of possible shape transitions.

\section{Optimal shapes}\label{sec:optimal_shapes}
Our study is confined to bipolar droplets, for which the anchoring at the interface is successfully holding up a tangential, albeit degenerate alignment of the nematic director. It is well-known \cite{virga:drops} that for sufficiently small droplets the nematic orientation inside them  tends to be uniform and the anchoring at their boundary is accordingly broken, so that the equilibrium shape is delivered by the classical Wulff's construction \cite{wulff:frage}. We need to make sure that the parameters are chosen in a range where such a configuration would be energetically disfavored. We shall see that this can be achieved provided that the reduced volume $\alpha$ in\eqref{eq:alpha} is sufficiently large.   

\subsection{Admissible  Volumes}\label{sec:admissible}
To identify the safeguard value of $\alpha$ below which it would be unwise to push our analysis, we perform here an energy comparison based on two simple estimates.

We begin by estimating the free energy $\free$ in \eqref{eq:free_energy_functional} for a uniformly aligned cylindrical drop with (constant) radius $R$ and height $L$ delivered by
\begin{equation}\label{eq:L}
	L=\frac43\frac{\Req^3}{R^2},
\end{equation}
for the constraint on the volume in \eqref{eq:volume_constraint} to be obeyed. Suppose that $\n$ is along the cylinder's axis; since $\nabla\n$ vanishes identically, no distortion energy is stored in the body $\body$ of the drop: all the free energy comes from the boundary $\boundary$; it is given by
\begin{equation}
	\label{eq:F_c}
	F_\mathrm{c}=2\pi\gamma\left(\frac43\frac{\Req^3}{R}+(1+\omega)R^2 \right),
\end{equation}
where use has been made of \eqref{eq:L} and account has been taken of the different orientation of $\n$ relative to $\normal$ on the lateral surface and on the bases of the cylinder. It is a very simple matter to see that $F_\mathrm{c}$ is minimized for
\begin{equation}
	\label{eq:R}
	R=\Req\sqrt[3]{\frac{2}{3(1+\omega)}}
\end{equation}
and that the corresponding value of $F_\mathrm{c}$ is
\begin{equation}\label{eq:F_c_min}
	F_\mathrm{c}=2\pi\gamma\Req^2\sqrt[3]{12(1+\omega)}.	
\end{equation}
This energy is to be compared with that estimated for a sphere with the bipolar director field emanating from the poles. Letting all constants $K_{11}$, $K_{22}$, and $K_{33}$  be equal to $K$ in the elastic energy computed in equation (2.18) of \cite{williams:transitions}, for the total free energy $F_\mathrm{s}$ of a spherical  drop of radius $\Req$ we obtain
\begin{equation}
	\label{eq:F_s}
F_\mathrm{s}=\left(7-\frac{\pi^2}{4}\right)\pi K\Req-4\pi K_{24}\Req+4\pi\gamma\Req^2.	
\end{equation}
The   demand that $F_\mathrm{s}<F_\mathrm{c}$ for all admissible $K_{24}$, which would make it unfavorable breaking the tangential surface anchoring, is thus reverted into an inequality for $\alpha$,
\begin{equation}
	\label{eq:alpha_safeguard}
\alpha>\alpha_\mathrm{s}(\omega):=\frac{7-\frac{\pi^2}{4}}{2(\sqrt[3]{12(1+\omega)}-2)},	
\end{equation}
which shows how the \emph{bipolar  safeguard} value $\alpha_\mathrm{s}$ for $\alpha$ depends on $\omega$. Clearly, the larger is $\omega$, the smaller is $\alpha_\mathrm{s}$. Estimating $\omega$ in the range $1-10$,\footnote{See, for example, \cite{puech:nematic} and \cite{kim:morphogenesis}.} we shall take $\alpha>\alpha_\mathrm{s}(5)\doteq1.0$.\footnote{This threshold is to some extent conventional, but cannot be too wrong, as $\alpha_\mathrm{s}(1)\doteq2.6$ and $\alpha_\mathrm{s}(10)\doteq0.7$.}

In view of \eqref{eq:alpha}, the latter inequality can be interpreted as a lower bound for the linear size $\Req$ of the drops admissible in our theory.\footnote{Thus making it clear in what sense this applies to sufficiently \emph{large} drops.} Taking $K\sim1$-$10\,\mathrm{pN}$ as typical value for all elastic constants\footnote{This estimate is supported by a number of contributions that span a long time interval, from  early works \cite{saupe:temperaturabhangigkeit,saupe:biegungselastizitat,orsay:recent,karat:elastic,karat:elasticity,maze:determination,skarp:measurements} to more recent ones \cite{bunning:frank,balzarini:high,hurd:field-indiced,taratuta:light-scattering,morris:measurements,lee:computations,lee:crossover,taratuta:anisotropic,itou:measurements}, both experimental and computational in nature, for liquid crystals ranging from thermomotropic to lyotropic, with both low and high molecular weight, see also \cite{williams:nematic} and \cite{prinsen:shape}. That  elastic constants are not too dissimilar for lyotropic and thermotropic liquid crystals  has also been confirmed by a recent study on chromonics \cite{zhou:elasticity}, see also \cite{kim:morphogenesis}.} and $\gamma\sim10^{-5}\,\mathrm{Nm}^{-1}$ as typical value for the interfacial energy of a nematic liquid crystal in contact with its melt,\footnote{This estimate is supported by the now classical experimental works \cite{faetti:nematic,faetti:measurements} on cyano-biphenyls  and a number of more recent works \cite{chen:measurement,chen:interfacial,chen:interfacial_1999,chen:interfacial_2002} on other materials, see also \cite{kahlweit:estimation,langevin:molecular,yokoyama:boundary} for further earlier sources.} the lower bound for $\alpha$ translates into  $\Req\gtrsim10^{-1}$-$1\,\mu\mathrm{m}$, which expresses in physical terms the appropriate range of validity of the theory presented here.\footnote{It is perhaps worth noting that such a range changes dramatically when the isotropic fluid surrounding the drop is not its melt. For example, the early measurements of Naggiar~\cite{naggiar:phenomenes} and  Schwartz~\cite{schwartz:surface} gave $\gamma\sim10^{-2}\mathrm{Nm}^{-1}$ for the surface tension of nematic liquid crystals in contact with its vapor. Correspondingly, an estimate for the admissible $\Req$ would then give $\Req\gtrsim10^{-11}\mathrm{m}$, which makes our theory applicable to drops of virtually all sizes in that environment.}

\subsection{Mininizing Trajectories}\label{sec:minimal}
Finding the minimum of the functional $F[\mu;R]$ in \eqref{eq:F_energy_functional} is not a problem that can be solved analytically, even in the class of shapes (and retracted meridian fields) described in \eqref{eq:profile} with $a$ and $b$ expressed as in \eqref{eq:a_b_representation} in terms of the configuration parameters $(\phi,\mu)$.

\begin{figure}
		\centering
		\begin{subfigure}[b]{0.4\linewidth}
			\centering
			\includegraphics[width=\linewidth]{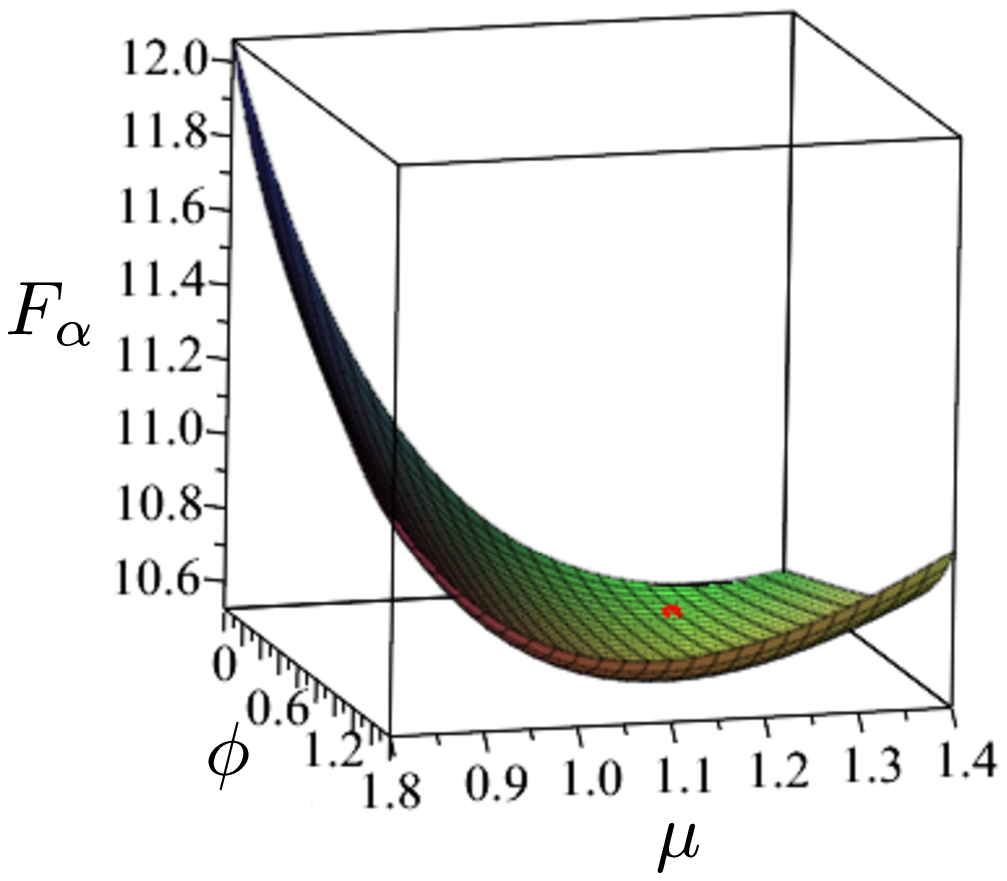}
			\caption{Graph of $\Fa$ against $\conf$ for $\alpha=10$. The red dot designates the minimum.}
			\label{fig:minima_a}
		\end{subfigure}
		\begin{subfigure}[b]{0.28\linewidth}
			\centering
			\includegraphics[width=.9\linewidth]{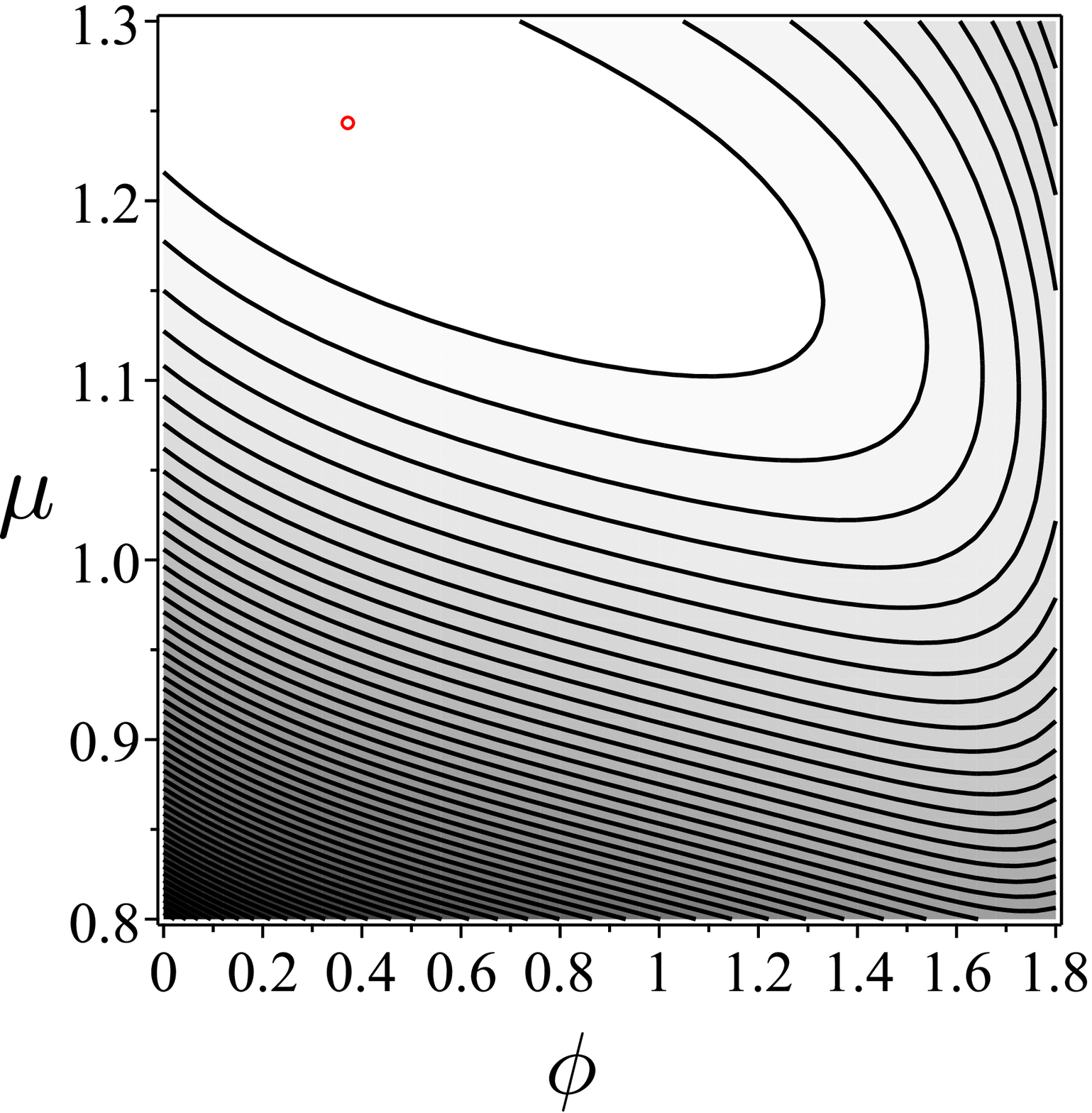}
			\caption{Contour plot of $\Fa$, for $\alpha=10$. The minimum is attained at the point where $\phi\doteq0.37$ and $\mu\doteq1.24$, marked by a red circle.}
			\label{fig:minima_b}
		\end{subfigure}
	\begin{subfigure}[b]{0.3\linewidth}
		\centering
		\includegraphics[width=.5\linewidth]{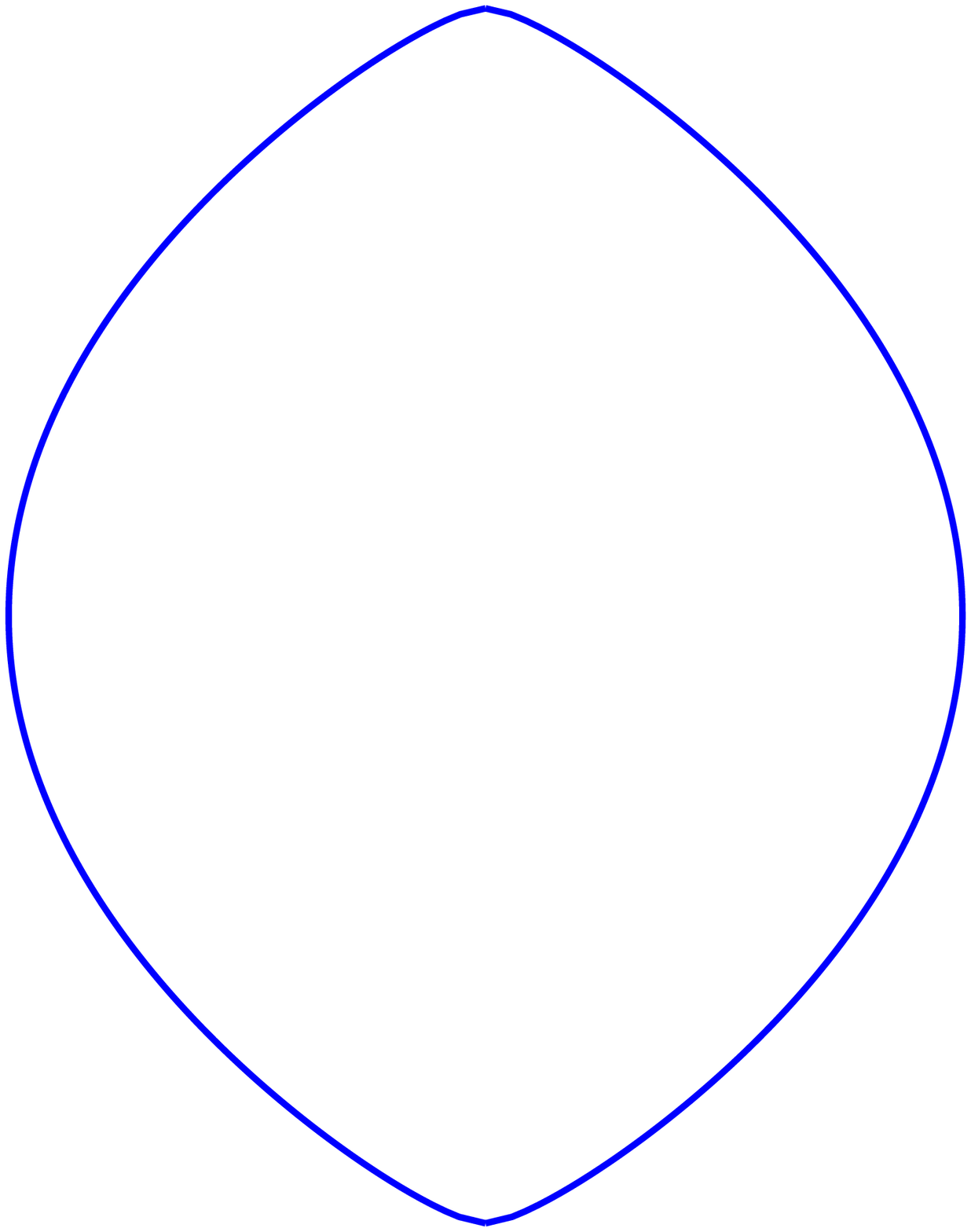}
		\caption{Cross-section of the equilibrium shape corresponding through \eqref{eq:profile} to the minimizer of $\Fa$ marked in Fig.~\ref{fig:minima_b}. Since $\frac{\pi}{16}<\phi<\frac{6\pi}{16}$, according to our conventional taxonomy in Sec.~\ref{sec:taxonomy}, it is a bumped spheroid (and looks indeed like one).}
		\label{fig:minima_c}
	\end{subfigure}
\caption{For given $\alpha>1$, the function $\Fa$ is defined in configuration space $\conf=\{(\phi,\mu):0\leqq\phi\leqq\frac{3\pi}{4},\mu>0\}$ by reducing the functional $F[\mu;R]$ in \eqref{eq:F_energy_functional} to the special families of shapes in \eqref{eq:profile}. Elastic constants are $k_3=1$, $k_{24}=\frac12$.}
\label{fig:minima}
\end{figure}

For a given choice of the elastic parameters $(k_3,k_{24})$, we evaluated numerically $F[\mu;R]$ for increasing values of $\alpha>1$ as a \emph{reduced} function $\Fa(\phi,\mu)$ on the configuration space $\conf$. Figure~\ref{fig:minima} illustrates the generic situation that we encountered. 
As shown in Fig.~\ref{fig:minima_a}, $\Fa$ has a convex graph and attains a single minimum in $\conf$, which is easily identified through the level sets of $\Fa$ depicted in Fig.~\ref{fig:minima_b}; the corresponding equilibrium shape, a bumped spheroid according to the taxonomy of Sec.~\ref{sec:taxonomy}, is illustrated in Fig.~\ref{fig:minima_c}.

\begin{figure}[h]
	\centering
	\begin{subfigure}[c]{0.25\linewidth}
		\centering
		\includegraphics[width=.9\linewidth]{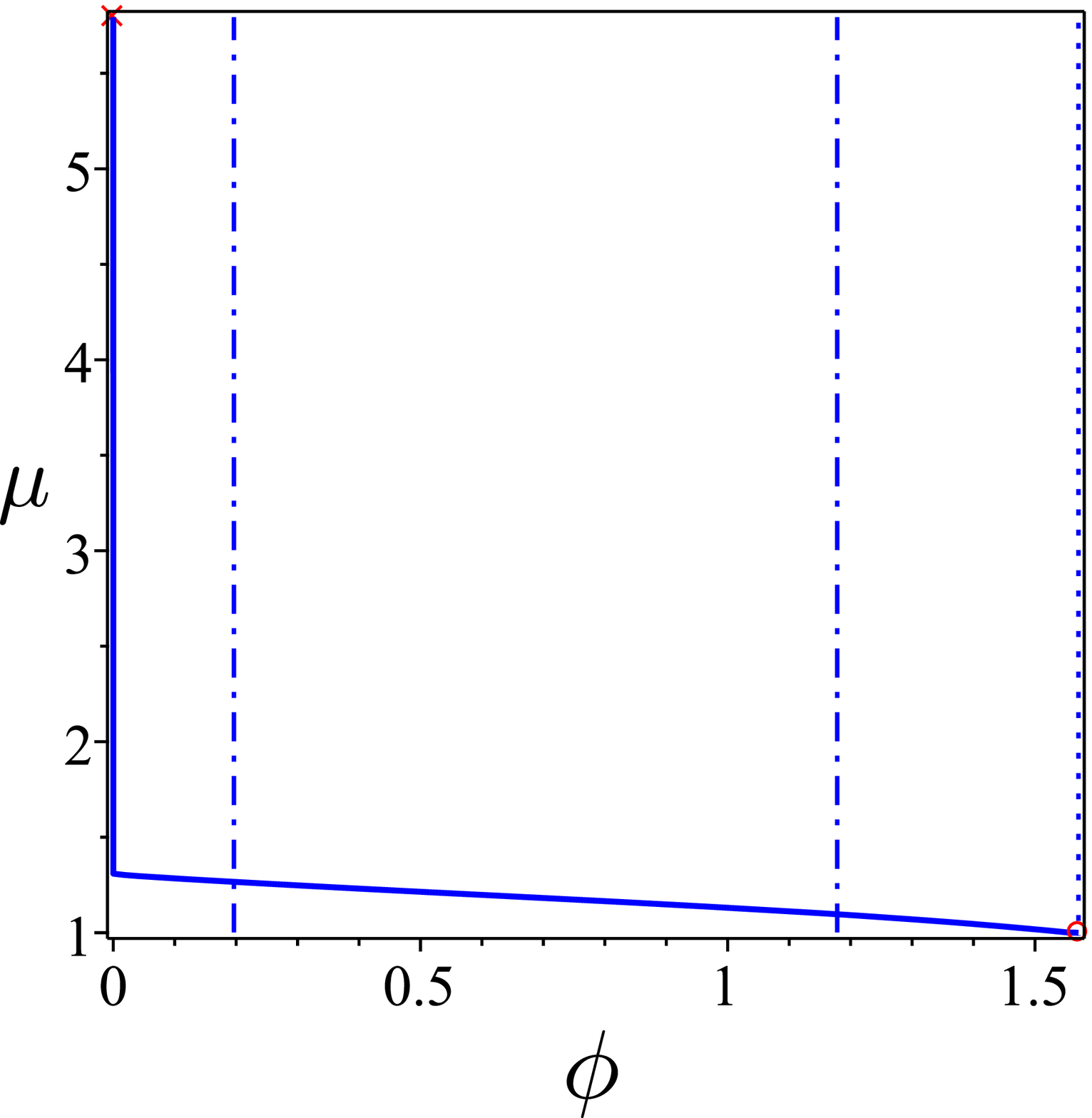}
		\caption{$k_{24}=0.1$}
		\label{fig:trajectories_a}
	\end{subfigure}
	\begin{subfigure}[c]{0.25\linewidth}
		\centering
		\includegraphics[width=.9\linewidth]{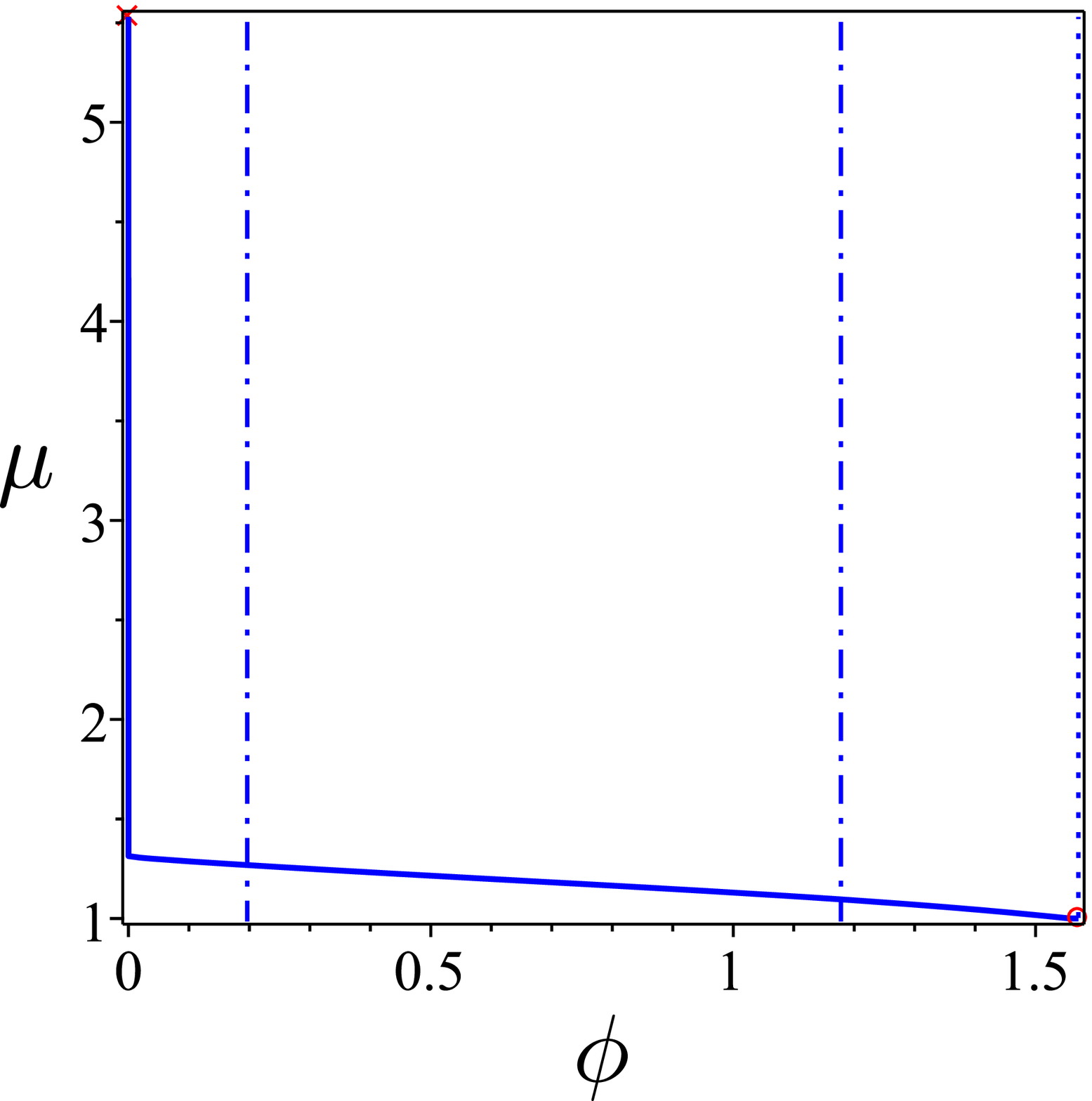}
		\caption{$k_{24}=0.2$}
		\label{fig:trajectoiries_b}
	\end{subfigure}
	\begin{subfigure}[c]{0.25\linewidth}
		\centering
		\includegraphics[width=.9\linewidth]{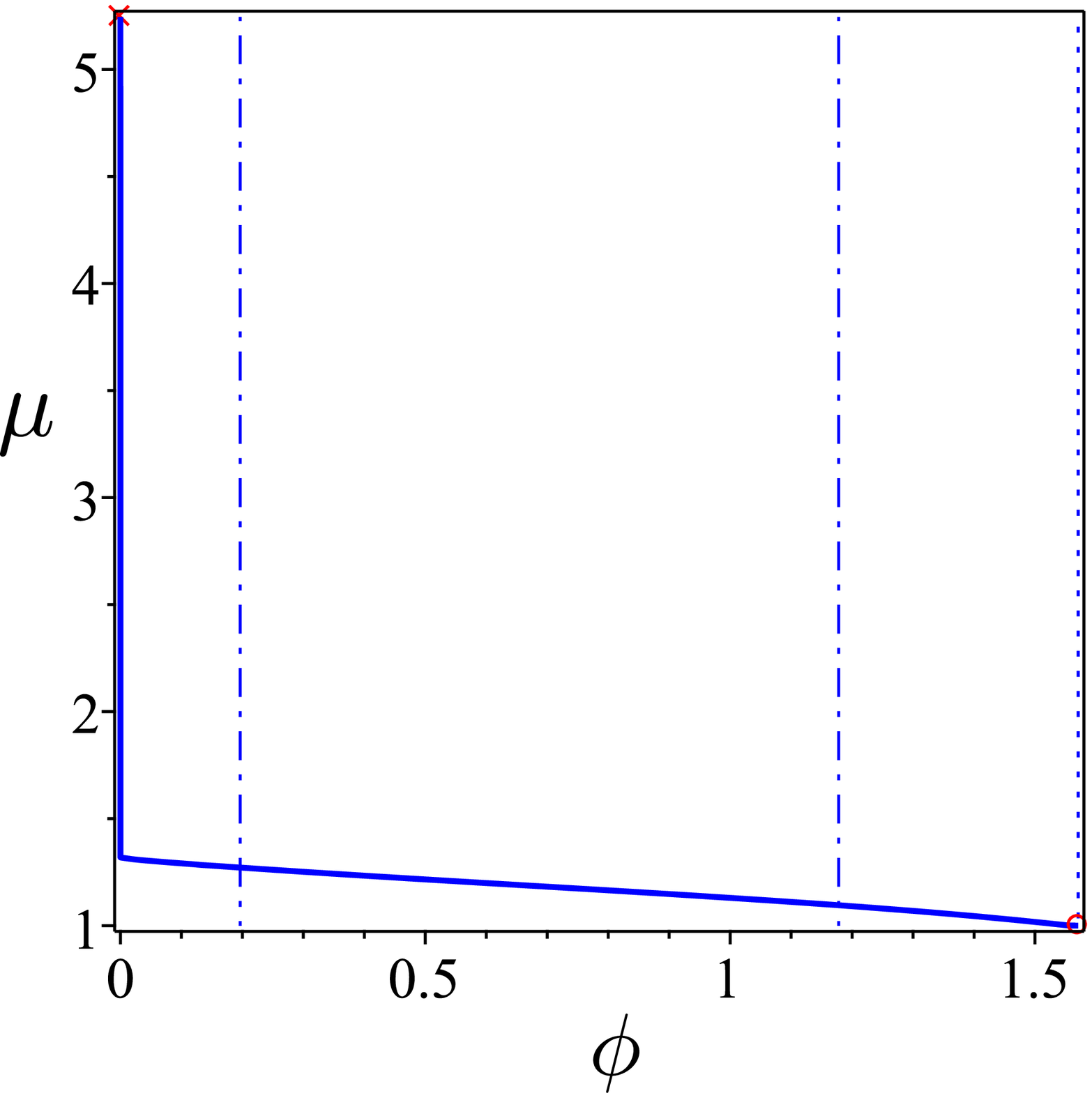}
		\caption{$k_{24}=0.3$}
		\label{fig:trajectoires_c}
	\end{subfigure}
	\begin{subfigure}[c]{0.25\linewidth}
	\centering
	\includegraphics[width=.9\linewidth]{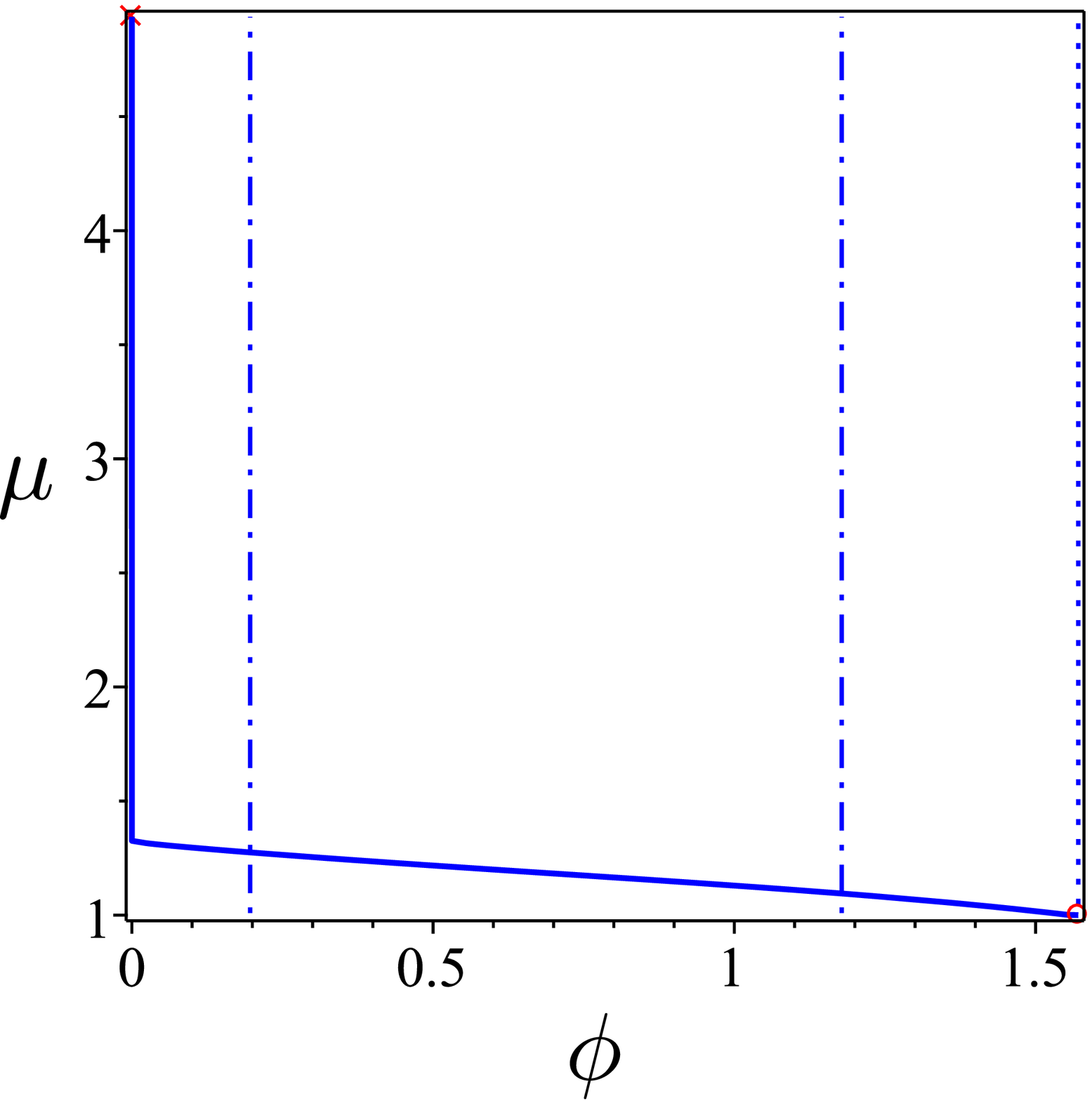}
	\caption{$k_{24}=0.4$}
	\label{fig:trajectories_d}
\end{subfigure}
\begin{subfigure}[c]{0.25\linewidth}
	\centering
	\includegraphics[width=.9\linewidth]{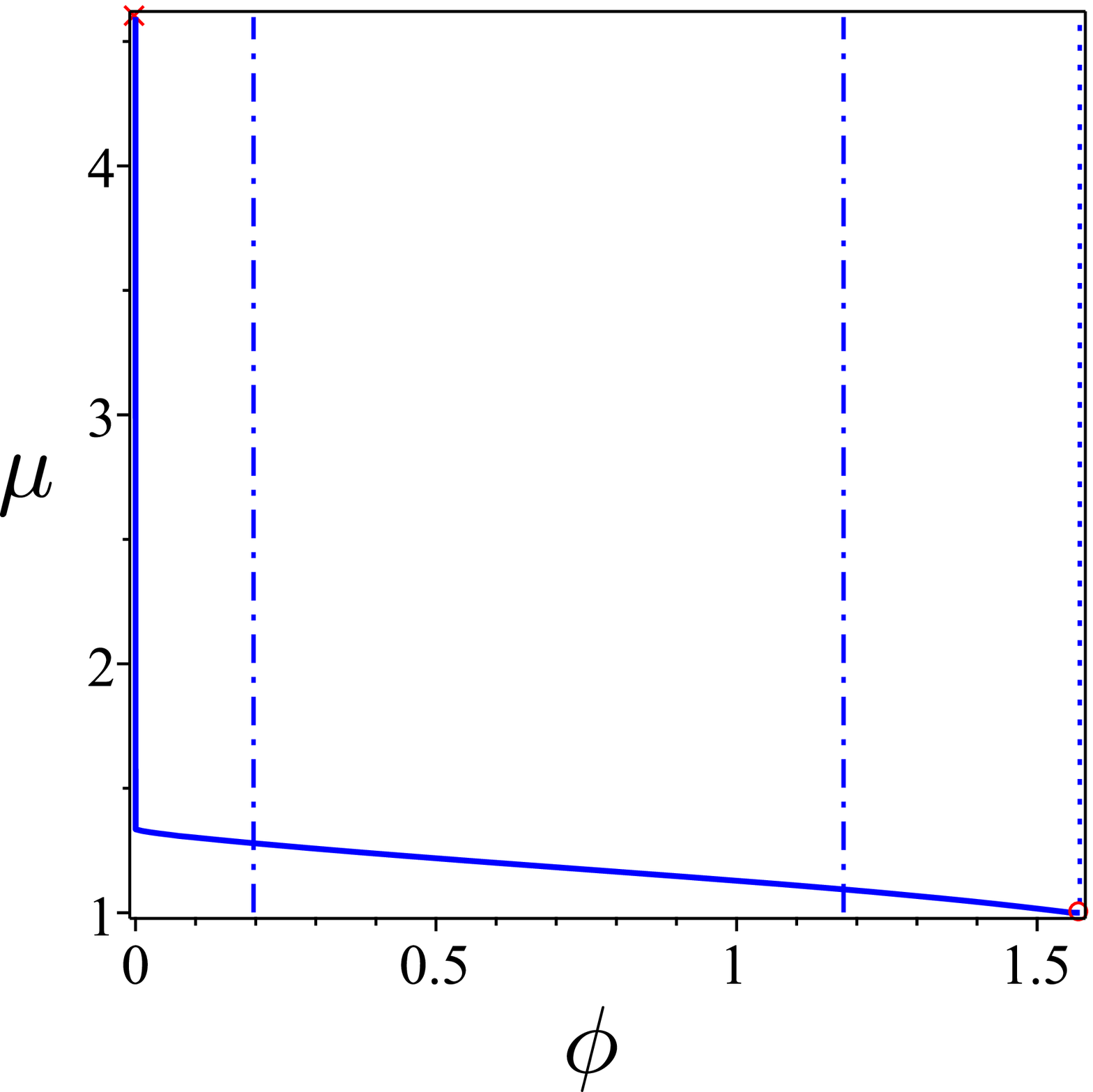}
	\caption{$k_{24}=0.5$}
	\label{fig:trajectoiries_e}
\end{subfigure}
\begin{subfigure}[c]{0.25\linewidth}
	\centering
	\includegraphics[width=.9\linewidth]{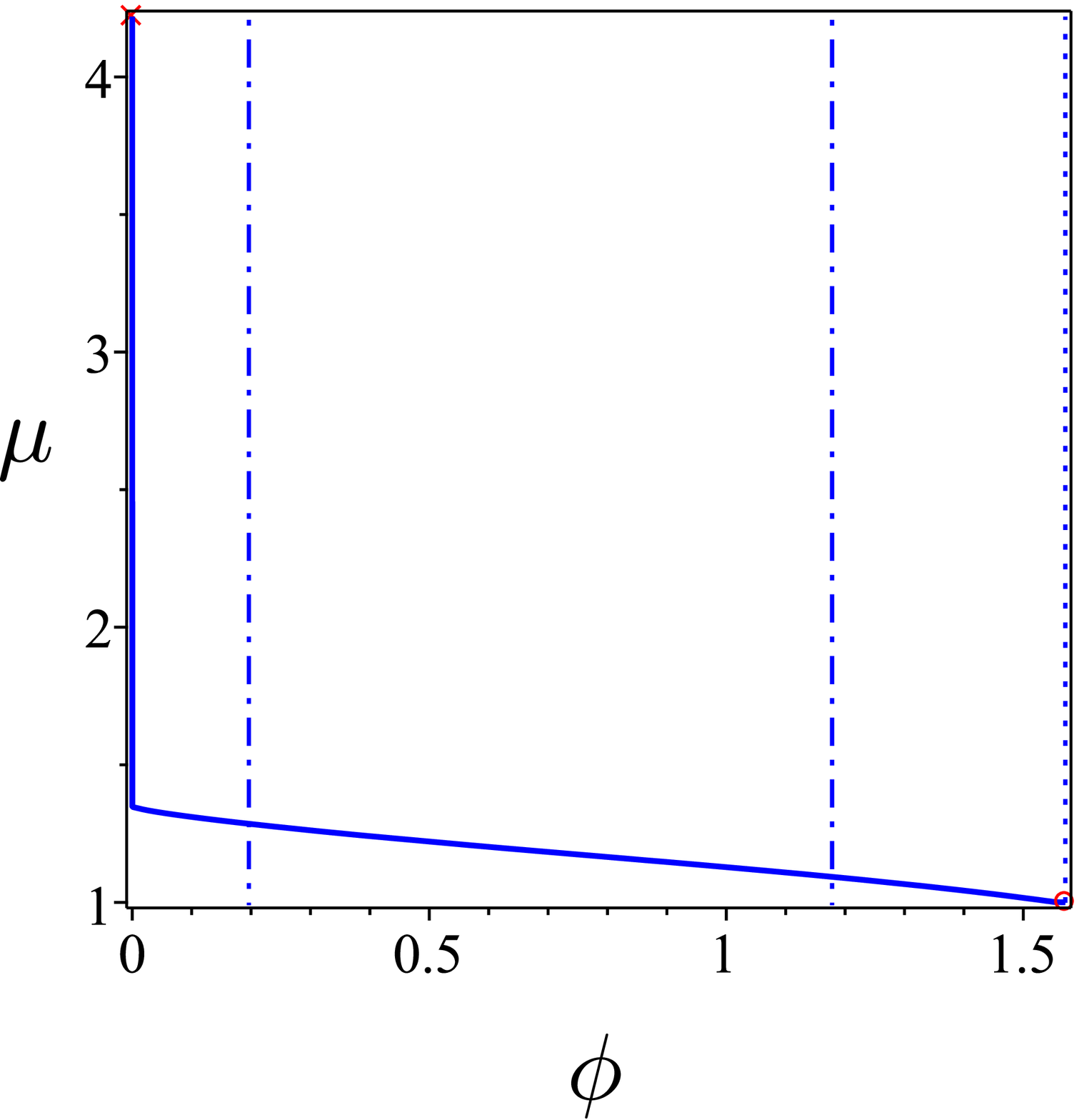}
	\caption{$k_{24}=0.6$}
	\label{fig:trajectoires_f}
\end{subfigure}
	\begin{subfigure}[c]{0.25\linewidth}
	\centering
	\includegraphics[width=.9\linewidth]{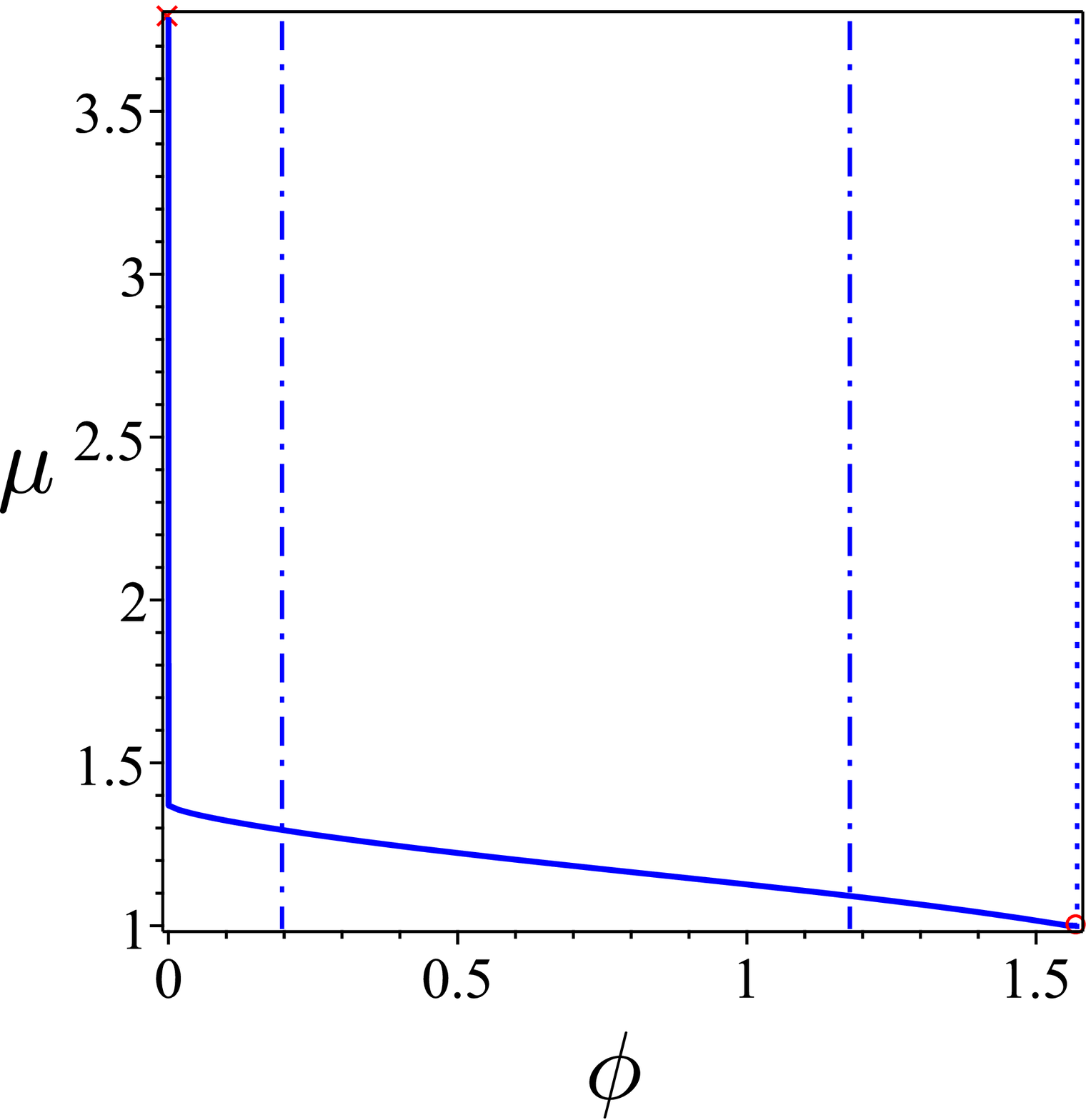}
	\caption{$k_{24}=0.7$}
	\label{fig:trajectories_g}
\end{subfigure}
\begin{subfigure}[c]{0.25\linewidth}
	\centering
	\includegraphics[width=.9\linewidth]{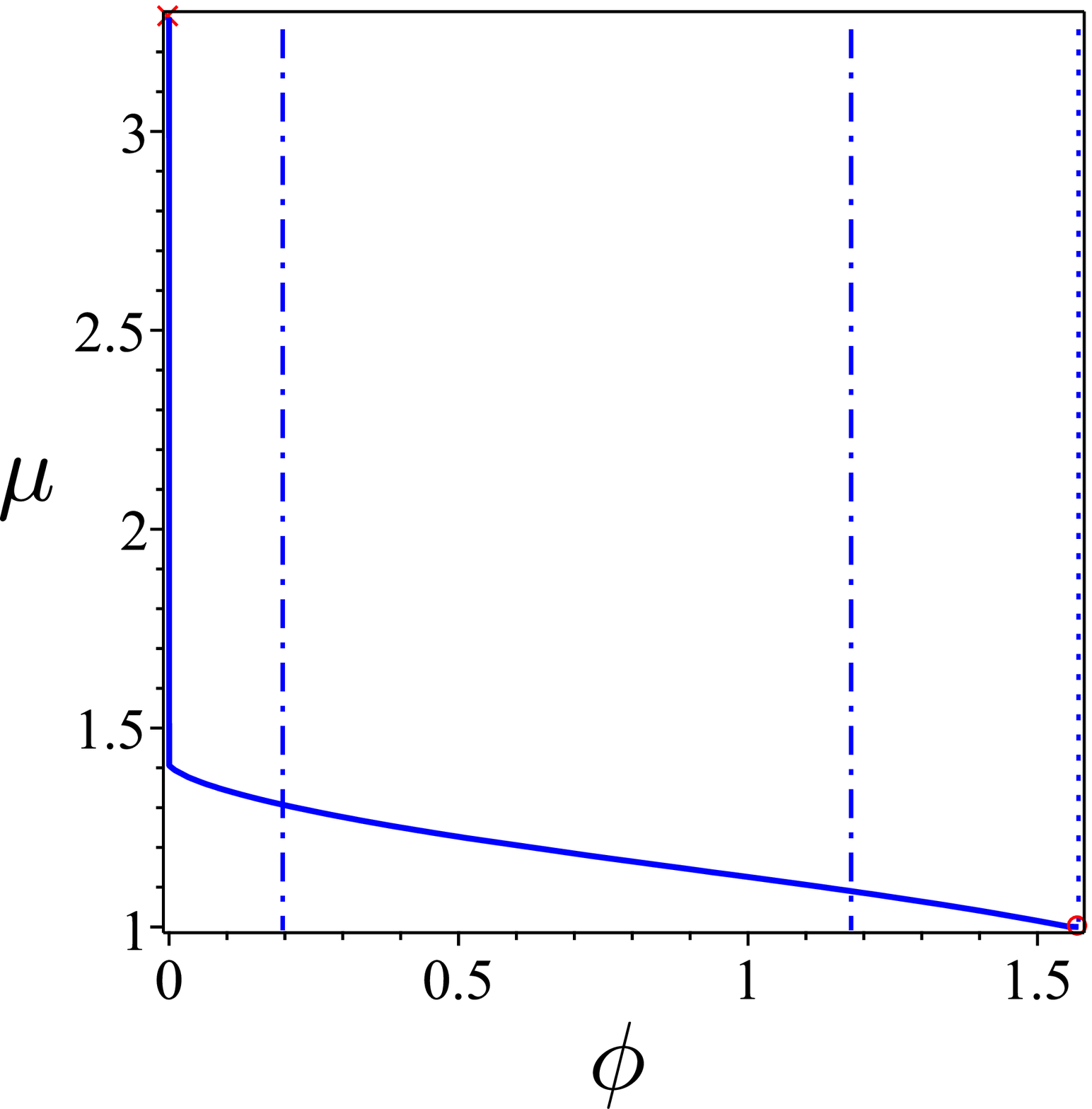}
	\caption{$k_{24}=0.8$}
	\label{fig:trajectoiries_h}
\end{subfigure}
\begin{subfigure}[c]{0.25\linewidth}
	\centering
	\includegraphics[width=.9\linewidth]{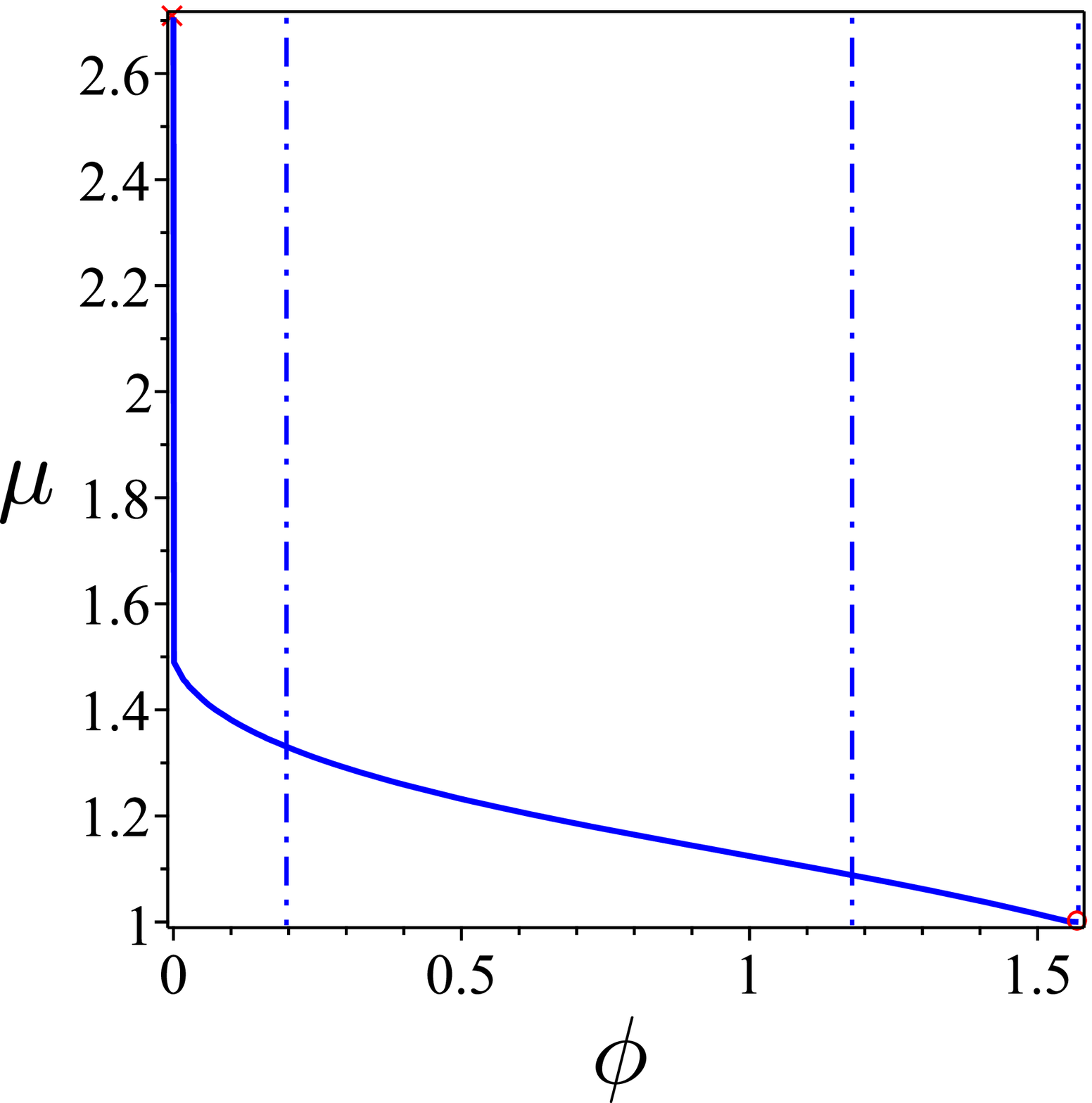}
	\caption{$k_{24}=0.9$}
	\label{fig:trajectoires_i}
\end{subfigure}
	\caption{Minimizing trajectories (solid lines) in the configuration space $\conf$ for $k_3=1$ and different values of $k_{24}$. All trajectories are parameterized in $\alpha$ and start from the configuration representing the minimum of $\Fa$ for $\alpha=1$ (marked by a red cross); they all converge to the point representing in $\conf$ the sphere of radius $\Req$ (marked by a red circle). The dashed lines (at $\phi=\frac{\pi}{6}$ and $\phi=\frac{6\pi}{16}$)  represent the barriers introduced in Sec.~\ref{sec:taxonomy} to delimit different families of shapes.}
	\label{fig:trajectories}
\end{figure}

We performed a systematic search for the minimizer of $\Fa$ upon increasing $\alpha>1$, for given elastic parameters $(k_3,k_{24})$. Each search delivered a path of minimizers in the configuration space $\conf$, parameterized in $\alpha$. These paths are shown in Fig.~\ref{fig:trajectories} for $k_3=1$ and a sequence of values of $k_{24}$ in the admissible interval $[0,1]$.
They all have a number of features in common.

First, they reside on the $\mu$-axis (for $\phi=0$) until $\alpha$ reaches a critical value, $\azero$, upon crossing which they leave the boundary of $\conf$ and dive into its interior. Clearly, for $1<\alpha<\azero$, the equilibrium shape of the drop is a \emph{genuine} tactoid (with sharply pointed tips). For $\alpha>\azero$, the minimizing trajectory traverses the domain of generic tactoids, until $\phi$ reaches the conventional barrier $\phi=\frac{\pi}{6}$. Such a crossing takes place for $\alpha=\aone$; there, the equilibrium drop undergoes a (smooth) shape transition, becoming a bumped spheroid. The whole territory of these latter shapes is then traversed by the minimizing trajectories, which enter the realm of spheroids for $\alpha=\atwo$ (where $\phi=\frac{6\pi}{11}$). Upon further increasing $\alpha$, all trajectories converge towards the point that in $\conf$ represents the sphere of radius $\Req$ (marked by a red dot in Fig.~\ref{fig:trajectories}). 

\begin{figure}[h] 
\includegraphics[width=.35\linewidth]{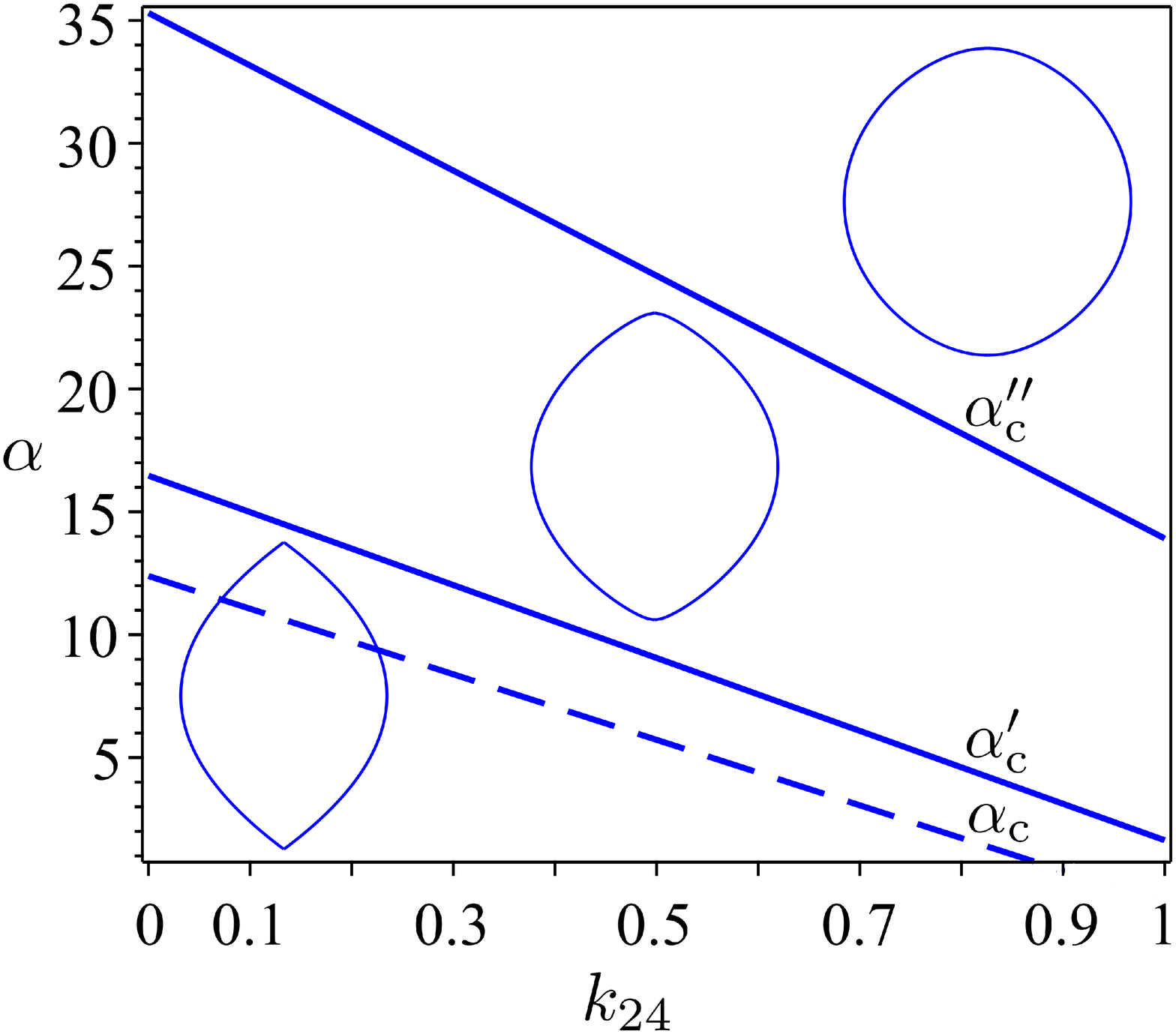}
\caption{Critical values of $\alpha$ plotted against $k_{24}$ for $k_3=1$. The lowest (dashed) line, $\azero$, designates the extinction of genuine (sharply pointed) tactoids; the middle line, $\aone$, marks the extinction of tactoids (pointed or not) and the onset of bumped spheroids; the upper line, $\atwo$, marks the extinction of bumped spheroids and the onset of spheroids.}
\label{fig:alpha_critical}
\end{figure}

Qualitatively, this scenario remains the same for different values of $k_{24}$. As shown by the panels in Fig.~\ref{fig:trajectories}, the only remarkable difference is that the minimizing trajectory resides on the line of genuine tactoids for a \emph{longer} stretch when $k_{24}$ is \emph{smaller}. This feature has two consequences. First, for a given (sufficiently small) value of $\alpha$, genuine tactoids are more \emph{slender} for smaller $k_{24}$. Second, the critical value $\azero$, which marks the extinction of genuine tactoids, decreases as $k_{24}$ increases. Actually, as shown in Fig.~\ref{fig:alpha_critical}, this is a property that $\azero$ shares with both $\aone$ and $\atwo$.
This means that as $k_{24}$ increases both tactoids and bumped spheroids persist only in smaller and smaller intervals for $\alpha$, giving way to larger colonies of  spheroids. In brief, we may say that $k_{24}$ is an  \emph{antidote} to slender shapes. In particular, the tactoidal population prospers only as $k_{24}$ decreases. 

\begin{figure}[h] 
	\includegraphics[width=.4\linewidth]{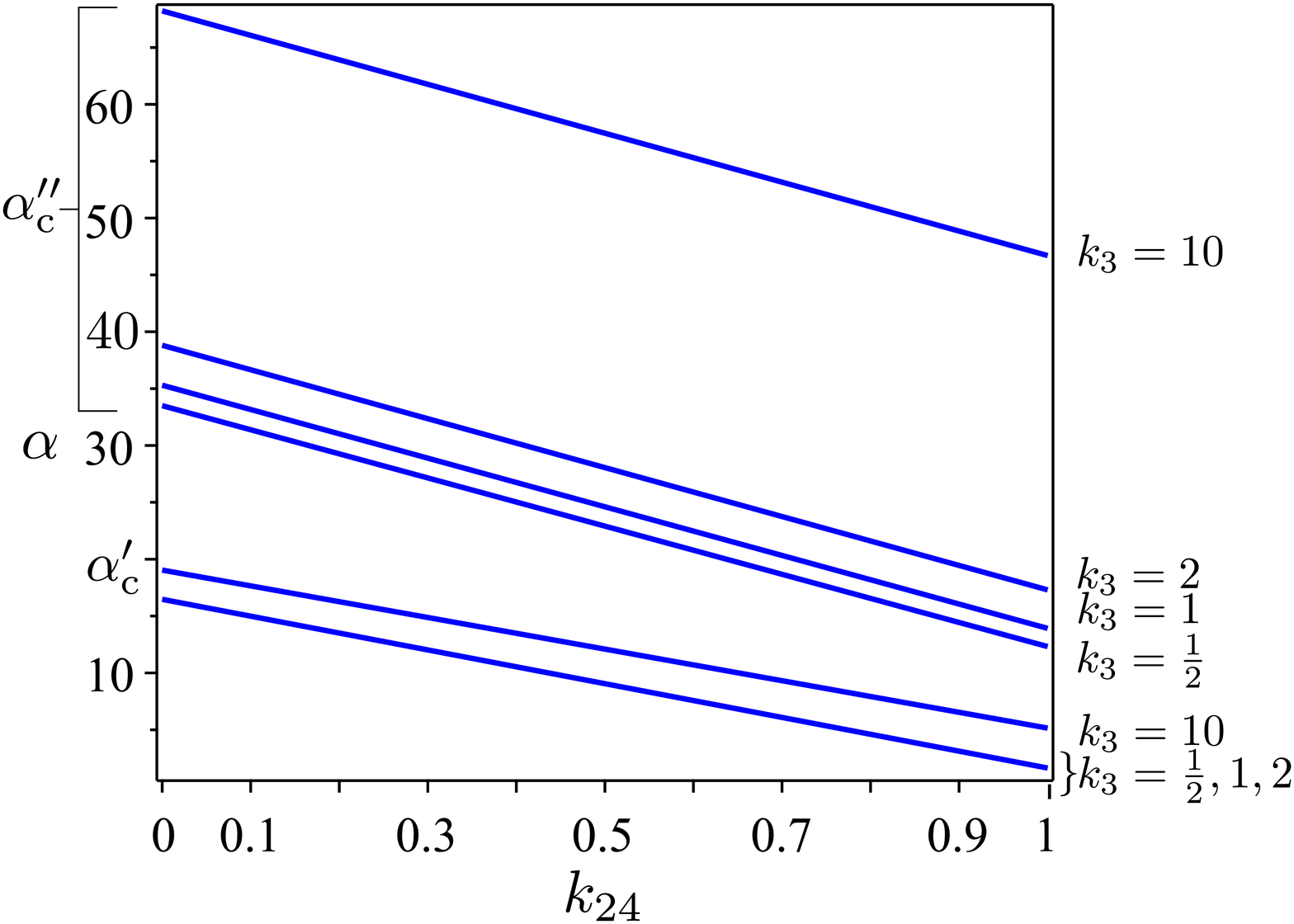}
	\caption{The critical values $\aone$ and $\atwo$ plotted against $k_{24}$ as in Fig.~\ref{fig:alpha_critical}, but for four different values of $k_3$, namely, $k_3=\frac12,1,2,10$. While the first three graphs of $\aone$ virtually coalesce on one another and are not discernible at this scale, the corresponding graphs of $\atwo$ are one on top of the other, ordered like the values of $k_3$.}
	\label{fig:alpha_critical_several_k3}
\end{figure}

Figure \ref{fig:alpha_critical_several_k3} shows how this characteristic is quantitatively affected by changes in $k_3$.  
While the graph of $\aone$ against $k_{24}$ is essentially the same for $k_3=\frac12,1,2$, it moves upward for $k_3=10$; correspondingly, all four graphs of $\atwo$ are orderly one on top of the other, keeping their features unchanged. A quantitative inspection shows that, for given $k_{24}$, both $\aone$ and $\atwo$ vary sub-linearly with $k_3$.\footnote{This was also confirmed by computations for $k_3=100$, not shown here.} This suggests that the latter just acts as a quantitative amplifier of the critical values. Thus, $\alpha$ and $k_{24}$ are the only  (dimensionless) parameters essential in describing how the droplet's shapes are distributed in the different families we have identified. 

In Sec.~\ref{sec:polupations}, we shall detail such a shape distribution, which is one distinctive feature of this work. Other works have illuminated the multiplicity of shapes exhibited by bipolar nematic droplets. To close this section, we show how these works relate to ours. 

\subsection{Comparison with Previous Work}\label{sec:comparison}
The variety of stable equilibrium shapes offered by nematic bipolar droplets has been the object of a remarkable series of theoretical papers, inspired by the seminal work of Williams~\cite{williams:transitions}. In particular, the papers \cite{prinsen:shape,prinsen:continuous,prinsen:parity,kaznacheev:nature,kaznacheev:influence} have followed in the same footsteps, sharing the original geometrical approach, which, as we shall see here, is unrelated to ours.

\begin{figure}[h!]  
	\includegraphics[width=.35\linewidth]{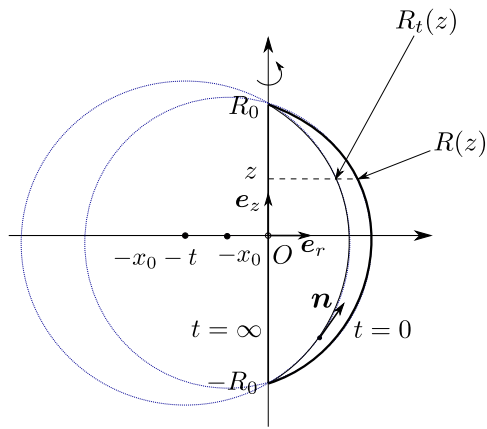}
	\caption{Apollonian family of shapes. The boundary $\boundary$ is obtained by rotating a circular segment about the $z$-axis, so that the profile $R(z)$ is described by \eqref{eq:apollonian_outer_circle}. The integral lines of the director field $\n$ are a family of circles represented by $R_t(z)$ in \eqref{eq:apollonian_inner_circles}. For $t\to\infty$, $R_t$ tends to the $z$-axis, whereas it represents the droplet's boundary for $t=0$.}
	\label{fig:apollonian}
\end{figure}

The class of admissible droplet shapes suggested by Williams includes spindles and spheres, all obtained by rotating about the $z$-axis a circular segment  hinged at the points  $z=\pm R_0$ (the poles of the drop), see Fig.~\ref{fig:apollonian}.
On a meridian cross-section of the drop, the integral lines of the family of admissible director fields are  Apollonian circles passing through both poles with radius increasing on approaching the $z$-axis (see, for example, $\S$\,2 of \cite{ogilvy:excursions}).  
Thus, in the parameterization introduced in Sec.~\ref{sec:retraction}, the boundary $\boundary$ of the drop is described by the function
\begin{equation}
	\label{eq:apollonian_outer_circle}
	R(z)=\sqrt{R_0^2+x_0^2-z^2}-x_0,\quad z\in[-R_0,R_0],
\end{equation}
where the point $(-x_0,0)$ in the $(\e_r,\e_z)$ plane  is the center of the bounding Apollonian circle (for $x_0=0$, the drop is spherical). Accordingly, the integral lines of  the director field $\n$ are circles in a family parameterized by $t\in[0,+\infty)$: 
\begin{equation}
	\label{eq:apollonian_inner_circles}
	R_t(z)=\sqrt{R_0^2+(x_0+t)^2-z^2}-(x_0+t),
\end{equation}
where the point $(-(x_0+t),0)$ in the $(\e_r,\e_z)$ plane is the center of a circle passing through the poles.
The curves of this family at $t=0$ and $t=+\infty$ correspond to the boundary of the drop and to its symmetry axis,  respectively.

It is now a simple exercise to show that this family of shapes does not fall within that introduced in Sec.~\ref{sec:retraction}, as there is no function $g(t)$ such that $R_t(z)$ in \eqref{eq:apollonian_inner_circles} could be expressed as $R_t(z)=g(t)R(z)$, with $R(z)$ as in \eqref{eq:apollonian_outer_circle}.
To afford a fair comparison between our approach and this one, we need to compare the minima of the total free energy $\free[\body]$ computed with the two methods.

The total free energy in \eqref{eq:F_energy_functional} associated with a droplet described by $R$ in \eqref{eq:apollonian_outer_circle} has been computed in \cite{prinsen:parity}, see in particular their equation (12), which we now transliterate in our language. Instead of $\mu$, defined in \eqref{eq:mu_definition}, Prinsen and van der Schoot~\cite{prinsen:parity} used the aspect ratio $\vae:=R(0)/R_0\leqq1$
to parameterize tactoids in their class of shapes. By letting the droplet's volume $V_0$ be expressed as in (A.14) of \cite{prinsen:parity} (where, incidentally, $R$ is to be identified with our $\Req$), we easily arrive at the relation
\begin{equation}\label{eq:mu_epsilon}
	\mu(\vae):=
	\frac{2}{\sqrt[3]{\left(\frac{1+\varepsilon^2}{\varepsilon}\right)^2\left(1-\frac{1-\varepsilon^2}{\varepsilon}\arctan\varepsilon\right)-4}},
\end{equation}
which shows how $\mu$ and $\vae$ are in a one-to-one correspondence, with $\mu(1)=1$ and $\mu(\vae)\to\infty$ as $\vae\to0$. In \cite{prinsen:parity}, the dimensionless strength of  surface energy is defined as
\begin{equation}
	\label{eq:upsilon}
	\upsilon:=V_0\left(\frac{\gamma\omega}{\widetilde{K}_{11}}\right)^3,
\end{equation}
where $\widetilde{K}_{11}:=K_{11}-K_{24}$ is the \emph{reduced} splay constant\footnote{As already recalled, in this theory, $K_{24}$ enters only through a renormalization of $K_{11}$.} and we replaced their $\tau$ with our $\gamma$ (with  the same physical meaning).  Making use of \eqref{eq:upsilon}, \eqref{eq:mu_epsilon}, and \eqref{eq:alpha} in (12) of \cite{prinsen:parity}, having noted that there $\widetilde{F}=\free/\gamma V_0^{2/3}$, we arrive at the following form for the reduced total free energy in terms of $\vae$,
\begin{equation}\label{eq:tactoid_free_energy_epsilon}
	F_t(\vae):=\frac{\free[\body]}{2\pi K_{11}\Req}=\mu(\vae)\bigg\{
    \left(2(1-k_{24})+\frac32k_3+\alpha\frac{1+\vae^2}{\vae}\mu(\vae)\right)\left(1-\frac{1-\vae^2}{\vae}\arctan\vae\right)- 2k_3\arctan^2\vae
	\bigg\}.
\end{equation}

It is not difficult to show that $F_t$ is a function that diverges like $1/\vae^{1/3}$ as $\vae\to0$ and has a single minimum for $0<\vae<1$, which approaches $\vae=1$ as $\alpha\to\infty$. Moreover, 
\begin{equation}
	\label{eq:F_t_1}
	F_t(1)=2+\left(\frac32-\frac{\pi^2}{8}\right)k_3+2(\alpha-k_{24}),
\end{equation}
which agrees with formula (2.18) of \cite{williams:transitions} for the reduced free energy of a bipolar sphere. Thus, in this theory, the minimum of the free energy is attained on a bipolar tactoid.\footnote{Leaving aside the possibility that it undergoes the twist instability first predicted in \cite{williams:transitions} for sufficiently small $K_{22}$.} Differently said, there is no critical value of $\alpha$ above which the equilibrium shape of the droplet becomes smooth, although remaining elongated, which is a feature of the theory presented in this paper.

To ascertain whether the smoothening transition that we predict is real or not, we need compare the minimum of $F_t(\vae)$ for $\vae$ in $[0,1]$ and the minimum of $\Fa(\phi,\mu)$ in $\conf$. Unfortunately, we do not have a general closed-form formula for $\Fa(\phi,\mu)$ to be compared with \eqref{eq:tactoid_free_energy_epsilon}, and so generically the comparison between minima is  to be  performed numerically.

There are two instances worth mentioning for which we can provide closed-form expressions for $\Fa(\phi,\mu)$. These are for $\phi=0$ and any $\mu$, corresponding to genuine tactoids, and for $\phi=\frac\pi2$ and $\mu=1$, corresponding to the sphere of radius $\Req$. We record both formulae in Appendix~\ref{sec:energies}, for the reader's convenience; the energy of the sphere in \eqref{eq:formula_sphere} is the one that especially interests us here. Contrasting it with \eqref{eq:F_t_1} shows that the asymptotic behavior as $\alpha\to\infty$ of $F_t(1)$ and $\Fa(\frac\pi2,1)$ is the same. Now, since the minimizing shapes for both $F_t$ and $\Fa$ converge to the sphere as $\alpha\to\infty$, we conclude that for sufficiently large $\alpha$ we cannot distinguish between the two theories. But we can for finite values of $\alpha$.

\begin{figure}[h] 
	\includegraphics[width=.4\linewidth]{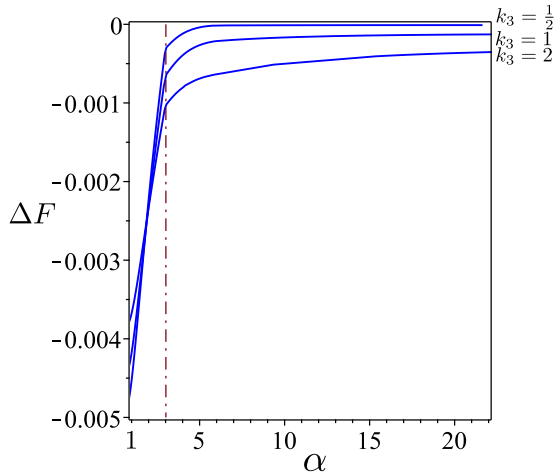}
	\caption{Plots of the relative energy difference $\Delta F$ against $\alpha$, according to \eqref{eq:energy_difference}, for $k_{24}=0.7$ and $k_3=\frac12,1,2$. The dashed straight line marks the critical value $\azero\doteq3.02$ (virtually identical in the three cases) where our theory predicts that the minimizing droplet's shape ceases to be a genuine tactoid.}
	\label{fig:energy_difference}
\end{figure}

We computed the relative energy difference $\Delta F$ defined as
\begin{equation}
	\label{eq:energy_difference}
	\Delta F(\alpha):=\frac{\min_{(\phi,\mu)}\Fa(\phi,\mu)-\min_{\vae}F_t(\vae)}{\min_{\vae}F_t(\vae)}.
\end{equation}
The graph of $\Delta F$ against $\alpha$ for $k_{24}=0.7$ and three values of $k_3$ is plotted in Fig.~\ref{fig:energy_difference};
it shows that $\Delta F$, although tiny in absolute value, is always \emph{negative}. This property has been confirmed for similar numerical computations performed for $k_{24}=0.2$ and $k_{24}=0.4$. We have thus good reasons to hold that smooth shapes, be they generic tactoids or bumped spheroids, are energetically more favorable than genuine tactoids. We shall substantiate this claim more quantitatively  in the following section.

\section{Shape populations}\label{sec:polupations}
Typical methods for generating liquid crystal droplets produce a wide range of droplet sizes. For simplicity, we assume that droplets are uniformly distributed in size within a certain volume interval $(V_0,V_0+\delta V)$. Correspondingly, in view of \eqref{eq:alpha}, for given isotropic interfacial tension and elastic constants, $\alpha$ ranges in an interval $(\alpha_0,\alpha_0+\delta\alpha)$.

We have already seen in Figs.~\ref{fig:alpha_critical} and \ref{fig:alpha_critical_several_k3} how the width of the strips in $(k_{24},\alpha)$ plane inhabited by different droplet's shapes depends on $k_{24}$. Assuming uniform distribution of droplets in a probe volume interval (parameterized in $\alpha$), we convert this information into a frequency of occurrence in the whole equilibrium shape population of the three distinctive shapes that we have identified as most easily 
recognizable, namely, tactoids (either genuine or not), bumped spheroids, and  spheroids. Formally, for given $k_{24}$, the frequency of occurrence $f$ of a shape is defined as the ratio of the  span of values of  $\alpha$ where the selected shape occurs at equilibrium over the whole explored range $\delta\alpha$. These frequencies  depend non-trivially on $k_{24}$; they suggest themselves as possible  statistical measures for $k_{24}$,  based on shape recurrence.

\begin{figure}[h] 
	\includegraphics[width=.3\linewidth]{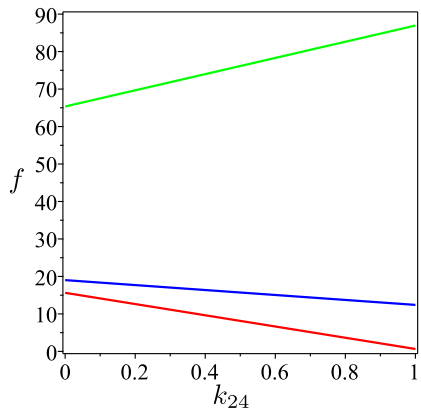}
	\caption{Relative frequencies (in percentage) for the occurrence of  tactoids (red), bumped spheroids (green) and spheroids (blue) in the population of equilibrium shapes. The three functions $f$ are plotted against  $k_{24}$; they have been computed for $k_3=1$ under the assumption that the droplet size is uniformly distributed in a range of volumes corresponding to $1<\alpha<100$.}
	\label{fig:freqk4}
\end{figure}

Figure \ref{fig:freqk4} shows the graphs of $f$ for the  three shape populations as functions of $k_{24}$, for $0<\alpha<100$.
It is clear that the population of tactoids is depleted as $k_{24}$ grows; the same trend (but with higher values) is exhibited by the population of bumped spheroids; the majority always lies with spheroids  when $\alpha$ ranges in an interval large enough to allow them to arise. Unlike tactoids and bumped spheroids (the elongated kin), spheroids are increased in number as $k_{24}$ increases. Thus,  $k_{24}$ depresses slim shapes, while fostering fat ones.

\begin{figure}[h]
	\centering
	\begin{subfigure}[t]{0.19\linewidth}
		\centering
		\includegraphics[width=\linewidth]{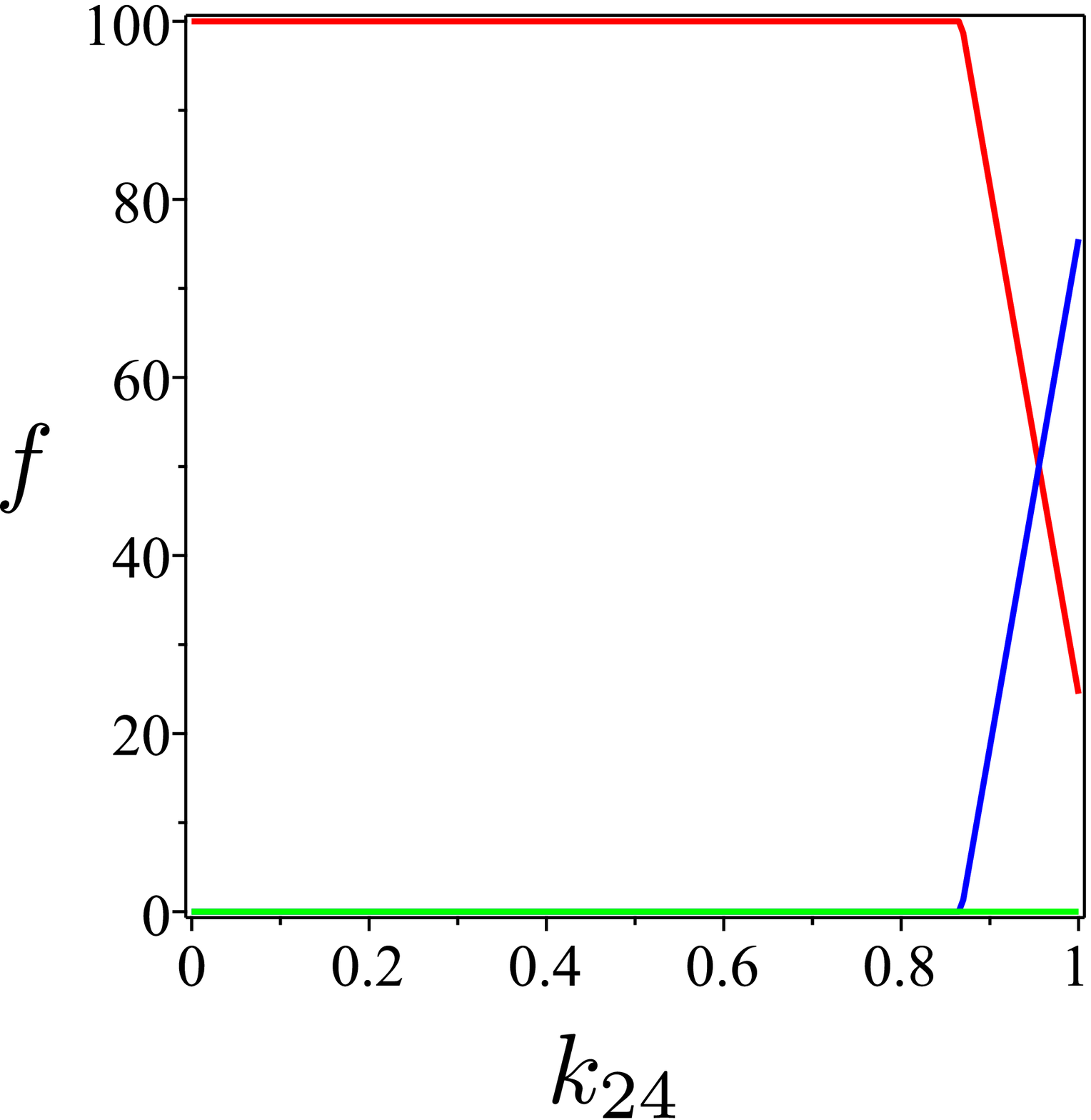}
		\caption{$1\leqq\alpha\leqq3.6$}
	\end{subfigure}
	\begin{subfigure}[t]{0.19\linewidth}
		\centering
		\includegraphics[width=\linewidth]{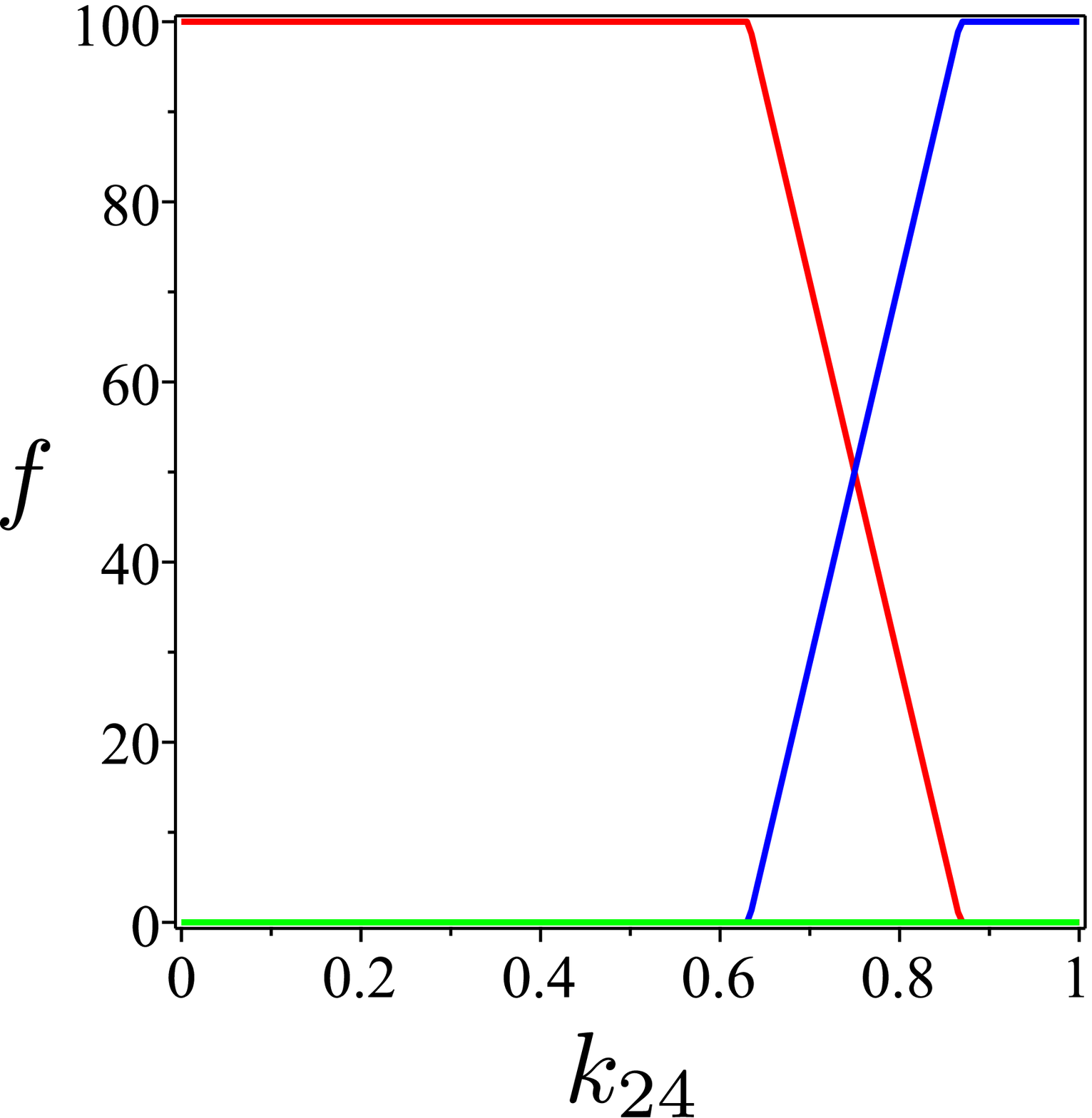}
		\caption{$3.6\leqq\alpha\leqq7.1$}
	\end{subfigure}
	\begin{subfigure}[t]{0.19\linewidth}
		\centering
		\includegraphics[width=\linewidth]{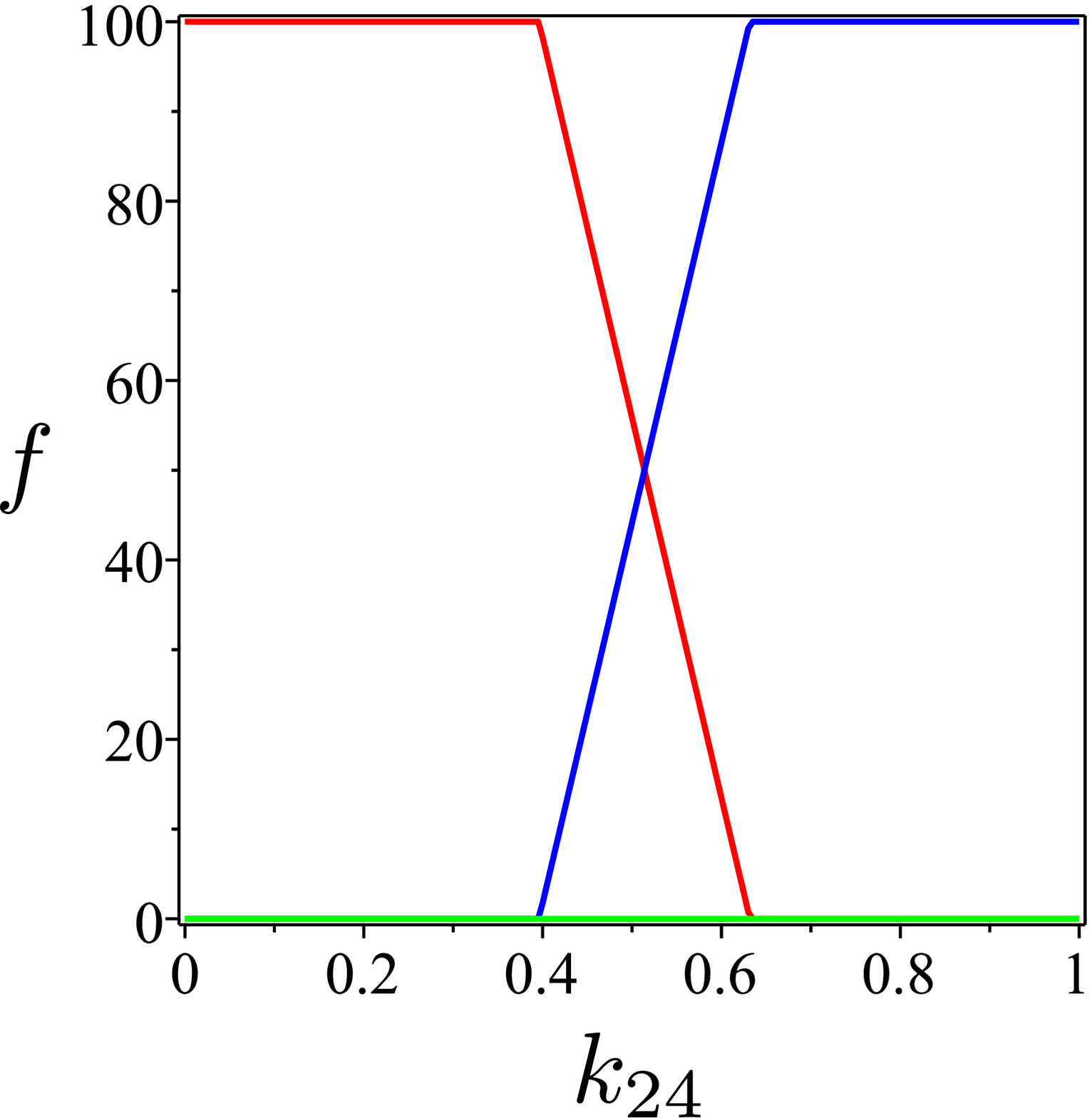}
		\caption{$7.1\leqq\alpha\leqq10.6$}
	\end{subfigure}
	\begin{subfigure}[t]{0.19\linewidth}
		\centering
		\includegraphics[width=\linewidth]{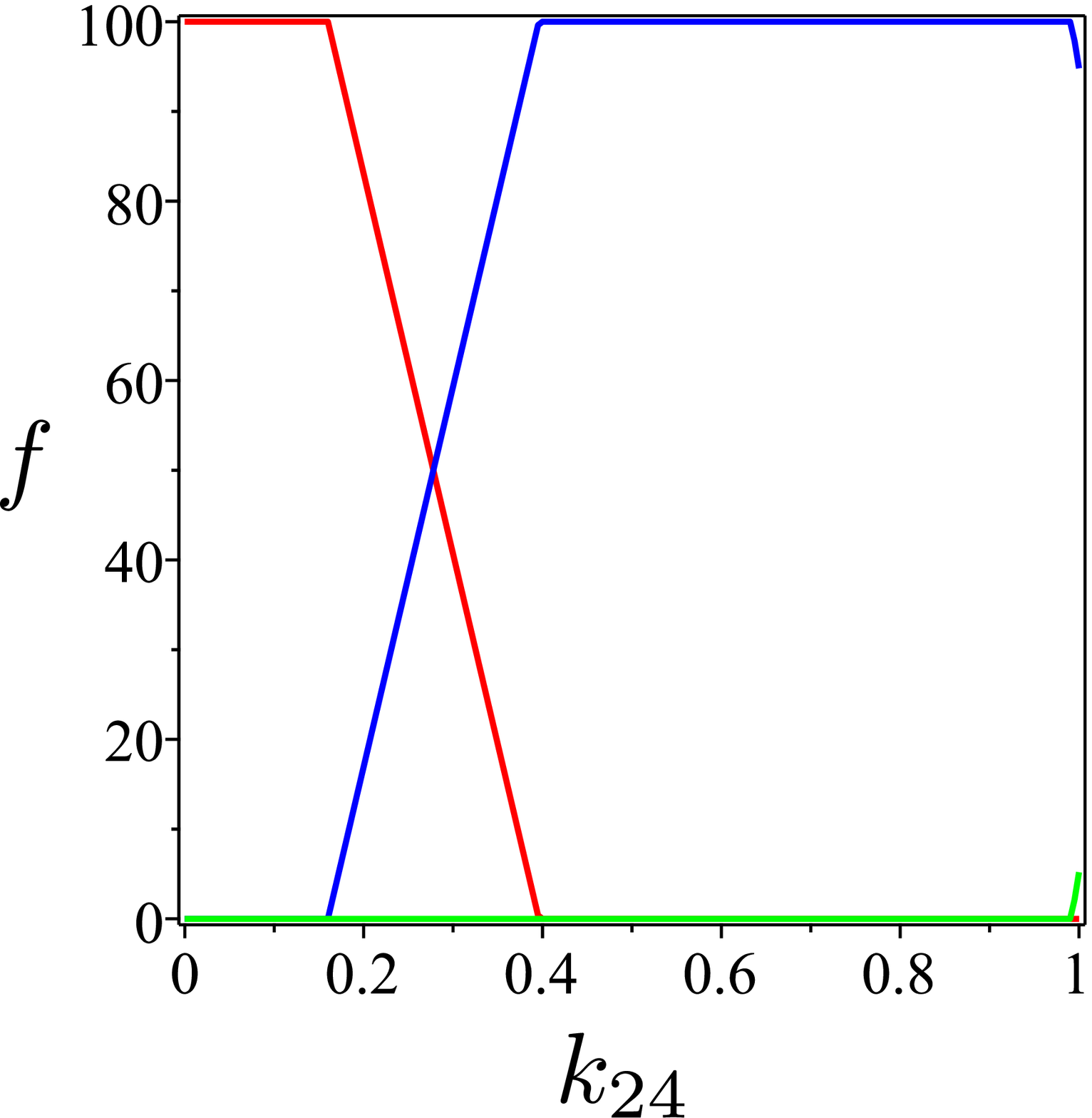}
		\caption{$10.6\leqq\alpha\leqq14.1$}
	\end{subfigure}
	\begin{subfigure}[t]{0.19\linewidth}
		\centering
		\includegraphics[width=\linewidth]{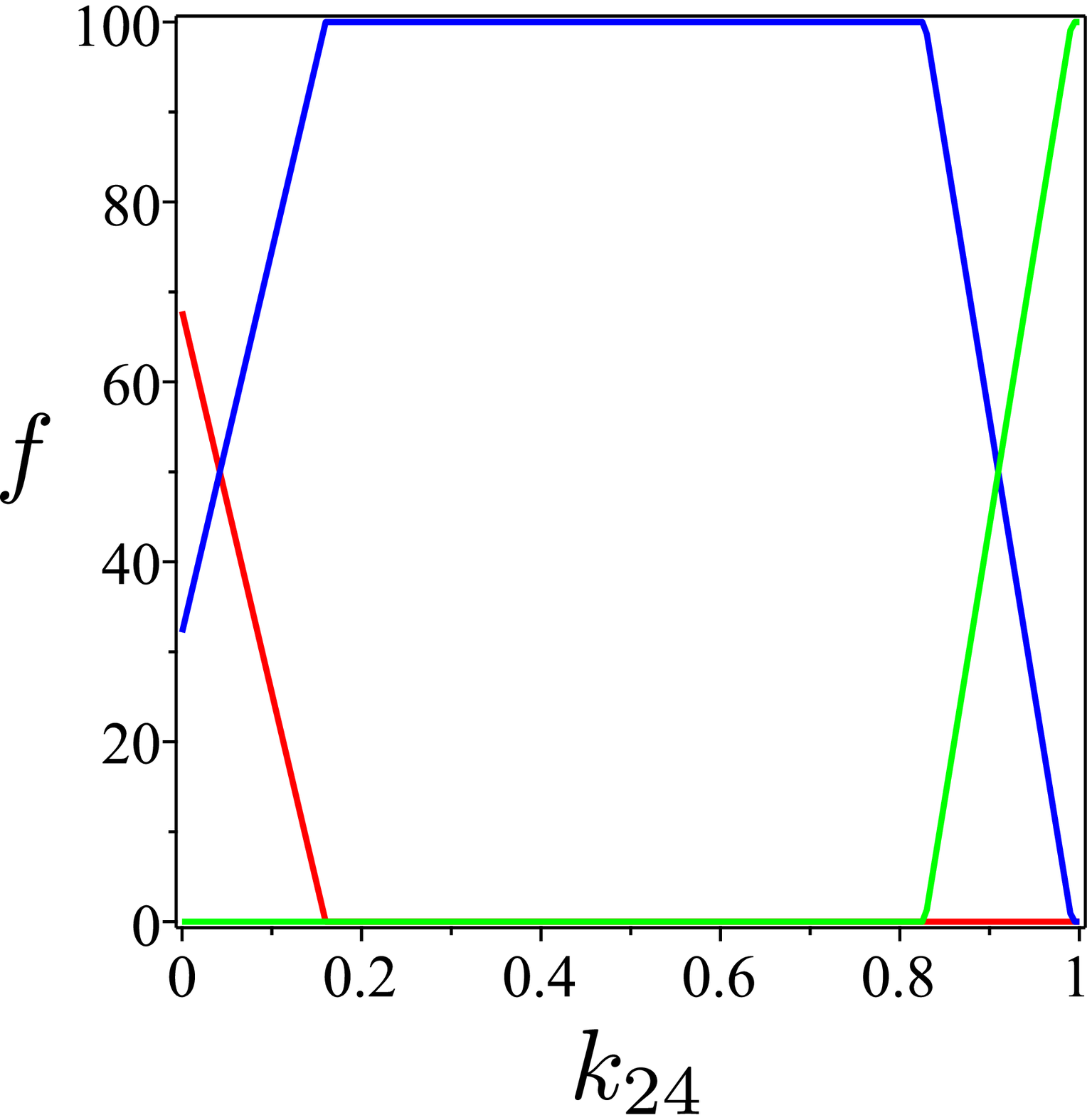}
		\caption{$14.1\leqq\alpha\leqq17.6$}
	\end{subfigure}
	\begin{subfigure}[t]{0.19\linewidth}
		\centering
		\includegraphics[width=\linewidth]{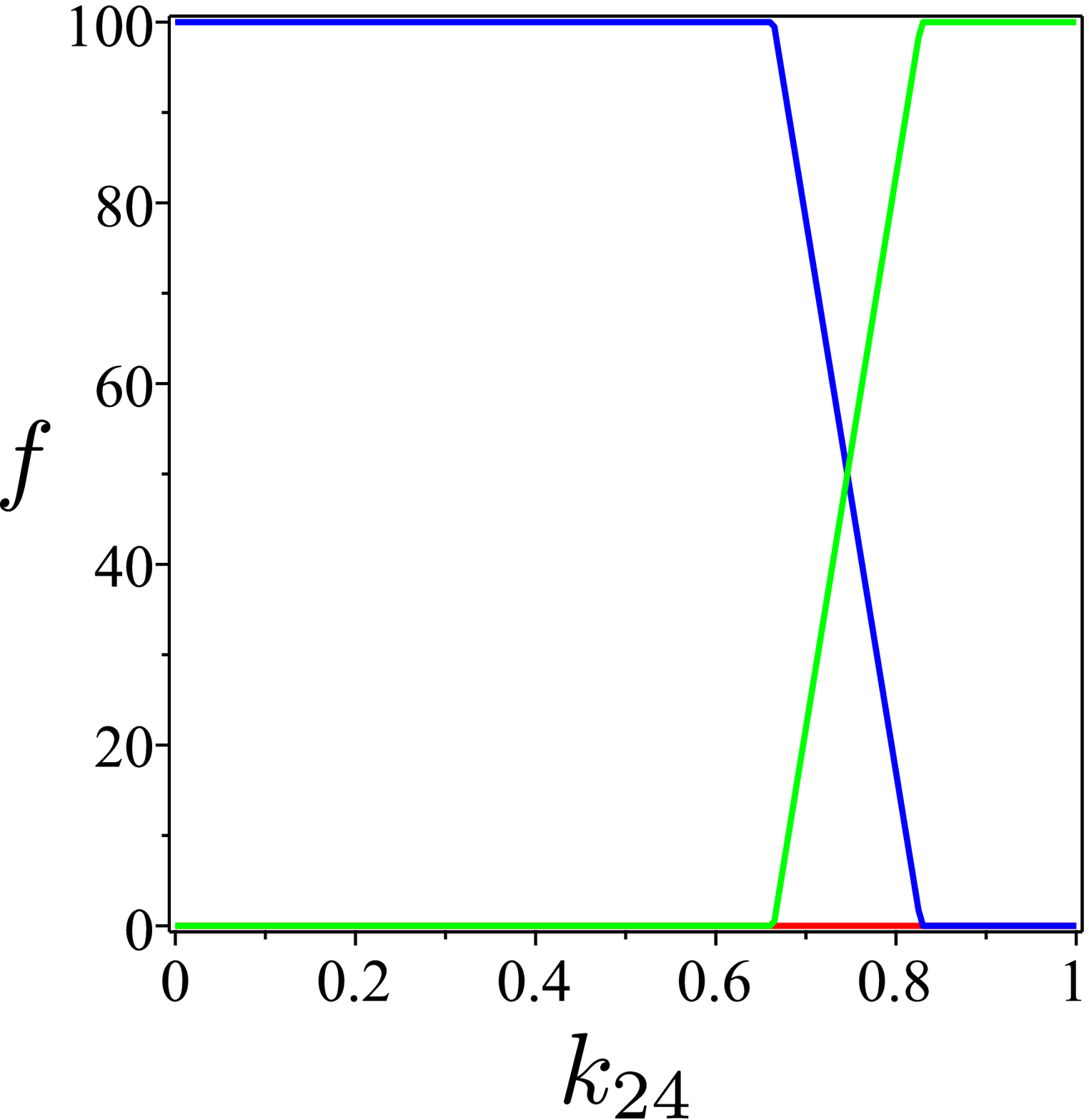}
		\caption{$17.6\leqq\alpha\leqq21.1$}
	\end{subfigure}
	\begin{subfigure}[t]{0.19\linewidth}
		\centering
		\includegraphics[width=\linewidth]{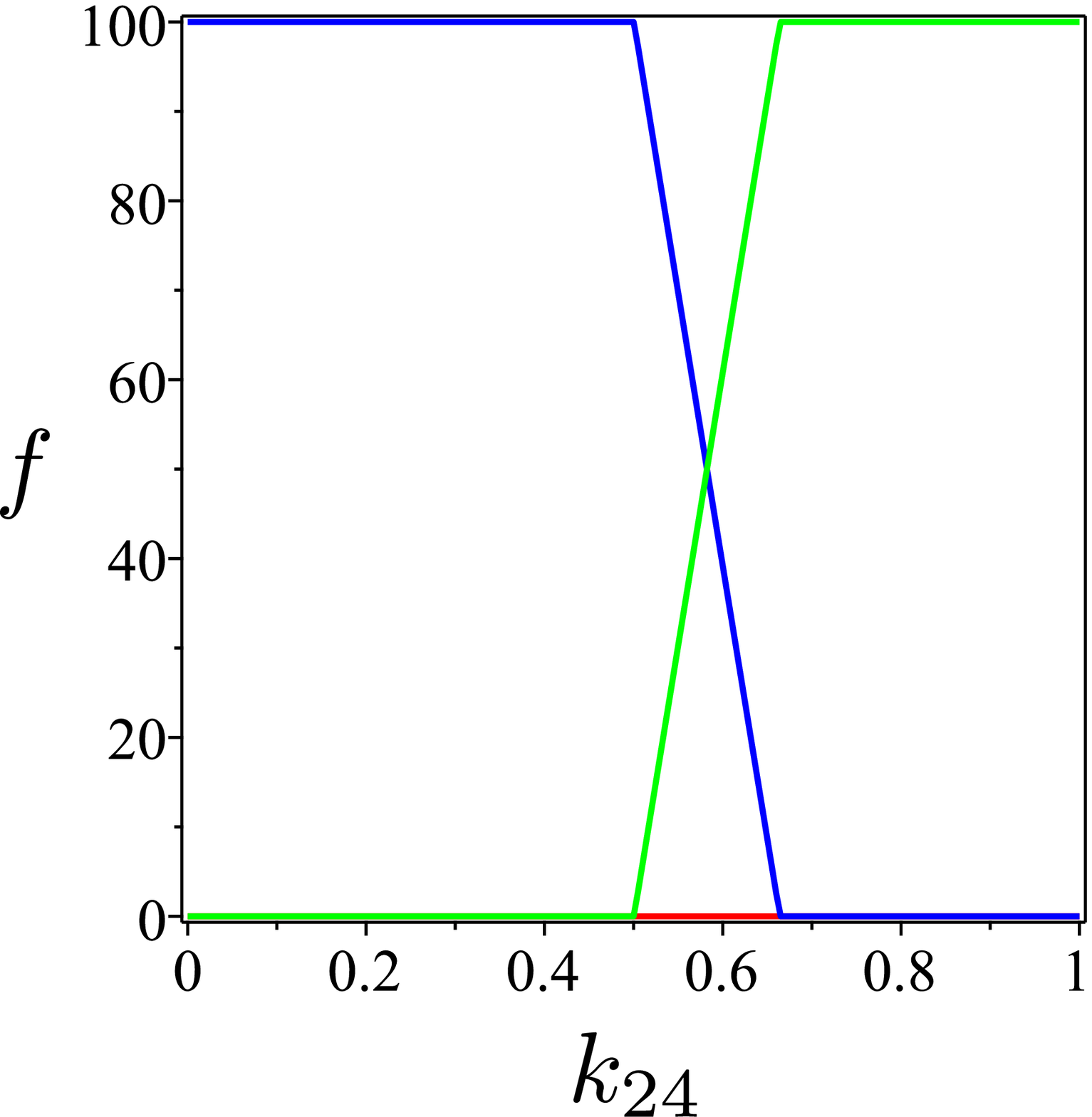}
		\caption{$21.1\leqq\alpha\leqq24.6$}
	\end{subfigure}
	\begin{subfigure}[t]{0.19\linewidth}
		\centering
		\includegraphics[width=\linewidth]{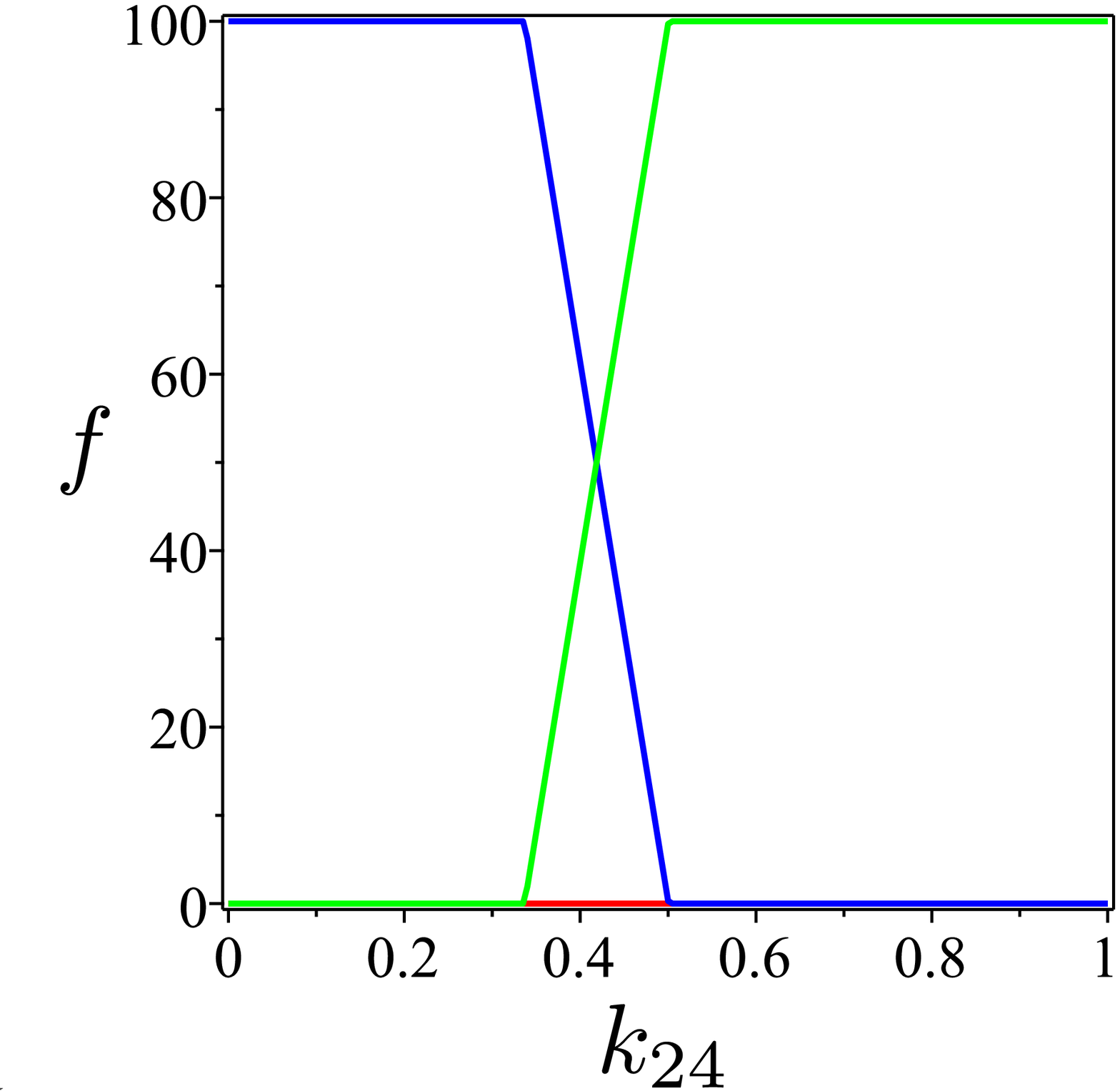}
		\caption{$24.6\leqq\alpha\leqq28.1$}
	\end{subfigure}
	\begin{subfigure}[t]{0.19\linewidth}
		\centering
		\includegraphics[width=\linewidth]{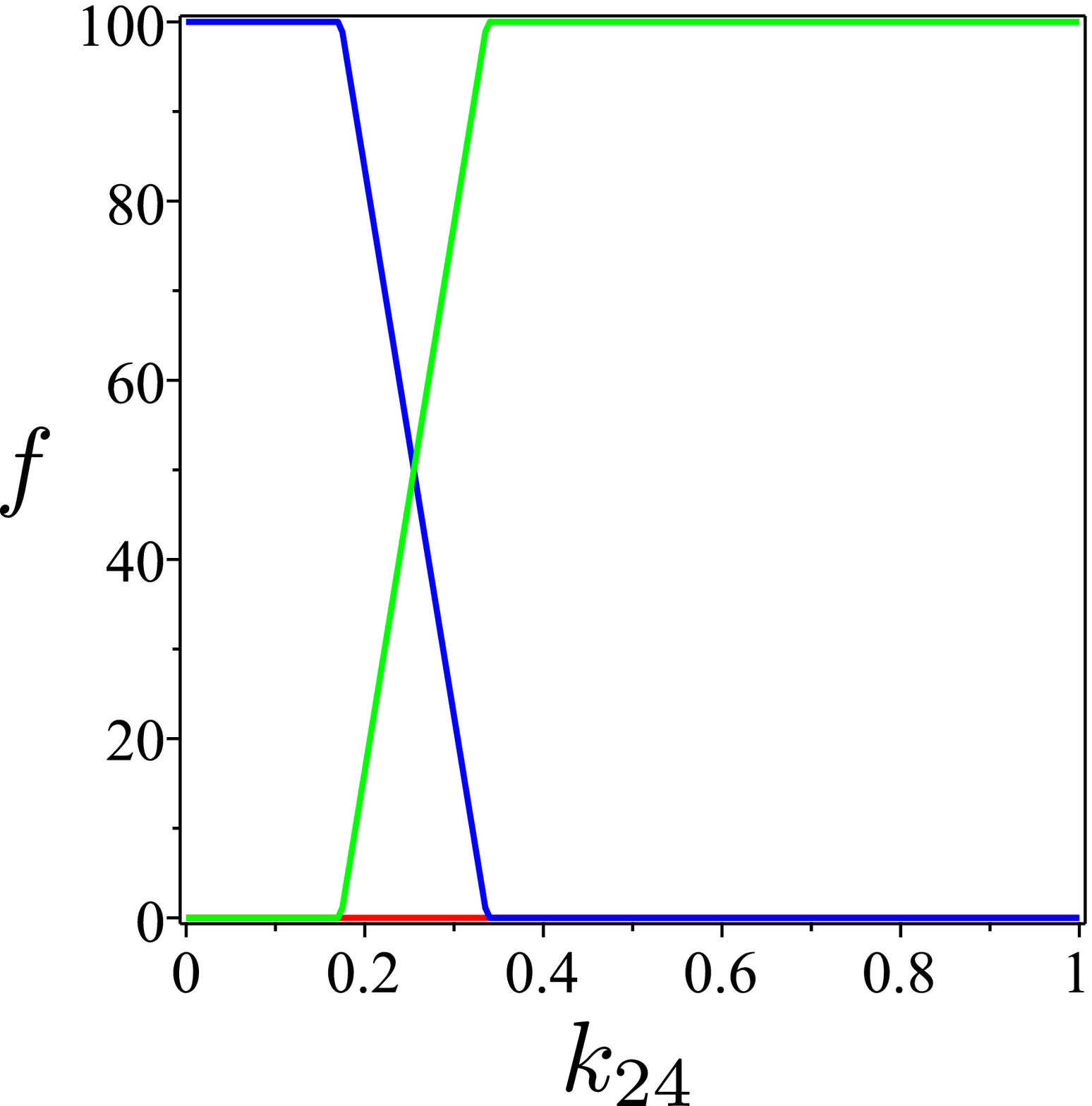}
		\caption{$28.1\leqq\alpha\leqq31.6$}
	\end{subfigure}
	\begin{subfigure}[t]{0.19\linewidth}
		\centering
		\includegraphics[width=\linewidth]{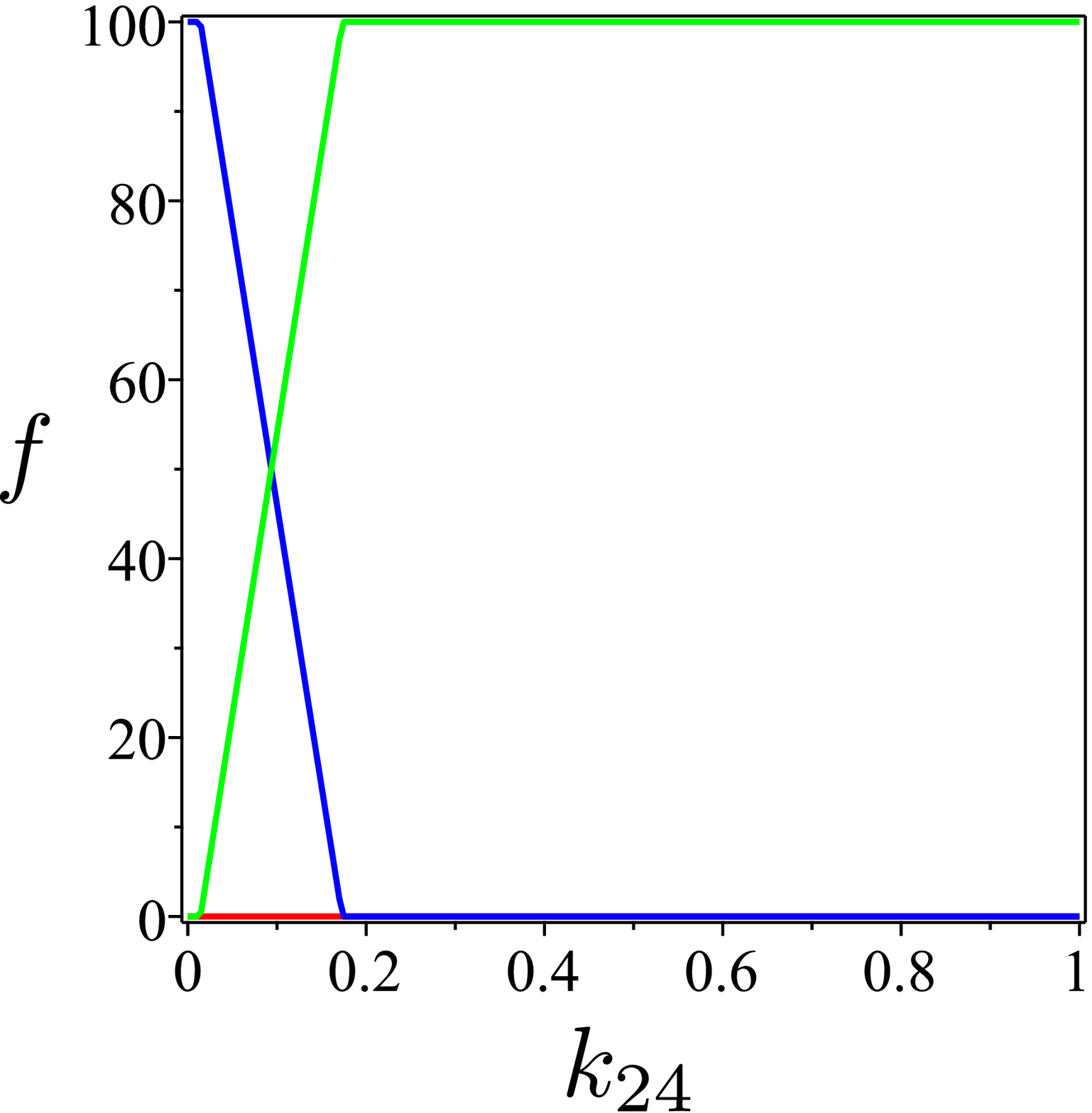}
		\caption{$31.6\leqq\alpha\leqq35$}
	\end{subfigure}
	\caption{Relative frequencies (in percentage) for the occurrence of tactoids (red), bumped spheroids (green) and spheroids (blue) as functions of $k_{24}$, for $k_3=1$ and increasing values of the average volume in the specified range, under the assumption of uniform distribution in droplet size.}
	\label{fig:averagefreq}
\end{figure}

In Fig.~\ref{fig:averagefreq}, we illustrate a finer analysis of the frequencies of shapes, performed on a sequence of elementary volume intervals of equal amplitude, $\delta\alpha=3.5$.
Such a splitting of the whole range of droplet volumes in smaller intervals around increasing values reveals different scenarios in shape populations. 
When the average volume is small, tactoids dominate over bumped spheroids, for a wide range in $k_{24}$, the \emph{equal population} point being close to $k_{24}=1$.  As the average volume increases, the equal population point decreases, until bumped spheroids displace tactoids completely. As the average volume further increases, bumped spheroids are challenged by spheroids, which first reach an equal population point with bumped spheroids close to $k_{24}=1$ and then eventually dominate the scene completely, as the volume is further increased.

The simple morale of the whole story is that in bipolar nematic droplets the population of tactoids (and elongated shapes, in general) is favored by small saddle-splay elastic constants (compared to the splay constant), provided that the droplet (dimensionless) volume is not too large.

\begin{figure}[h] 
	\includegraphics[width=.3\linewidth]{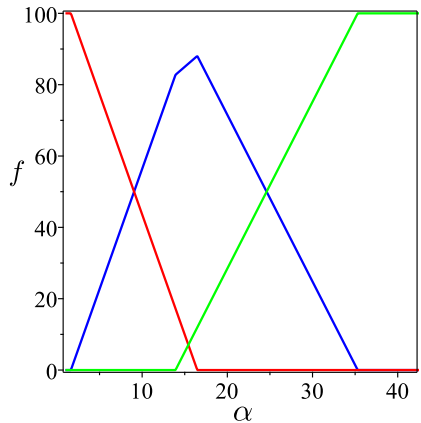}
	\caption{Relative frequencies (in percentage) for the occurrence of tactoids (red), bumped spheroids (green) and  spheroids (blue) as functions of the (reduced) volume $\alpha$, for $k_3=1$ and under the assumption of uniform distribution in the (dimensionless) saddle-splay constant $k_{24}$.}
	\label{fig:freqalpha}
\end{figure}

To isolate the role played by the droplet's volume in the distribution of equilibrium shapes, in Fig.~\ref{fig:freqalpha} we plot the relative frequency of the three shapes as functions of $\alpha$.  
These graphs are extracted from Fig.~\ref{fig:alpha_critical}, as were those in Fig.~\ref{fig:freqk4}, but assuming uniformity in the distribution of $k_{24}$. Again, we may say that small, intermediate, large  volumes promote tactoids, bumped spheroids, and spheroids, respectively. But, perhaps, the most interesting feature shown in Fig.~\ref{fig:freqalpha} is the coexistence of all shapes for moderate volumes. 

\section{Conclusions}\label{sec:conclusions}
This paper took a census of all possible shapes that a bipolar droplet of nematic liquid crystal can have upon varying its volume and the elastic constants of the material that constitutes it. In the adopted class of shapes, we found that either tactoids (genuine or not), bumped spheroids, and spheroids can be optimal. The prevalence in population of one shape is determined by the volume $V_0$ and the saddle-splay constant $K_{24}$ (appropriately scaled). One may say that tactoids prevail when both volume and saddle-splay  constant are small (with the bend constant $K_{33}$ acting as an amplifying factor). But there is more to it: for a given average volume $V_0$, the prevalence in shape population changes upon increasing $K_{24}$, shifting first from tactoids to bumped spheroids, and then from the latter to spheroids, as $V_0$ is increased. It may be a stretch to think that our ``demographic'' analysis has the potential to indicate the ballpark where to find the ratio $K_{24}/K_{11}$ of a specific material, provided we can produce droplets in a range of wide enough volumes.

In comparing our work with others, we saw that the optimal shapes we find in our class may have slightly less energy than shapes found in other classes, but the qualitative difference in shapes was substantial even if the gain in energy was marginal. This adds to the difficulty of the problem tackled here, indicating that the energy minimum is rather shallow.

Williams~\cite{williams:transitions} studied the stability of a bipolar spherical droplet against twisting distortions. This study was sharpened and extended to tactoids by Prinsen and van der Schoot~\cite{prinsen:parity}. As expected, small values of the twist constant $K_{22}$  (relative to $K_{11}$ and $K_{33}$) promote a twisting instability in the director field, which start exhibiting a chiral pattern around the symmetry axis. It would be desirable to find the critical value of $K_{22}$ above which the bipolar droplets studied in this paper are stable, as our conclusions are valid only in this regime. In light of the role played by $K_{24}$ in determining the optimal bipolar droplet, we expect that their range of stability would also be affected in novel ways. This is likely to shed new light on the chiral symmetry breaking exhibited by tactoids in some chromonic liquid crystals \cite{tortora:chiral,jeong:chiral_2014,peng:chirality}. 

\appendix
\section{Retracted Meridian Field}\label{sec:retracted_coordinates}
Our aim here is to justify the expression  \eqref{eq:nabla_n} for the gradient of the the retracted meridian field $\n$ in \eqref{eq:n}.

First, we remark that differentiating $\n$ along the smooth curve $\xi\mapsto(t(\xi),\vt(\xi),z(\xi))$ introduced in Sec.~\ref{sec:shape_class} we easily obtain from \eqref{eq:n} that
\begin{equation}
\label{eq:dot_n}
\dot{\n}=\frac{g'R'\dot{t}+gR''\dot{z}}{1+(gR')^2}\nper+\frac{gR'\dot{\vt}}{\sqrt{1+(gR')^2}}\e_\vt,
\end{equation}
where a superimposed dot denotes differentiation with respect to $\xi$ and use also been made of \eqref{eq:n_perp}.

Now, $\nabla\n$ must be such that 
\begin{equation}
\label{eq:dot_n_=}
\dot{\n}=(\nabla\n)\dot{\p},
\end{equation}
where $\dot{\p}$ is as in \eqref{eq:dot_curve_t_rewritten} for arbitrary $(\dot{t},\dot{\vt},\dot{z})$. Since $\n$ is a unit vector field and we wish to express its gradient $\nabla\n$ in the orthonormal frame $\framen$, we can write
\begin{equation}
\label{eq:nabla_n_representation}
\nabla\n=\nper\otimes\bm{a}+\e_\vt\otimes\bm{b},
\end{equation}
where $\bm{a}=a_1\n+a_2\nper+a_3\e_\vt$ and $\bm{b}=b_1\n+b_2\nper+b_3\e_\vt$, with $a_i$ and $b_i$ scalar components to be determined. Thus, \eqref{eq:dot_n_=} also reads as
\begin{equation}
\label{eq:dot_n_==}
\dot{\n}=(\bm{a}\cdot\dot{\p})\nper+(\bm{b}\cdot\dot{\p})\e_\vt.
\end{equation}
Making use of \eqref{eq:dot_curve_t_rewritten} and both \eqref{eq:n} and \eqref{eq:n_perp}, we readily see that 
\begin{equation}\label{eq:x_dotted_with_dot_p}
\bm{x}\cdot\dot{\p}=\frac{g'R(gR'x_1+x_2)\dot{t}}{\sqrt{1+(gR')^2}}+gRx_3\dot{\vt}+\sqrt{1+(gR')^2}x_1\dot{z},	
\end{equation}
for any vector $\bm{x}=x_1\n+x_2\nper+x_3\e_\vt$. Specializing \eqref{eq:x_dotted_with_dot_p} for $\bm{x}=\bm{a}$ and $\bm{x}=\bm{b}$ and inserting both resulting equations  in \eqref{eq:dot_n_==} alongside with 
\eqref{eq:dot_n}, we obtain an identity for arbitrary $(\dot{t},\dot{\vt},\dot{z})$ only if the components of $\bm{a}$ and $\bm{b}$ in the frame $\framen$ are given by
\begin{subequations}\label{eq:a_b_components}
\begin{align}
a_1&=\frac{gR''}{\big(1+(gR')^2\big)^{3/2}},\quad a_2=\frac{R'}{R}\frac{1}{\sqrt{1+(gR')^2}}-\frac{g^2R'R''}{\big(1+(gR')^2\big)^{3/2}},\quad a_3=0,\\
b_1&=b_2=0,\quad b_3=\frac{R'}{R}\dfrac{1}{\sqrt{1+(gR')^2}},
\end{align}
\end{subequations}
which with the aid of \eqref{eq:nabla_n_representation} deliver \eqref{eq:nabla_n} in the main text.

\section{Tactoidal Measure}\label{sec:tau}
In this Appendix, we introduce a tactoidal measure to justify the conventional choices made in Sec.~\ref{sec:taxonomy} to classify the different shapes that inhabit  the special family represented by \eqref{eq:profile}. 

\begin{figure}[h] 
	\includegraphics[width=.25\linewidth]{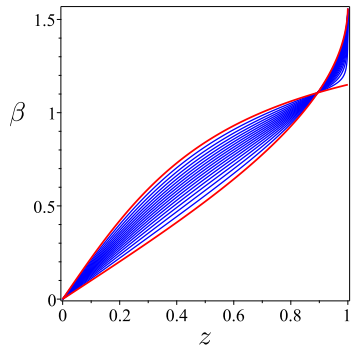}
	\caption{Graphs of $\beta$ against $z$ according to \eqref{eq:beta_against_z} for several values of $0\leqq\phi\leqq\frac\pi2$. The concave red curve corresponds to $\phi=0$, whereas the convex red curve corresponds to $\phi=\frac\pi2$. All other (blue) curves in the pencil interpolating the red curves have an inflection point (besides that at $z=0$), which defines $\tau$.}
	\label{fig:beta_against_z}
\end{figure}

Consider the angle $\beta$ that the tangent to the drop's profile makes with the symmetry axis (see Fig.~\ref{fig:shape_cross_section}). If for $\mu=1$ we draw the graph of $\beta$ as a function of $z$, we observe a drastic difference in the two cases $\phi=0$ and $\phi=\frac\pi2$, corresponding to a shape $\body$ that is a genuine tactoid and the round sphere, respectively. In the former case, the graph is concave, whereas it is convex in the latter. There is indeed more to that: as also shown in Fig.~\ref{fig:beta_against_z}, as soon as $\phi>0$, the graph of $\beta$ exhibits an inflection point at $z=\tau>0$, which slides gradually towards $z=0$ (corresponding to the equator of the drop) as $\phi$ increases towards $\frac\pi2$.\footnote{It is perhaps in order to remark that $z=0$ is an inflection point for $\beta(z)$, for all values of $\phi$, see also \eqref{eq:beta_against_z}.} 

It is precisely $\tau$ that we take as a \emph{tactoidal measure}. The closer is $\tau$ to unity, the more likely is $\body$ to look like a tactoid (even if its outer unit normal $\normal$ is continuous throughout $\boundary$). Formally,
\begin{equation}
\label{eq:beta_against_z}
\beta=\arctan\left(\frac{2\cos\phi}{\sqrt{h(\phi)}}z+\frac{\sin\phi}{\sqrt{h(\phi)}}\frac{z}{\sqrt{1-z^2}}\right)
\end{equation}
and $\tau$ is defined as the positive root of the equation $\beta''(z)=0$.

\begin{figure}[h] 
	\includegraphics[width=.25\linewidth]{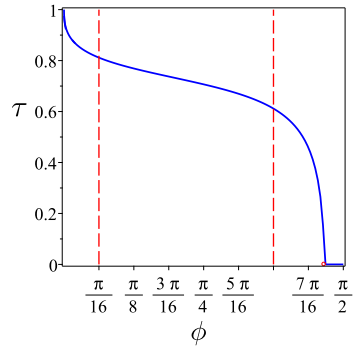}
	\caption{The graph of the tactodial measure $\tau$ against $\phi$. The circle marks the point at $\phi\doteq1.46\,\rad$, where $\tau$ bifurcates off the trivial inflection point for $\beta$ at $z=0$. The two dashed vertical (red) lines delimit the nearly linear behavior of $\tau$ between two ``knees'' (one upward and the other downward); they are placed approximately at $\phi=\frac{\pi}{16}$ and $\phi=\frac{6\pi}{16}$, which are precisely the barriers conventionally introduced in configuration space to delimit the range of bumped spheroids, see Fig.~\ref{fig:configuration_space}.}
	\label{fig:Tau_against_phi}
\end{figure}

Figure~\ref{fig:Tau_against_phi} illustrates how $\tau$ depends on $\phi$.
A simple asymptotic analysis shows that $1-\tau=O(\phi^{2/5})$ as $\phi\to0$ and that a bifurcation of $\tau$ occurs out of the trivial inflection point of $\beta$ at $z=0$ for $\phi\doteq1.46\,\rad$. It is remarkable how the graph of $\tau$ in Fig.~\ref{fig:Tau_against_phi} exhibits a nearly linear behavior between two ``knees'', the first upward and the second downward, which are approximately placed at $\phi=\frac{\pi}{16}$ and $\phi=\frac{6\pi}{16}$. In Sec.~\ref{sec:taxonomy}, we interpreted the former as the upper limit for a tactoidal shape and the latter as the upper limit for what we called a bumped spheroid.

We fully appreciate that a good deal of conventionality remains attached to this choice of ours and to the taxonomy of shapes that ensued in Sec.~\ref{sec:taxonomy}. Perhaps, the best way to convince the reader that it has some merit is to see it at work in Fig.~\ref{fig:gallery}. In any event, here we have recounted the (possibly meandering) path that we took to identify the barriers that delimit the range of bumped spheroids in the configuration space shown in Fig.~\ref{fig:configuration_space}. 
 
\section{The energy of genuine tactoids and sphere}\label{sec:energies}
Here we record the closed-form formulae that can be obtained by performing the integrals in \eqref{eq:F_energy_functional} for special values of the parameters $(\phi,\mu)$ featuring in \eqref{eq:profile} via \eqref{eq:a_b_representation}.

For genuine tactoids, corresponding to $\phi=0$ and any $\mu$,
\begin{equation}
	\label{eq:formula_genuine_tactoids}
\begin{split}		
\Fa(0,\mu)&=\sqrt{5\mu}\arctan\sqrt{\frac{5}{\mu^3}}\left[\frac15\left(\frac{47}{32}-\frac{3}{32}k_3-k_{24}\right)\mu^2+\left(\frac{1}{16}(13-k_3)-k_{24} \right)\frac{1}{\mu}+\frac{5}{32}\left(\frac13+k_3\right)\frac{1}{\mu^4} \right]\\
&+\left(\frac13\ln\left(1+\frac{5}{\mu^3}\right)+\frac{1}{32}(3k_3-47)+k_{24}\right)\mu+\frac{5}{32}\left(\frac13+k_3\right)\frac{1}{\mu^2}\\
&+\frac12\alpha\left[\ln\left(\sqrt{\frac{5}{\mu^3}}+\sqrt{1+\frac{5}{\mu^3}}\right)\left(1+\frac{1}{20}\mu^3\right)\mu^2+\frac{\sqrt{5}}{2}\sqrt{\mu^3+5}\left(\frac{1}{\mu}-\frac{\mu^2}{10}\right)\right].
\end{split}
\end{equation}
In complete analogy to formula \eqref{eq:tactoid_free_energy_epsilon} for $F_t$ in the main text, this function has a unique minimum for $\mu$ in $[0,\infty)$; it diverges to $+\infty$ like $1/\mu^{7/2}$ as $\mu\to0$, and like $\sqrt{\mu}$ as $\mu\to\infty$.

For a sphere of radius $\Req$, corresponding to $\phi=\frac\pi2$ and $\mu=1$,
\begin{equation}
	\label{eq:formula_sphere}
	\Fa({\textstyle\frac\pi2},1)=
	\frac13\ln2+\frac{23}{12}+\frac14k_3+2(\alpha-k_{24}).
\end{equation}


\begin{thebibliography}{79}%
	\makeatletter
	\providecommand \@ifxundefined [1]{%
		\@ifx{#1\undefined}
	}%
	\providecommand \@ifnum [1]{%
		\ifnum #1\expandafter \@firstoftwo
		\else \expandafter \@secondoftwo
		\fi
	}%
	\providecommand \@ifx [1]{%
		\ifx #1\expandafter \@firstoftwo
		\else \expandafter \@secondoftwo
		\fi
	}%
	\providecommand \natexlab [1]{#1}%
	\providecommand \enquote  [1]{``#1''}%
	\providecommand \bibnamefont  [1]{#1}%
	\providecommand \bibfnamefont [1]{#1}%
	\providecommand \citenamefont [1]{#1}%
	\providecommand \href@noop [0]{\@secondoftwo}%
	\providecommand \href [0]{\begingroup \@sanitize@url \@href}%
	\providecommand \@href[1]{\@@startlink{#1}\@@href}%
	\providecommand \@@href[1]{\endgroup#1\@@endlink}%
	\providecommand \@sanitize@url [0]{\catcode `\\12\catcode `\$12\catcode
		`\&12\catcode `\#12\catcode `\^12\catcode `\_12\catcode `\%12\relax}%
	\providecommand \@@startlink[1]{}%
	\providecommand \@@endlink[0]{}%
	\providecommand \url  [0]{\begingroup\@sanitize@url \@url }%
	\providecommand \@url [1]{\endgroup\@href {#1}{\urlprefix }}%
	\providecommand \urlprefix  [0]{URL }%
	\providecommand \Eprint [0]{\href }%
	\providecommand \doibase [0]{https://doi.org/}%
	\providecommand \selectlanguage [0]{\@gobble}%
	\providecommand \bibinfo  [0]{\@secondoftwo}%
	\providecommand \bibfield  [0]{\@secondoftwo}%
	\providecommand \translation [1]{[#1]}%
	\providecommand \BibitemOpen [0]{}%
	\providecommand \bibitemStop [0]{}%
	\providecommand \bibitemNoStop [0]{.\EOS\space}%
	\providecommand \EOS [0]{\spacefactor3000\relax}%
	\providecommand \BibitemShut  [1]{\csname bibitem#1\endcsname}%
	\let\auto@bib@innerbib\@empty
	\bibitem [{\citenamefont {Zocher}\ and\ \citenamefont
		{Jacobsohn}(1929)}]{zocher:taktosole}%
	\BibitemOpen
	\bibfield  {author} {\bibinfo {author} {\bibfnamefont {H.}~\bibnamefont
			{Zocher}}\ and\ \bibinfo {author} {\bibfnamefont {K.}~\bibnamefont
			{Jacobsohn}},\ }\bibfield  {title} {\bibinfo {title} {\"{U}ber {T}aktosole},\
	}\href@noop {} {\bibfield  {journal} {\bibinfo  {journal} {Kolloidchem.
				Beihefte}\ }\textbf {\bibinfo {volume} {28}},\ \bibinfo {pages} {167}
		(\bibinfo {year} {1929})}\BibitemShut {NoStop}%
	\bibitem [{\citenamefont {Sonin}(1998)}]{sonin:inorganic}%
	\BibitemOpen
	\bibfield  {author} {\bibinfo {author} {\bibfnamefont {A.~S.}\ \bibnamefont
			{Sonin}},\ }\bibfield  {title} {\bibinfo {title} {Inorganic lyotropic liquid
			crystals},\ }\href@noop {} {\bibfield  {journal} {\bibinfo  {journal} {J.
				Mater. Chem.}\ }\textbf {\bibinfo {volume} {8}},\ \bibinfo {pages} {2557}
		(\bibinfo {year} {1998})}\BibitemShut {NoStop}%
	\bibitem [{\citenamefont {Zocher}(1925)}]{zocher:freiwillige}%
	\BibitemOpen
	\bibfield  {author} {\bibinfo {author} {\bibfnamefont {H.}~\bibnamefont
			{Zocher}},\ }\bibfield  {title} {\bibinfo {title} {\"{U}ber freiwillige
			{S}trukturbildung in {S}olen. ({E}ine neue {A}rt anisotrop fl\"ussiger
			{M}edien.)},\ }\href@noop {} {\bibfield  {journal} {\bibinfo  {journal} {Z.
				anorg. allg. Chem.}\ }\textbf {\bibinfo {volume} {147}},\ \bibinfo {pages}
		{91} (\bibinfo {year} {1925})}\BibitemShut {NoStop}%
	\bibitem [{\citenamefont {Watson}\ \emph {et~al.}(1949)\citenamefont {Watson},
		\citenamefont {Heller},\ and\ \citenamefont
		{Wojtowicz}}]{watson:comparative}%
	\BibitemOpen
	\bibfield  {author} {\bibinfo {author} {\bibfnamefont {J.~H.~L.}\
			\bibnamefont {Watson}}, \bibinfo {author} {\bibfnamefont {W.}~\bibnamefont
			{Heller}},\ and\ \bibinfo {author} {\bibfnamefont {W.}~\bibnamefont
			{Wojtowicz}},\ }\bibfield  {title} {\bibinfo {title} {Comparative electron
			and light microscopic investigations of tactoid structures in
			$\mathrm{V}_2\mathrm{O}_5$-sols},\ }\href@noop {} {\bibfield  {journal}
		{\bibinfo  {journal} {Science}\ }\textbf {\bibinfo {volume} {109}},\ \bibinfo
		{pages} {274} (\bibinfo {year} {1949})}\BibitemShut {NoStop}%
	\bibitem [{\citenamefont {Stanley}(1935)}]{stanley:isolation}%
	\BibitemOpen
	\bibfield  {author} {\bibinfo {author} {\bibfnamefont {W.~M.}\ \bibnamefont
			{Stanley}},\ }\bibfield  {title} {\bibinfo {title} {Isolation of a
			crystalline protein possessing the properties of tobacco-mosaic virus},\
	}\href@noop {} {\bibfield  {journal} {\bibinfo  {journal} {Science}\ }\textbf
		{\bibinfo {volume} {81}},\ \bibinfo {pages} {644} (\bibinfo {year}
		{1935})}\BibitemShut {NoStop}%
	\bibitem [{\citenamefont {Bawden}\ \emph {et~al.}(1936)\citenamefont {Bawden},
		\citenamefont {Pirie}, \citenamefont {Bernal},\ and\ \citenamefont
		{Fankuchen}}]{bawden:liquid}%
	\BibitemOpen
	\bibfield  {author} {\bibinfo {author} {\bibfnamefont {F.~C.}\ \bibnamefont
			{Bawden}}, \bibinfo {author} {\bibfnamefont {N.~W.}\ \bibnamefont {Pirie}},
		\bibinfo {author} {\bibfnamefont {J.~D.}\ \bibnamefont {Bernal}},\ and\
		\bibinfo {author} {\bibfnamefont {I.}~\bibnamefont {Fankuchen}},\ }\bibfield
	{title} {\bibinfo {title} {Liquid crystalline substances from virus-infected
			plants},\ }\href@noop {} {\bibfield  {journal} {\bibinfo  {journal} {Nature}\
		}\textbf {\bibinfo {volume} {138}},\ \bibinfo {pages} {1051} (\bibinfo {year}
		{1936})}\BibitemShut {NoStop}%
	\bibitem [{\citenamefont {Fraden}\ \emph {et~al.}(1985)\citenamefont {Fraden},
		\citenamefont {Hurd}, \citenamefont {Meyer}, \citenamefont {Cahoon},\ and\
		\citenamefont {Caspar}}]{fraden:magnetic-field}%
	\BibitemOpen
	\bibfield  {author} {\bibinfo {author} {\bibfnamefont {S.}~\bibnamefont
			{Fraden}}, \bibinfo {author} {\bibfnamefont {A.~J.}\ \bibnamefont {Hurd}},
		\bibinfo {author} {\bibfnamefont {R.~B.}\ \bibnamefont {Meyer}}, \bibinfo
		{author} {\bibfnamefont {M.}~\bibnamefont {Cahoon}},\ and\ \bibinfo {author}
		{\bibfnamefont {D.~L.~D.}\ \bibnamefont {Caspar}},\ }\bibfield  {title}
	{\bibinfo {title} {Magnetic-field-induced alignment and instabilities in
			ordered colloids of tobacco mosaic virus},\ }\href@noop {} {\bibfield
		{journal} {\bibinfo  {journal} {J. Phys. ({P}aris) Colloq.}\ }\textbf
		{\bibinfo {volume} {46}},\ \bibinfo {pages} {C3/85} (\bibinfo {year}
		{1985})}\BibitemShut {NoStop}%
	\bibitem [{\citenamefont {Bernal}\ and\ \citenamefont
		{Fankuchen}(1941)}]{bernal:x-ray}%
	\BibitemOpen
	\bibfield  {author} {\bibinfo {author} {\bibfnamefont {J.~D.}\ \bibnamefont
			{Bernal}}\ and\ \bibinfo {author} {\bibfnamefont {I.}~\bibnamefont
			{Fankuchen}},\ }\bibfield  {title} {\bibinfo {title} {X-ray and
			crystallographic studies of plant virus preparations: {I.} {I}ntroduction and
			preparation of specimens. {II.} {M}odes of aggregation of the virus
			particles},\ }\href@noop {} {\bibfield  {journal} {\bibinfo  {journal} {J.
				Gen. Physiol.}\ }\textbf {\bibinfo {volume} {25}},\ \bibinfo {pages} {111}
		(\bibinfo {year} {1941})}\BibitemShut {NoStop}%
	\bibitem [{\citenamefont {Onsager}(1949)}]{onsager:effects}%
	\BibitemOpen
	\bibfield  {author} {\bibinfo {author} {\bibfnamefont {L.}~\bibnamefont
			{Onsager}},\ }\bibfield  {title} {\bibinfo {title} {The effects of shape on
			the interaction of colloidal particles},\ }\href@noop {} {\bibfield
		{journal} {\bibinfo  {journal} {Ann. N. Y. Acad. Sci.}\ }\textbf {\bibinfo
			{volume} {51}},\ \bibinfo {pages} {627} (\bibinfo {year} {1949})}\BibitemShut
	{NoStop}%
	\bibitem [{\citenamefont {Zocher}(1954)}]{zocher:taktosole_meso}%
	\BibitemOpen
	\bibfield  {author} {\bibinfo {author} {\bibfnamefont {H.}~\bibnamefont
			{Zocher}},\ }\bibfield  {title} {\bibinfo {title} {Taktosole und
			{M}esophasen},\ }\href@noop {} {\bibfield  {journal} {\bibinfo  {journal}
			{Kolloid-Z.}\ }\textbf {\bibinfo {volume} {139}},\ \bibinfo {pages} {81}
		(\bibinfo {year} {1954})}\BibitemShut {NoStop}%
	\bibitem [{\citenamefont {Zocher}\ and\ \citenamefont
		{T\"or\"ok}(1960)}]{zocher:neuere}%
	\BibitemOpen
	\bibfield  {author} {\bibinfo {author} {\bibfnamefont {H.}~\bibnamefont
			{Zocher}}\ and\ \bibinfo {author} {\bibfnamefont {C.}~\bibnamefont
			{T\"or\"ok}},\ }\bibfield  {title} {\bibinfo {title} {Neuere {B}eitr\"age zur
			{K}enntnis der {T}aktosole},\ }\href@noop {} {\bibfield  {journal} {\bibinfo
			{journal} {Kolloid-Zeit.}\ }\textbf {\bibinfo {volume} {170}},\ \bibinfo
		{pages} {140} (\bibinfo {year} {1960})}\BibitemShut {NoStop}%
	\bibitem [{\citenamefont {Zocher}(1969)}]{zocher:nematic}%
	\BibitemOpen
	\bibfield  {author} {\bibinfo {author} {\bibfnamefont {H.}~\bibnamefont
			{Zocher}},\ }\bibfield  {title} {\bibinfo {title} {{II}. {N}ematic and
			smectic phases of higher order},\ }\href@noop {} {\bibfield  {journal}
		{\bibinfo  {journal} {Mol. Cryst.}\ }\textbf {\bibinfo {volume} {7}},\
		\bibinfo {pages} {177} (\bibinfo {year} {1969})}\BibitemShut {NoStop}%
	\bibitem [{\citenamefont {Dickinson}\ \emph {et~al.}(2009)\citenamefont
		{Dickinson}, \citenamefont {La{R}acuente}, \citenamefont {Mc{K}itterick},\
		and\ \citenamefont {Collings}}]{dickinson:aggregate}%
	\BibitemOpen
	\bibfield  {author} {\bibinfo {author} {\bibfnamefont {A.~J.}\ \bibnamefont
			{Dickinson}}, \bibinfo {author} {\bibfnamefont {N.~D.}\ \bibnamefont
			{La{R}acuente}}, \bibinfo {author} {\bibfnamefont {C.~B.}\ \bibnamefont
			{Mc{K}itterick}},\ and\ \bibinfo {author} {\bibfnamefont {P.~J.}\
			\bibnamefont {Collings}},\ }\bibfield  {title} {\bibinfo {title} {Aggregate
			structure and free energy changes in chromonic liquid crystals},\ }\href@noop
	{} {\bibfield  {journal} {\bibinfo  {journal} {Mol. Cryst. Liq. Cryst.}\
		}\textbf {\bibinfo {volume} {509}},\ \bibinfo {pages} {751} (\bibinfo {year}
		{2009})}\BibitemShut {NoStop}%
	\bibitem [{\citenamefont {Tam-Chang}\ and\ \citenamefont
		{Huang}(2008)}]{tam-chang:chromonic}%
	\BibitemOpen
	\bibfield  {author} {\bibinfo {author} {\bibfnamefont {S.-W.}\ \bibnamefont
			{Tam-Chang}}\ and\ \bibinfo {author} {\bibfnamefont {L.}~\bibnamefont
			{Huang}},\ }\bibfield  {title} {\bibinfo {title} {Chromonic liquid crystals:
			properties and applications as functional materials},\ }\href@noop {}
	{\bibfield  {journal} {\bibinfo  {journal} {Chem. Commun.}\ }\textbf
		{\bibinfo {volume} {44}} (\bibinfo {year} {2008})}\BibitemShut {NoStop}%
	\bibitem [{\citenamefont {Mariani}\ \emph {et~al.}(2009)\citenamefont
		{Mariani}, \citenamefont {Spinozzi}, \citenamefont {Federiconi},
		\citenamefont {Amenitsch}, \citenamefont {Spindler},\ and\ \citenamefont
		{Drevensek-Olenik}}]{mariani:small}%
	\BibitemOpen
	\bibfield  {author} {\bibinfo {author} {\bibfnamefont {P.}~\bibnamefont
			{Mariani}}, \bibinfo {author} {\bibfnamefont {F.}~\bibnamefont {Spinozzi}},
		\bibinfo {author} {\bibfnamefont {F.}~\bibnamefont {Federiconi}}, \bibinfo
		{author} {\bibfnamefont {H.}~\bibnamefont {Amenitsch}}, \bibinfo {author}
		{\bibfnamefont {L.}~\bibnamefont {Spindler}},\ and\ \bibinfo {author}
		{\bibfnamefont {I.}~\bibnamefont {Drevensek-Olenik}},\ }\bibfield  {title}
	{\bibinfo {title} {Small angle {X}-ray scattering analysis of deoxyguanosine
			5'-monophosphate self-assembing in solution: {N}ucleation and growth of
			{G}-quadruplexes},\ }\href@noop {} {\bibfield  {journal} {\bibinfo  {journal}
			{J. Phys. Chem. B}\ }\textbf {\bibinfo {volume} {113}},\ \bibinfo {pages}
		{7934} (\bibinfo {year} {2009})}\BibitemShut {NoStop}%
	\bibitem [{\citenamefont {Zanchetta}\ \emph {et~al.}(2008)\citenamefont
		{Zanchetta}, \citenamefont {Nakata}, \citenamefont {Buscaglia}, \citenamefont
		{Bellini},\ and\ \citenamefont {Clark}}]{zanchetta:phase}%
	\BibitemOpen
	\bibfield  {author} {\bibinfo {author} {\bibfnamefont {G.}~\bibnamefont
			{Zanchetta}}, \bibinfo {author} {\bibfnamefont {M.}~\bibnamefont {Nakata}},
		\bibinfo {author} {\bibfnamefont {M.}~\bibnamefont {Buscaglia}}, \bibinfo
		{author} {\bibfnamefont {T.}~\bibnamefont {Bellini}},\ and\ \bibinfo {author}
		{\bibfnamefont {N.~A.}\ \bibnamefont {Clark}},\ }\bibfield  {title} {\bibinfo
		{title} {Phase separation and liquid crystallization of complementary
			sequences in mixtures of nanodna oligomers},\ }\href@noop {} {\bibfield
		{journal} {\bibinfo  {journal} {Proc. Natl. Acad. Sci. USA}\ }\textbf
		{\bibinfo {volume} {105}},\ \bibinfo {pages} {1111} (\bibinfo {year}
		{2008})}\BibitemShut {NoStop}%
	\bibitem [{\citenamefont {Nakata}\ \emph {et~al.}(2007)\citenamefont {Nakata},
		\citenamefont {Zanchetta}, \citenamefont {Chapman}, \citenamefont {Jones},
		\citenamefont {Cross}, \citenamefont {Pindak}, \citenamefont {Bellini},\ and\
		\citenamefont {Clark}}]{nakata:end-to-end}%
	\BibitemOpen
	\bibfield  {author} {\bibinfo {author} {\bibfnamefont {M.}~\bibnamefont
			{Nakata}}, \bibinfo {author} {\bibfnamefont {G.}~\bibnamefont {Zanchetta}},
		\bibinfo {author} {\bibfnamefont {B.~D.}\ \bibnamefont {Chapman}}, \bibinfo
		{author} {\bibfnamefont {C.~D.}\ \bibnamefont {Jones}}, \bibinfo {author}
		{\bibfnamefont {J.~O.}\ \bibnamefont {Cross}}, \bibinfo {author}
		{\bibfnamefont {R.}~\bibnamefont {Pindak}}, \bibinfo {author} {\bibfnamefont
			{T.}~\bibnamefont {Bellini}},\ and\ \bibinfo {author} {\bibfnamefont {N.~A.}\
			\bibnamefont {Clark}},\ }\bibfield  {title} {\bibinfo {title} {End-to-end
			stacking and liquid crystal condensation of 6- to 20-base pair {DNA}
			duplexes},\ }\href@noop {} {\bibfield  {journal} {\bibinfo  {journal}
			{Science}\ }\textbf {\bibinfo {volume} {318}},\ \bibinfo {pages} {1276}
		(\bibinfo {year} {2007})}\BibitemShut {NoStop}%
	\bibitem [{\citenamefont {Lydon}(2011)}]{lydon:chromonic}%
	\BibitemOpen
	\bibfield  {author} {\bibinfo {author} {\bibfnamefont {J.}~\bibnamefont
			{Lydon}},\ }\bibfield  {title} {\bibinfo {title} {Chromonic liquid
			crystalline phases},\ }\href@noop {} {\bibfield  {journal} {\bibinfo
			{journal} {Liq. Cryst.}\ }\textbf {\bibinfo {volume} {38}},\ \bibinfo {pages}
		{1663} (\bibinfo {year} {2011})}\BibitemShut {NoStop}%
	\bibitem [{\citenamefont {Zhou}(2017)}]{zhou:lyotropic}%
	\BibitemOpen
	\bibfield  {author} {\bibinfo {author} {\bibfnamefont {S.}~\bibnamefont
			{Zhou}},\ }\href@noop {} {\emph {\bibinfo {title} {Lyotropic Chromonic Liquid
				Crystals}}},\ Springer Theses\ (\bibinfo  {publisher} {Springer},\ \bibinfo
	{address} {Cham, Switzerland},\ \bibinfo {year} {2017})\BibitemShut {NoStop}%
	\bibitem [{\citenamefont {Nastishin}\ \emph {et~al.}(2005)\citenamefont
		{Nastishin}, \citenamefont {Liu}, \citenamefont {Schneider}, \citenamefont
		{Nazarenko}, \citenamefont {Vasyuta}, \citenamefont {Shiyanovskii},\ and\
		\citenamefont {Lavrentovich}}]{nastishin:optical}%
	\BibitemOpen
	\bibfield  {author} {\bibinfo {author} {\bibfnamefont {Y.~A.}\ \bibnamefont
			{Nastishin}}, \bibinfo {author} {\bibfnamefont {H.}~\bibnamefont {Liu}},
		\bibinfo {author} {\bibfnamefont {T.}~\bibnamefont {Schneider}}, \bibinfo
		{author} {\bibfnamefont {V.}~\bibnamefont {Nazarenko}}, \bibinfo {author}
		{\bibfnamefont {R.}~\bibnamefont {Vasyuta}}, \bibinfo {author} {\bibfnamefont
			{S.~V.}\ \bibnamefont {Shiyanovskii}},\ and\ \bibinfo {author} {\bibfnamefont
			{O.~D.}\ \bibnamefont {Lavrentovich}},\ }\bibfield  {title} {\bibinfo {title}
		{Optical characterization of the nematic lyotropic chromonic liquid crystals:
			Light absorption, birefringence, and scalar order parameter},\ }\href@noop {}
	{\bibfield  {journal} {\bibinfo  {journal} {Phys. Rev. E}\ }\textbf {\bibinfo
			{volume} {72}},\ \bibinfo {pages} {041711} (\bibinfo {year}
		{2005})}\BibitemShut {NoStop}%
	\bibitem [{\citenamefont {Tortora}\ and\ \citenamefont
		{Lavrentovich}(2011)}]{tortora:chiral}%
	\BibitemOpen
	\bibfield  {author} {\bibinfo {author} {\bibfnamefont {L.}~\bibnamefont
			{Tortora}}\ and\ \bibinfo {author} {\bibfnamefont {O.~D.}\ \bibnamefont
			{Lavrentovich}},\ }\bibfield  {title} {\bibinfo {title} {Chiral symmetry
			breaking by spatial confinement in tactoidal droplets of lyotropic chromonic
			liquid crystals},\ }\href@noop {} {\bibfield  {journal} {\bibinfo  {journal}
			{Proc. Natl. Acad. Sci. USA}\ }\textbf {\bibinfo {volume} {108}},\ \bibinfo
		{pages} {5163} (\bibinfo {year} {2011})}\BibitemShut {NoStop}%
	\bibitem [{\citenamefont {Kim}\ \emph {et~al.}(2013)\citenamefont {Kim},
		\citenamefont {Shiyanovskii},\ and\ \citenamefont
		{Lavrentovich}}]{kim:morphogenesis}%
	\BibitemOpen
	\bibfield  {author} {\bibinfo {author} {\bibfnamefont {Y.-K.}\ \bibnamefont
			{Kim}}, \bibinfo {author} {\bibfnamefont {S.~V.}\ \bibnamefont
			{Shiyanovskii}},\ and\ \bibinfo {author} {\bibfnamefont {O.~D.}\ \bibnamefont
			{Lavrentovich}},\ }\bibfield  {title} {\bibinfo {title} {Morphogenesis of
			defects and tactoids during isotropic-nematic phase transition in
			self-assembled lyotropic chromonic liquid crystals},\ }\href@noop {}
	{\bibfield  {journal} {\bibinfo  {journal} {J. Phys.: Condens. Matter}\
		}\textbf {\bibinfo {volume} {25}},\ \bibinfo {pages} {404202} (\bibinfo
		{year} {2013})}\BibitemShut {NoStop}%
	\bibitem [{\citenamefont {Jeong}\ \emph {et~al.}(2014)\citenamefont {Jeong},
		\citenamefont {Davidson}, \citenamefont {Collings}, \citenamefont
		{Lubensky},\ and\ \citenamefont {Yodh}}]{jeong:chiral_2014}%
	\BibitemOpen
	\bibfield  {author} {\bibinfo {author} {\bibfnamefont {J.}~\bibnamefont
			{Jeong}}, \bibinfo {author} {\bibfnamefont {Z.~S.}\ \bibnamefont {Davidson}},
		\bibinfo {author} {\bibfnamefont {P.~J.}\ \bibnamefont {Collings}}, \bibinfo
		{author} {\bibfnamefont {T.~C.}\ \bibnamefont {Lubensky}},\ and\ \bibinfo
		{author} {\bibfnamefont {A.~G.}\ \bibnamefont {Yodh}},\ }\bibfield  {title}
	{\bibinfo {title} {Chiral symmetry breaking and surface faceting in chromonic
			liquid crystal droplets with giant elastic anisotropy},\ }\href@noop {}
	{\bibfield  {journal} {\bibinfo  {journal} {Proc. Natl. Acad. Sci. USA}\
		}\textbf {\bibinfo {volume} {111}},\ \bibinfo {pages} {1742} (\bibinfo {year}
		{2014})}\BibitemShut {NoStop}%
	\bibitem [{\citenamefont {Peng}\ and\ \citenamefont
		{Lavrentovich}(2015)}]{peng:chirality}%
	\BibitemOpen
	\bibfield  {author} {\bibinfo {author} {\bibfnamefont {C.}~\bibnamefont
			{Peng}}\ and\ \bibinfo {author} {\bibfnamefont {O.~D.}\ \bibnamefont
			{Lavrentovich}},\ }\bibfield  {title} {\bibinfo {title} {Chirality
			amplification and detection by tactoids of lyotropic chromonic liquid
			crystals},\ }\href@noop {} {\bibfield  {journal} {\bibinfo  {journal} {Soft
				Matter}\ }\textbf {\bibinfo {volume} {11}},\ \bibinfo {pages} {7221}
		(\bibinfo {year} {2015})}\BibitemShut {NoStop}%
	\bibitem [{\citenamefont {Oakes}\ \emph {et~al.}(2007)\citenamefont {Oakes},
		\citenamefont {Viamontes},\ and\ \citenamefont {Tang}}]{oakes:growth}%
	\BibitemOpen
	\bibfield  {author} {\bibinfo {author} {\bibfnamefont {P.~W.}\ \bibnamefont
			{Oakes}}, \bibinfo {author} {\bibfnamefont {J.}~\bibnamefont {Viamontes}},\
		and\ \bibinfo {author} {\bibfnamefont {J.~X.}\ \bibnamefont {Tang}},\
	}\bibfield  {title} {\bibinfo {title} {Growth of tactoidal droplets during
			the first-order isotropic to nematic phase transition of f-actin},\
	}\href@noop {} {\bibfield  {journal} {\bibinfo  {journal} {Phys. Rev. E}\
		}\textbf {\bibinfo {volume} {75}},\ \bibinfo {pages} {061902} (\bibinfo
		{year} {2007})}\BibitemShut {NoStop}%
	\bibitem [{\citenamefont {Verhoeff}\ \emph {et~al.}(2011)\citenamefont
		{Verhoeff}, \citenamefont {Bakelaar}, \citenamefont {Otten}, \citenamefont
		{{van der Schoot}},\ and\ \citenamefont {Lekkerkerker}}]{verhoeff:tactoids}%
	\BibitemOpen
	\bibfield  {author} {\bibinfo {author} {\bibfnamefont {A.~A.}\ \bibnamefont
			{Verhoeff}}, \bibinfo {author} {\bibfnamefont {I.~A.}\ \bibnamefont
			{Bakelaar}}, \bibinfo {author} {\bibfnamefont {R.~H.~J.}\ \bibnamefont
			{Otten}}, \bibinfo {author} {\bibfnamefont {P.}~\bibnamefont {{van der
					Schoot}}},\ and\ \bibinfo {author} {\bibfnamefont {H.~N.~W.}\ \bibnamefont
			{Lekkerkerker}},\ }\bibfield  {title} {\bibinfo {title} {Tactoids of
			plate-like particles: Size, shape, and director field},\ }\href@noop {}
	{\bibfield  {journal} {\bibinfo  {journal} {Langmuir}\ }\textbf {\bibinfo
			{volume} {27}},\ \bibinfo {pages} {116} (\bibinfo {year} {2011})}\BibitemShut
	{NoStop}%
	\bibitem [{\citenamefont {Oseen}(1933)}]{oseen:theory}%
	\BibitemOpen
	\bibfield  {author} {\bibinfo {author} {\bibfnamefont {C.~W.}\ \bibnamefont
			{Oseen}},\ }\bibfield  {title} {\bibinfo {title} {The theory of liquid
			crystals},\ }\href@noop {} {\bibfield  {journal} {\bibinfo  {journal} {Trans.
				Faraday Soc.}\ }\textbf {\bibinfo {volume} {29}},\ \bibinfo {pages} {883}
		(\bibinfo {year} {1933})}\BibitemShut {NoStop}%
	\bibitem [{\citenamefont {Frank}(1958)}]{frank:theory}%
	\BibitemOpen
	\bibfield  {author} {\bibinfo {author} {\bibfnamefont {F.~C.}\ \bibnamefont
			{Frank}},\ }\bibfield  {title} {\bibinfo {title} {On the theory of liquid
			crystals},\ }\href@noop {} {\bibfield  {journal} {\bibinfo  {journal}
			{Discuss. Faraday Soc.}\ }\textbf {\bibinfo {volume} {25}},\ \bibinfo {pages}
		{19} (\bibinfo {year} {1958})}\BibitemShut {NoStop}%
	\bibitem [{\citenamefont {Parsons}(1976)}]{parsons:molecular}%
	\BibitemOpen
	\bibfield  {author} {\bibinfo {author} {\bibfnamefont {J.}~\bibnamefont
			{Parsons}},\ }\bibfield  {title} {\bibinfo {title} {A molecular theory of
			surface tension in nematic liquid crystals},\ }\href@noop {} {\bibfield
		{journal} {\bibinfo  {journal} {J. Phys. France}\ }\textbf {\bibinfo {volume}
			{37}},\ \bibinfo {pages} {1187} (\bibinfo {year} {1976})}\BibitemShut
	{NoStop}%
	\bibitem [{\citenamefont {Candau}\ \emph {et~al.}(1973)\citenamefont {Candau},
		\citenamefont {Roy},\ and\ \citenamefont {Debeauvais}}]{candau:magnetic}%
	\BibitemOpen
	\bibfield  {author} {\bibinfo {author} {\bibfnamefont {S.}~\bibnamefont
			{Candau}}, \bibinfo {author} {\bibfnamefont {P.~L.}\ \bibnamefont {Roy}},\
		and\ \bibinfo {author} {\bibfnamefont {F.}~\bibnamefont {Debeauvais}},\
	}\bibfield  {title} {\bibinfo {title} {Magnetic field effects in nematic and
			cholesteric droplets suspended in a isotropic liquid},\ }\href@noop {}
	{\bibfield  {journal} {\bibinfo  {journal} {Mol. Cryst. Liq. Cryst.}\
		}\textbf {\bibinfo {volume} {23}},\ \bibinfo {pages} {283} (\bibinfo {year}
		{1973})}\BibitemShut {NoStop}%
	\bibitem [{\citenamefont {Volovik}\ and\ \citenamefont
		{Lavrentovich}(1983)}]{volovik:topological}%
	\BibitemOpen
	\bibfield  {author} {\bibinfo {author} {\bibfnamefont {G.~E.}\ \bibnamefont
			{Volovik}}\ and\ \bibinfo {author} {\bibfnamefont {D.}~\bibnamefont
			{Lavrentovich}},\ }\bibfield  {title} {\bibinfo {title} {Topological dynamics
			of defects: boojums in nematic drops},\ }\href@noop {} {\bibfield  {journal}
		{\bibinfo  {journal} {Sov. Phys. JETP}\ }\textbf {\bibinfo {volume} {58}},\
		\bibinfo {pages} {1159} (\bibinfo {year} {1983})},\ \bibinfo {note} {zh.
		Eksp. Teor. Fiz. \textbf{85}(1983), 1997--2010}\BibitemShut {NoStop}%
	\bibitem [{\citenamefont {Kurik}\ and\ \citenamefont
		{Lavrentovich}(1982)}]{kurik:negative-positive}%
	\BibitemOpen
	\bibfield  {author} {\bibinfo {author} {\bibfnamefont {M.~V.}\ \bibnamefont
			{Kurik}}\ and\ \bibinfo {author} {\bibfnamefont {D.}~\bibnamefont
			{Lavrentovich}},\ }\bibfield  {title} {\bibinfo {title} {Negative-positive
			monople transitions in cholesteric liquid crystals},\ }\href@noop {}
	{\bibfield  {journal} {\bibinfo  {journal} {JETP Lett.}\ }\textbf {\bibinfo
			{volume} {35}},\ \bibinfo {pages} {444} (\bibinfo {year} {1982})},\ \bibinfo
	{note} {pis'ma Zh. Eksp. Teor. Fiz. \textbf{35}(1982), 362--365}\BibitemShut
	{NoStop}%
	\bibitem [{\citenamefont {Chandrasekhar}(1966)}]{chandrasekhar:surface}%
	\BibitemOpen
	\bibfield  {author} {\bibinfo {author} {\bibfnamefont {S.}~\bibnamefont
			{Chandrasekhar}},\ }\bibfield  {title} {\bibinfo {title} {Surface tension of
			liquid crystals},\ }\href@noop {} {\bibfield  {journal} {\bibinfo  {journal}
			{Mol. Cryst.}\ }\textbf {\bibinfo {volume} {2}},\ \bibinfo {pages} {71}
		(\bibinfo {year} {1966})}\BibitemShut {NoStop}%
	\bibitem [{\citenamefont {Dubois-Violette}\ and\ \citenamefont
		{Parodi}(1969)}]{dubois-violette:emulsions}%
	\BibitemOpen
	\bibfield  {author} {\bibinfo {author} {\bibfnamefont {E.}~\bibnamefont
			{Dubois-Violette}}\ and\ \bibinfo {author} {\bibfnamefont {O.}~\bibnamefont
			{Parodi}},\ }\bibfield  {title} {\bibinfo {title} {\'{E}mulsions
			n\'ematiques. {E}ffects de champ magn\'etiques et effects
			pi\'ezo\'electriques},\ }\href@noop {} {\bibfield  {journal} {\bibinfo
			{journal} {J. Phys. Colloq.}\ }\textbf {\bibinfo {volume} {30}},\ \bibinfo
		{pages} {C4.57} (\bibinfo {year} {1969})}\BibitemShut {NoStop}%
	\bibitem [{\citenamefont {Williams}(1985)}]{williams:nematic}%
	\BibitemOpen
	\bibfield  {author} {\bibinfo {author} {\bibfnamefont {R.~D.}\ \bibnamefont
			{Williams}},\ }\href@noop {} {\emph {\bibinfo {title} {Nematic liquid crystal
				droplets}}},\ \bibinfo {type} {Tech. Rep.}\ \bibinfo {number} {{RAL}-85-028}\
	(\bibinfo  {institution} {Rutherford Appleton Laboratory},\ \bibinfo
	{address} {Chilton, UK},\ \bibinfo {year} {1985})\ \bibinfo {note}
	{\url{https://lib-extopc.kek.jp/preprints/PDF/1986/8606/8606437.pdf}}\BibitemShut
	{NoStop}%
	\bibitem [{\citenamefont {Williams}(1986)}]{williams:transitions}%
	\BibitemOpen
	\bibfield  {author} {\bibinfo {author} {\bibfnamefont {R.~D.}\ \bibnamefont
			{Williams}},\ }\bibfield  {title} {\bibinfo {title} {Two transitions in
			tangentially anchored nematic droplets},\ }\href@noop {} {\bibfield
		{journal} {\bibinfo  {journal} {J. Phys. A: Math. Gen.}\ }\textbf {\bibinfo
			{volume} {19}},\ \bibinfo {pages} {3211} (\bibinfo {year}
		{1986})}\BibitemShut {NoStop}%
	\bibitem [{\citenamefont {Kaznacheev}\ \emph {et~al.}(2002)\citenamefont
		{Kaznacheev}, \citenamefont {Bogdanov},\ and\ \citenamefont
		{A.Taraskin}}]{kaznacheev:nature}%
	\BibitemOpen
	\bibfield  {author} {\bibinfo {author} {\bibfnamefont {A.~V.}\ \bibnamefont
			{Kaznacheev}}, \bibinfo {author} {\bibfnamefont {M.~M.}\ \bibnamefont
			{Bogdanov}},\ and\ \bibinfo {author} {\bibfnamefont {S.}~\bibnamefont
			{A.Taraskin}},\ }\bibfield  {title} {\bibinfo {title} {The nature of prolate
			shape of tactoids in lyotropic inorganic liquid crystals},\ }\href@noop {}
	{\bibfield  {journal} {\bibinfo  {journal} {J. Exp. Theor. Phys.}\ }\textbf
		{\bibinfo {volume} {95}},\ \bibinfo {pages} {57} (\bibinfo {year}
		{2002})}\BibitemShut {NoStop}%
	\bibitem [{\citenamefont {Prinsen}\ and\ \citenamefont {van~der
			Schoot}(2003)}]{prinsen:shape}%
	\BibitemOpen
	\bibfield  {author} {\bibinfo {author} {\bibfnamefont {P.}~\bibnamefont
			{Prinsen}}\ and\ \bibinfo {author} {\bibfnamefont {P.}~\bibnamefont {van~der
				Schoot}},\ }\bibfield  {title} {\bibinfo {title} {Shape and director-field
			transformation of tactoids},\ }\href@noop {} {\bibfield  {journal} {\bibinfo
			{journal} {Phys. Rev. E}\ }\textbf {\bibinfo {volume} {68}},\ \bibinfo
		{pages} {021701} (\bibinfo {year} {2003})}\BibitemShut {NoStop}%
	\bibitem [{\citenamefont {Kaznacheev}\ \emph {et~al.}(2003)\citenamefont
		{Kaznacheev}, \citenamefont {Bogdanov},\ and\ \citenamefont
		{Sonin}}]{kaznacheev:influence}%
	\BibitemOpen
	\bibfield  {author} {\bibinfo {author} {\bibfnamefont {A.~V.}\ \bibnamefont
			{Kaznacheev}}, \bibinfo {author} {\bibfnamefont {M.~M.}\ \bibnamefont
			{Bogdanov}},\ and\ \bibinfo {author} {\bibfnamefont {A.~S.}\ \bibnamefont
			{Sonin}},\ }\bibfield  {title} {\bibinfo {title} {The influence of anchoring
			energy on the prolate shape of tactoids in lyotropic inorganic liquid
			crystals},\ }\href@noop {} {\bibfield  {journal} {\bibinfo  {journal} {J.
				Exp. Theor. Phys.}\ }\textbf {\bibinfo {volume} {97}},\ \bibinfo {pages}
		{1159} (\bibinfo {year} {2003})}\BibitemShut {NoStop}%
	\bibitem [{\citenamefont {Prinsen}\ and\ \citenamefont {{van der
				Schoot}}(2004)}]{prinsen:continuous}%
	\BibitemOpen
	\bibfield  {author} {\bibinfo {author} {\bibfnamefont {P.}~\bibnamefont
			{Prinsen}}\ and\ \bibinfo {author} {\bibfnamefont {P.}~\bibnamefont {{van der
					Schoot}}},\ }\bibfield  {title} {\bibinfo {title} {Continuous director-field
			transformation of nematic tactoids},\ }\href@noop {} {\bibfield  {journal}
		{\bibinfo  {journal} {Euro. Phys. J. E}\ }\textbf {\bibinfo {volume} {13}},\
		\bibinfo {pages} {35} (\bibinfo {year} {2004})}\BibitemShut {NoStop}%
	\bibitem [{\citenamefont {Prinsen}\ and\ \citenamefont {van~der
			Schoot}(2004)}]{prinsen:parity}%
	\BibitemOpen
	\bibfield  {author} {\bibinfo {author} {\bibfnamefont {P.}~\bibnamefont
			{Prinsen}}\ and\ \bibinfo {author} {\bibfnamefont {P.}~\bibnamefont {van~der
				Schoot}},\ }\bibfield  {title} {\bibinfo {title} {Parity breaking in nematic
			tactoids},\ }\href@noop {} {\bibfield  {journal} {\bibinfo  {journal} {J. P.:
				Condens. Matter}\ }\textbf {\bibinfo {volume} {16}},\ \bibinfo {pages} {8835}
		(\bibinfo {year} {2004})}\BibitemShut {NoStop}%
	\bibitem [{\citenamefont {{de~G}ennes}\ and\ \citenamefont
		{Prost}(1993)}]{degennes:physics}%
	\BibitemOpen
	\bibfield  {author} {\bibinfo {author} {\bibfnamefont {P.~G.}\ \bibnamefont
			{{de~G}ennes}}\ and\ \bibinfo {author} {\bibfnamefont {J.}~\bibnamefont
			{Prost}},\ }\href@noop {} {\emph {\bibinfo {title} {The Physics of Liquid
				Crystals}}},\ \bibinfo {edition} {2nd}\ ed.,\ \bibinfo {series} {The
		International Series of Monographs on Physics}, Vol.~\bibinfo {volume} {83}\
	(\bibinfo  {publisher} {Clarendon Press},\ \bibinfo {address} {Oxford},\
	\bibinfo {year} {1993})\BibitemShut {NoStop}%
	\bibitem [{\citenamefont {Virga}(1994)}]{virga:variational}%
	\BibitemOpen
	\bibfield  {author} {\bibinfo {author} {\bibfnamefont {E.~G.}\ \bibnamefont
			{Virga}},\ }\href@noop {} {\emph {\bibinfo {title} {Variational Theories for
				Liquid Crystals}}},\ \bibinfo {series} {Applied Mathematics and Mathematical
		Computation}, Vol.~\bibinfo {volume} {8}\ (\bibinfo  {publisher} {Chapman \&
		Hall},\ \bibinfo {address} {London},\ \bibinfo {year} {1994})\BibitemShut
	{NoStop}%
	\bibitem [{\citenamefont {Selinger}(2018)}]{selinger:interpretation}%
	\BibitemOpen
	\bibfield  {author} {\bibinfo {author} {\bibfnamefont {J.~V.}\ \bibnamefont
			{Selinger}},\ }\bibfield  {title} {\bibinfo {title} {Interpretation of
			saddle-splay and the {O}seen-{F}rank free energy in liquid crystals},\
	}\href@noop {} {\bibfield  {journal} {\bibinfo  {journal} {Liq. Cryst. Rev.}\
		}\textbf {\bibinfo {volume} {6}},\ \bibinfo {pages} {129} (\bibinfo {year}
		{2018})}\BibitemShut {NoStop}%
	\bibitem [{\citenamefont {Pedrini}\ and\ \citenamefont
		{Virga}(2020)}]{pedrini:liquid}%
	\BibitemOpen
	\bibfield  {author} {\bibinfo {author} {\bibfnamefont {A.}~\bibnamefont
			{Pedrini}}\ and\ \bibinfo {author} {\bibfnamefont {E.~G.}\ \bibnamefont
			{Virga}},\ }\bibfield  {title} {\bibinfo {title} {Liquid crystal distortions
			revealed by an octupolar tensor},\ }\href@noop {} {\bibfield  {journal}
		{\bibinfo  {journal} {Phys. Rev. E}\ }\textbf {\bibinfo {volume} {101}},\
		\bibinfo {pages} {012703} (\bibinfo {year} {2020})}\BibitemShut {NoStop}%
	\bibitem [{\citenamefont {Ericksen}(1962)}]{ericksen:nilpotent}%
	\BibitemOpen
	\bibfield  {author} {\bibinfo {author} {\bibfnamefont {J.~L.}\ \bibnamefont
			{Ericksen}},\ }\bibfield  {title} {\bibinfo {title} {Nilpotent energies in
			liquid crystal theory},\ }\href@noop {} {\bibfield  {journal} {\bibinfo
			{journal} {Arch. Rational Mech. Anal.}\ }\textbf {\bibinfo {volume} {10}},\
		\bibinfo {pages} {189} (\bibinfo {year} {1962})}\BibitemShut {NoStop}%
	\bibitem [{\citenamefont {Ericksen}(1966)}]{ericksen:inequalities}%
	\BibitemOpen
	\bibfield  {author} {\bibinfo {author} {\bibfnamefont {J.~L.}\ \bibnamefont
			{Ericksen}},\ }\bibfield  {title} {\bibinfo {title} {Inequalities in liquid
			crystal theory},\ }\href@noop {} {\bibfield  {journal} {\bibinfo  {journal}
			{Phys. Fluids}\ }\textbf {\bibinfo {volume} {9}},\ \bibinfo {pages} {1205}
		(\bibinfo {year} {1966})}\BibitemShut {NoStop}%
	\bibitem [{\citenamefont {Virga}(1989)}]{virga:drops}%
	\BibitemOpen
	\bibfield  {author} {\bibinfo {author} {\bibfnamefont {E.~G.}\ \bibnamefont
			{Virga}},\ }\bibfield  {title} {\bibinfo {title} {Drops of nematic liquid
			crystals},\ }\href@noop {} {\bibfield  {journal} {\bibinfo  {journal} {Arch.
				Rational Mech. Anal.}\ }\textbf {\bibinfo {volume} {107}},\ \bibinfo {pages}
		{371} (\bibinfo {year} {1989})}\BibitemShut {NoStop}%
	\bibitem [{\citenamefont {Wulff}(1901)}]{wulff:frage}%
	\BibitemOpen
	\bibfield  {author} {\bibinfo {author} {\bibfnamefont {G.}~\bibnamefont
			{Wulff}},\ }\bibfield  {title} {\bibinfo {title} {Zur {F}rage der
			{G}eschwindigkeit des {W}achsthums und der {A}ufl\"osung der
			{K}ristallfl\"achen},\ }\href@noop {} {\bibfield  {journal} {\bibinfo
			{journal} {Z. Kristallographie und Mineralogie}\ }\textbf {\bibinfo {volume}
			{34}},\ \bibinfo {pages} {449} (\bibinfo {year} {1901})}\BibitemShut
	{NoStop}%
	\bibitem [{\citenamefont {Puech}\ \emph {et~al.}(2010)\citenamefont {Puech},
		\citenamefont {Grelet}, \citenamefont {Poulin}, \citenamefont {Blanc},\ and\
		\citenamefont {van~der Schoot}}]{puech:nematic}%
	\BibitemOpen
	\bibfield  {author} {\bibinfo {author} {\bibfnamefont {N.}~\bibnamefont
			{Puech}}, \bibinfo {author} {\bibfnamefont {E.}~\bibnamefont {Grelet}},
		\bibinfo {author} {\bibfnamefont {P.}~\bibnamefont {Poulin}}, \bibinfo
		{author} {\bibfnamefont {C.}~\bibnamefont {Blanc}},\ and\ \bibinfo {author}
		{\bibfnamefont {P.}~\bibnamefont {van~der Schoot}},\ }\bibfield  {title}
	{\bibinfo {title} {Nematic droplets in aqueous dispersions of carbon
			nanotubes},\ }\href@noop {} {\bibfield  {journal} {\bibinfo  {journal} {Phys.
				Rev. E}\ }\textbf {\bibinfo {volume} {82}},\ \bibinfo {pages} {020702(R)}
		(\bibinfo {year} {2010})}\BibitemShut {NoStop}%
	\bibitem [{\citenamefont
		{Saupe}(1960{\natexlab{a}})}]{saupe:temperaturabhangigkeit}%
	\BibitemOpen
	\bibfield  {author} {\bibinfo {author} {\bibfnamefont {A.}~\bibnamefont
			{Saupe}},\ }\bibfield  {title} {\bibinfo {title} {Temperaturabh\"angigkeit
			und {G}r\"o{\ss}e der {D}eformationskonstanten nematischer
			{F}l\"ussigkeiten},\ }\href@noop {} {\bibfield  {journal} {\bibinfo
			{journal} {Z. Naturforschg.}\ }\textbf {\bibinfo {volume} {15a}},\ \bibinfo
		{pages} {810} (\bibinfo {year} {1960}{\natexlab{a}})}\BibitemShut {NoStop}%
	\bibitem [{\citenamefont
		{Saupe}(1960{\natexlab{b}})}]{saupe:biegungselastizitat}%
	\BibitemOpen
	\bibfield  {author} {\bibinfo {author} {\bibfnamefont {A.}~\bibnamefont
			{Saupe}},\ }\bibfield  {title} {\bibinfo {title} {Die {B}iegungselastizit\"at
			der nematischen {P}hase von {A}zoxyanisol},\ }\href@noop {} {\bibfield
		{journal} {\bibinfo  {journal} {Z. Naturforschg.}\ }\textbf {\bibinfo
			{volume} {15a}},\ \bibinfo {pages} {815} (\bibinfo {year}
		{1960}{\natexlab{b}})}\BibitemShut {NoStop}%
	\bibitem [{\citenamefont {{Orsay Liquid Crystal Group}}(1970)}]{orsay:recent}%
	\BibitemOpen
	\bibfield  {author} {\bibinfo {author} {\bibnamefont {{Orsay Liquid Crystal
					Group}}},\ }\bibfield  {title} {\bibinfo {title} {Recent experimental
			investigations in nematic and cholesteric mesophases},\ }in\ \href@noop {}
	{\emph {\bibinfo {booktitle} {Liquid Crystals and Ordered Fluids:
				{P}roceedings of an {A}merican {C}hemical {S}ociety {S}ymposium on {O}rdered
				{F}luids and {L}iquid {C}rystals, held in {N}ew {Y}ork {C}ity, {S}eptember
				10-12, 1969}}},\ \bibinfo {editor} {edited by\ \bibinfo {editor}
		{\bibfnamefont {J.~F.}\ \bibnamefont {Johnson}}\ and\ \bibinfo {editor}
		{\bibfnamefont {R.~S.}\ \bibnamefont {Porter}}}\ (\bibinfo  {publisher}
	{Plenum Press},\ \bibinfo {address} {New York},\ \bibinfo {year} {1970})\
	pp.\ \bibinfo {pages} {447--453}\BibitemShut {NoStop}%
	\bibitem [{\citenamefont {Karat}\ and\ \citenamefont
		{Madhusudana}(1976)}]{karat:elastic}%
	\BibitemOpen
	\bibfield  {author} {\bibinfo {author} {\bibfnamefont {P.~P.}\ \bibnamefont
			{Karat}}\ and\ \bibinfo {author} {\bibfnamefont {N.~V.}\ \bibnamefont
			{Madhusudana}},\ }\bibfield  {title} {\bibinfo {title} {Elastic and optical
			properties of some 4'-n-{A}lkyl-4-{C}yanobiphenyls},\ }\href@noop {}
	{\bibfield  {journal} {\bibinfo  {journal} {Mol. Cryst. Liq. Cryst.}\
		}\textbf {\bibinfo {volume} {36}},\ \bibinfo {pages} {51} (\bibinfo {year}
		{1976})}\BibitemShut {NoStop}%
	\bibitem [{\citenamefont {Karat}\ and\ \citenamefont
		{Madhusudana}(1977)}]{karat:elasticity}%
	\BibitemOpen
	\bibfield  {author} {\bibinfo {author} {\bibfnamefont {P.~P.}\ \bibnamefont
			{Karat}}\ and\ \bibinfo {author} {\bibfnamefont {N.~V.}\ \bibnamefont
			{Madhusudana}},\ }\bibfield  {title} {\bibinfo {title} {Elasticity and
			orientational order in some 4'-n-{A}lkyl-4-{C}yanobiphenyls: {P}art {II}},\
	}\href@noop {} {\bibfield  {journal} {\bibinfo  {journal} {Mol. Cryst. Liq.
				Cryst.}\ }\textbf {\bibinfo {volume} {40}},\ \bibinfo {pages} {239} (\bibinfo
		{year} {1977})}\BibitemShut {NoStop}%
	\bibitem [{\citenamefont {Maze}(1978)}]{maze:determination}%
	\BibitemOpen
	\bibfield  {author} {\bibinfo {author} {\bibfnamefont {C.}~\bibnamefont
			{Maze}},\ }\bibfield  {title} {\bibinfo {title} {Determination of nematic
			liquid crystal elastic and dielectric properties from the shape of a
			capacitance-voltage curve},\ }\href@noop {} {\bibfield  {journal} {\bibinfo
			{journal} {Molecular Crystals and Liquid Crystals}\ }\textbf {\bibinfo
			{volume} {48}},\ \bibinfo {pages} {273} (\bibinfo {year} {1978})}\BibitemShut
	{NoStop}%
	\bibitem [{\citenamefont {Skarp}\ \emph {et~al.}(1980)\citenamefont {Skarp},
		\citenamefont {Lagerwall},\ and\ \citenamefont
		{Stebler}}]{skarp:measurements}%
	\BibitemOpen
	\bibfield  {author} {\bibinfo {author} {\bibfnamefont {K.}~\bibnamefont
			{Skarp}}, \bibinfo {author} {\bibfnamefont {S.~T.}\ \bibnamefont
			{Lagerwall}},\ and\ \bibinfo {author} {\bibfnamefont {B.}~\bibnamefont
			{Stebler}},\ }\bibfield  {title} {\bibinfo {title} {Measurements of
			hydrodynamic parameters for nematic 5{CB}},\ }\href@noop {} {\bibfield
		{journal} {\bibinfo  {journal} {Mol. Cryst. Liq. Cryst.}\ }\textbf {\bibinfo
			{volume} {60}},\ \bibinfo {pages} {215} (\bibinfo {year} {1980})}\BibitemShut
	{NoStop}%
	\bibitem [{\citenamefont {Bunning}\ \emph {et~al.}(1981)\citenamefont
		{Bunning}, \citenamefont {Faber},\ and\ \citenamefont
		{Sherrell}}]{bunning:frank}%
	\BibitemOpen
	\bibfield  {author} {\bibinfo {author} {\bibfnamefont {J.~D.}\ \bibnamefont
			{Bunning}}, \bibinfo {author} {\bibfnamefont {T.~E.}\ \bibnamefont {Faber}},\
		and\ \bibinfo {author} {\bibfnamefont {P.~L.}\ \bibnamefont {Sherrell}},\
	}\bibfield  {title} {\bibinfo {title} {The {F}rank constants of nematic 5{CB}
			at atmospheric pressure},\ }\href@noop {} {\bibfield  {journal} {\bibinfo
			{journal} {J. de Phys.}\ }\textbf {\bibinfo {volume} {42}},\ \bibinfo {pages}
		{1175} (\bibinfo {year} {1981})}\BibitemShut {NoStop}%
	\bibitem [{\citenamefont {Balzarini}\ \emph {et~al.}(1984)\citenamefont
		{Balzarini}, \citenamefont {Dunmur},\ and\ \citenamefont
		{Palffy-Muhoray}}]{balzarini:high}%
	\BibitemOpen
	\bibfield  {author} {\bibinfo {author} {\bibfnamefont {D.~A.}\ \bibnamefont
			{Balzarini}}, \bibinfo {author} {\bibfnamefont {D.~A.}\ \bibnamefont
			{Dunmur}},\ and\ \bibinfo {author} {\bibfnamefont {P.}~\bibnamefont
			{Palffy-Muhoray}},\ }\bibfield  {title} {\bibinfo {title} {High voltage
			birefringence measurements of elastic constants},\ }\href@noop {} {\bibfield
		{journal} {\bibinfo  {journal} {Mol. Cryst. Liq. Cryst.}\ }\textbf {\bibinfo
			{volume} {102}},\ \bibinfo {pages} {35} (\bibinfo {year} {1984})}\BibitemShut
	{NoStop}%
	\bibitem [{\citenamefont {Hurd}\ \emph {et~al.}(1985)\citenamefont {Hurd},
		\citenamefont {Fraden}, \citenamefont {Lonberg},\ and\ \citenamefont
		{Meyer}}]{hurd:field-indiced}%
	\BibitemOpen
	\bibfield  {author} {\bibinfo {author} {\bibfnamefont {A.~J.}\ \bibnamefont
			{Hurd}}, \bibinfo {author} {\bibfnamefont {S.}~\bibnamefont {Fraden}},
		\bibinfo {author} {\bibfnamefont {F.}~\bibnamefont {Lonberg}},\ and\ \bibinfo
		{author} {\bibfnamefont {R.~B.}\ \bibnamefont {Meyer}},\ }\bibfield  {title}
	{\bibinfo {title} {Field-induced transient periodic structures in nematic
			liquid crystals: the splay {F}rederiks transition},\ }\href@noop {}
	{\bibfield  {journal} {\bibinfo  {journal} {J. Phys. France}\ }\textbf
		{\bibinfo {volume} {46}},\ \bibinfo {pages} {905} (\bibinfo {year}
		{1985})}\BibitemShut {NoStop}%
	\bibitem [{\citenamefont {Taratuta}\ \emph {et~al.}(1985)\citenamefont
		{Taratuta}, \citenamefont {Hurd},\ and\ \citenamefont
		{Meyer}}]{taratuta:light-scattering}%
	\BibitemOpen
	\bibfield  {author} {\bibinfo {author} {\bibfnamefont {V.~G.}\ \bibnamefont
			{Taratuta}}, \bibinfo {author} {\bibfnamefont {A.~J.}\ \bibnamefont {Hurd}},\
		and\ \bibinfo {author} {\bibfnamefont {R.~B.}\ \bibnamefont {Meyer}},\
	}\bibfield  {title} {\bibinfo {title} {Light-scattering study of a polymer
			nematic liquid crystal},\ }\href@noop {} {\bibfield  {journal} {\bibinfo
			{journal} {Phys. Rev. Lett.}\ }\textbf {\bibinfo {volume} {55}},\ \bibinfo
		{pages} {246} (\bibinfo {year} {1985})}\BibitemShut {NoStop}%
	\bibitem [{\citenamefont {Morris}\ \emph {et~al.}(1986)\citenamefont {Morris},
		\citenamefont {Palffy-Muhoray},\ and\ \citenamefont
		{Balzarini}}]{morris:measurements}%
	\BibitemOpen
	\bibfield  {author} {\bibinfo {author} {\bibfnamefont {S.~W.}\ \bibnamefont
			{Morris}}, \bibinfo {author} {\bibfnamefont {P.}~\bibnamefont
			{Palffy-Muhoray}},\ and\ \bibinfo {author} {\bibfnamefont {D.~A.}\
			\bibnamefont {Balzarini}},\ }\bibfield  {title} {\bibinfo {title}
		{Measurements of the bend and splay elastic constants of
			octyl-cyanobiphenyl},\ }\href@noop {} {\bibfield  {journal} {\bibinfo
			{journal} {Mol. Cryst. Liq. Cryst.}\ }\textbf {\bibinfo {volume} {139}},\
		\bibinfo {pages} {263} (\bibinfo {year} {1986})}\BibitemShut {NoStop}%
	\bibitem [{\citenamefont {Lee}\ and\ \citenamefont
		{Meyer}(1986)}]{lee:computations}%
	\BibitemOpen
	\bibfield  {author} {\bibinfo {author} {\bibfnamefont {S.}~\bibnamefont
			{Lee}}\ and\ \bibinfo {author} {\bibfnamefont {R.~B.}\ \bibnamefont
			{Meyer}},\ }\bibfield  {title} {\bibinfo {title} {Computations of the phase
			equilibrium, elastic constants, and viscosities of a hard‐rod nematic
			liquid crystal},\ }\href@noop {} {\bibfield  {journal} {\bibinfo  {journal}
			{J. Chem. Phys.}\ }\textbf {\bibinfo {volume} {84}},\ \bibinfo {pages} {3443}
		(\bibinfo {year} {1986})}\BibitemShut {NoStop}%
	\bibitem [{\citenamefont {Lee}\ and\ \citenamefont
		{Meyer}(1988)}]{lee:crossover}%
	\BibitemOpen
	\bibfield  {author} {\bibinfo {author} {\bibfnamefont {S.-D.}\ \bibnamefont
			{Lee}}\ and\ \bibinfo {author} {\bibfnamefont {R.~B.}\ \bibnamefont
			{Meyer}},\ }\bibfield  {title} {\bibinfo {title} {Crossover behavior of the
			elastic coefficients and viscosities of a polymer nematic liquid crystal},\
	}\href@noop {} {\bibfield  {journal} {\bibinfo  {journal} {Phys. Rev. Lett.}\
		}\textbf {\bibinfo {volume} {61}},\ \bibinfo {pages} {2217} (\bibinfo {year}
		{1988})}\BibitemShut {NoStop}%
	\bibitem [{\citenamefont {Taratuta}\ \emph {et~al.}(1988)\citenamefont
		{Taratuta}, \citenamefont {Lonberg},\ and\ \citenamefont
		{Meyer}}]{taratuta:anisotropic}%
	\BibitemOpen
	\bibfield  {author} {\bibinfo {author} {\bibfnamefont {V.~G.}\ \bibnamefont
			{Taratuta}}, \bibinfo {author} {\bibfnamefont {F.}~\bibnamefont {Lonberg}},\
		and\ \bibinfo {author} {\bibfnamefont {R.~B.}\ \bibnamefont {Meyer}},\
	}\bibfield  {title} {\bibinfo {title} {Anisotropic mechanical properties of a
			polymer nematic liquid crystal},\ }\href@noop {} {\bibfield  {journal}
		{\bibinfo  {journal} {Phys. Rev. A}\ }\textbf {\bibinfo {volume} {37}},\
		\bibinfo {pages} {1831} (\bibinfo {year} {1988})}\BibitemShut {NoStop}%
	\bibitem [{\citenamefont {Itou}\ \emph {et~al.}(1991)\citenamefont {Itou},
		\citenamefont {Tozaki},\ and\ \citenamefont {Komatsu}}]{itou:measurements}%
	\BibitemOpen
	\bibfield  {author} {\bibinfo {author} {\bibfnamefont {S.}~\bibnamefont
			{Itou}}, \bibinfo {author} {\bibfnamefont {K.}~\bibnamefont {Tozaki}},\ and\
		\bibinfo {author} {\bibfnamefont {N.}~\bibnamefont {Komatsu}},\ }\bibfield
	{title} {\bibinfo {title} {Lyotropic liquid crystalline structures of
			synthetic polypeptide 2. {M}easurements of elastic constants and coupling
			constants of {P}oly($\gamma$-{B}enzyl {L}-{G}lutamate) solutions by the
			{F}reedericksz transition},\ }\href@noop {} {\bibfield  {journal} {\bibinfo
			{journal} {Jpn. J. Appl. Phys.}\ }\textbf {\bibinfo {volume} {30}},\ \bibinfo
		{pages} {1230} (\bibinfo {year} {1991})}\BibitemShut {NoStop}%
	\bibitem [{\citenamefont {Zhou}\ \emph {et~al.}(2012)\citenamefont {Zhou},
		\citenamefont {Nastishin}, \citenamefont {Omelchenko}, \citenamefont
		{Tortora}, \citenamefont {Nazarenko}, \citenamefont {Boiko}, \citenamefont
		{Ostapenko}, \citenamefont {Hu}, \citenamefont {Almasan}, \citenamefont
		{Sprunt}, \citenamefont {Gleeson},\ and\ \citenamefont
		{Lavrentovich}}]{zhou:elasticity}%
	\BibitemOpen
	\bibfield  {author} {\bibinfo {author} {\bibfnamefont {S.}~\bibnamefont
			{Zhou}}, \bibinfo {author} {\bibfnamefont {Y.~A.}\ \bibnamefont {Nastishin}},
		\bibinfo {author} {\bibfnamefont {M.~M.}\ \bibnamefont {Omelchenko}},
		\bibinfo {author} {\bibfnamefont {L.}~\bibnamefont {Tortora}}, \bibinfo
		{author} {\bibfnamefont {V.~G.}\ \bibnamefont {Nazarenko}}, \bibinfo {author}
		{\bibfnamefont {O.~P.}\ \bibnamefont {Boiko}}, \bibinfo {author}
		{\bibfnamefont {T.}~\bibnamefont {Ostapenko}}, \bibinfo {author}
		{\bibfnamefont {T.}~\bibnamefont {Hu}}, \bibinfo {author} {\bibfnamefont
			{C.~C.}\ \bibnamefont {Almasan}}, \bibinfo {author} {\bibfnamefont {S.~N.}\
			\bibnamefont {Sprunt}}, \bibinfo {author} {\bibfnamefont {J.~T.}\
			\bibnamefont {Gleeson}},\ and\ \bibinfo {author} {\bibfnamefont {O.~D.}\
			\bibnamefont {Lavrentovich}},\ }\bibfield  {title} {\bibinfo {title}
		{Elasticity of lyotropic chromonic liquid crystals probed by director
			reorientation in a magnetic field},\ }\href@noop {} {\bibfield  {journal}
		{\bibinfo  {journal} {Phys. Rev. Lett.}\ }\textbf {\bibinfo {volume} {109}},\
		\bibinfo {pages} {037801} (\bibinfo {year} {2012})}\BibitemShut {NoStop}%
	\bibitem [{\citenamefont {Faetti}\ and\ \citenamefont
		{Palleschi}(1984{\natexlab{a}})}]{faetti:nematic}%
	\BibitemOpen
	\bibfield  {author} {\bibinfo {author} {\bibfnamefont {S.}~\bibnamefont
			{Faetti}}\ and\ \bibinfo {author} {\bibfnamefont {V.}~\bibnamefont
			{Palleschi}},\ }\bibfield  {title} {\bibinfo {title} {Nematic-isotropic
			interface of some members of the homologous series of
			4-cyano-4\ensuremath{'}-($n$-alkyl)biphenyl liquid crystals},\ }\href@noop {}
	{\bibfield  {journal} {\bibinfo  {journal} {Phys. Rev. A}\ }\textbf {\bibinfo
			{volume} {30}},\ \bibinfo {pages} {3241} (\bibinfo {year}
		{1984}{\natexlab{a}})}\BibitemShut {NoStop}%
	\bibitem [{\citenamefont {Faetti}\ and\ \citenamefont
		{Palleschi}(1984{\natexlab{b}})}]{faetti:measurements}%
	\BibitemOpen
	\bibfield  {author} {\bibinfo {author} {\bibfnamefont {S.}~\bibnamefont
			{Faetti}}\ and\ \bibinfo {author} {\bibfnamefont {V.}~\bibnamefont
			{Palleschi}},\ }\bibfield  {title} {\bibinfo {title} {Measurements of the
			interfacial tension between nematic and isotropic phase of some
			cyanobiphenyls},\ }\href@noop {} {\bibfield  {journal} {\bibinfo  {journal}
			{J. Chem. Phys.}\ }\textbf {\bibinfo {volume} {81}},\ \bibinfo {pages} {6254}
		(\bibinfo {year} {1984}{\natexlab{b}})}\BibitemShut {NoStop}%
	\bibitem [{\citenamefont {Chen}\ \emph {et~al.}(1996)\citenamefont {Chen},
		\citenamefont {Sato},\ and\ \citenamefont {Teramoto}}]{chen:measurement}%
	\BibitemOpen
	\bibfield  {author} {\bibinfo {author} {\bibfnamefont {W.}~\bibnamefont
			{Chen}}, \bibinfo {author} {\bibfnamefont {T.}~\bibnamefont {Sato}},\ and\
		\bibinfo {author} {\bibfnamefont {A.}~\bibnamefont {Teramoto}},\ }\bibfield
	{title} {\bibinfo {title} {Measurement of the interfacial tension between
			coexisting isotropic and nematic phases of a lyotropic polymer liquid
			crystal},\ }\href@noop {} {\bibfield  {journal} {\bibinfo  {journal}
			{Macromolecules}\ }\textbf {\bibinfo {volume} {29}},\ \bibinfo {pages} {4283}
		(\bibinfo {year} {1996})}\BibitemShut {NoStop}%
	\bibitem [{\citenamefont {Chen}\ \emph {et~al.}(1998)\citenamefont {Chen},
		\citenamefont {Sato},\ and\ \citenamefont {Teramoto}}]{chen:interfacial}%
	\BibitemOpen
	\bibfield  {author} {\bibinfo {author} {\bibfnamefont {W.}~\bibnamefont
			{Chen}}, \bibinfo {author} {\bibfnamefont {T.}~\bibnamefont {Sato}},\ and\
		\bibinfo {author} {\bibfnamefont {A.}~\bibnamefont {Teramoto}},\ }\bibfield
	{title} {\bibinfo {title} {Interfacial tension between coexisting isotropic
			and nematic phases for a lyotropic polymer liquid crystal: {P}oly(n-hexyl
			isocyanate) solutions},\ }\href@noop {} {\bibfield  {journal} {\bibinfo
			{journal} {Macromolecules}\ }\textbf {\bibinfo {volume} {31}},\ \bibinfo
		{pages} {6506} (\bibinfo {year} {1998})}\BibitemShut {NoStop}%
	\bibitem [{\citenamefont {Chen}\ \emph {et~al.}(1999)\citenamefont {Chen},
		\citenamefont {Sato},\ and\ \citenamefont
		{Teramoto}}]{chen:interfacial_1999}%
	\BibitemOpen
	\bibfield  {author} {\bibinfo {author} {\bibfnamefont {W.}~\bibnamefont
			{Chen}}, \bibinfo {author} {\bibfnamefont {T.}~\bibnamefont {Sato}},\ and\
		\bibinfo {author} {\bibfnamefont {A.}~\bibnamefont {Teramoto}},\ }\bibfield
	{title} {\bibinfo {title} {Interfacial tension between coexisting isotropic
			and cholesteric phases for aqueous solutions of {S}chizophyllan},\
	}\href@noop {} {\bibfield  {journal} {\bibinfo  {journal} {Macromolecules}\
		}\textbf {\bibinfo {volume} {32}},\ \bibinfo {pages} {1549} (\bibinfo {year}
		{1999})}\BibitemShut {NoStop}%
	\bibitem [{\citenamefont {Chen}\ and\ \citenamefont
		{Gray}(2002)}]{chen:interfacial_2002}%
	\BibitemOpen
	\bibfield  {author} {\bibinfo {author} {\bibfnamefont {W.}~\bibnamefont
			{Chen}}\ and\ \bibinfo {author} {\bibfnamefont {D.~G.}\ \bibnamefont
			{Gray}},\ }\bibfield  {title} {\bibinfo {title} {Interfacial tension between
			isotropic and anisotropic phases of a suspension of rodlike particles},\
	}\href@noop {} {\bibfield  {journal} {\bibinfo  {journal} {Langmuir}\
		}\textbf {\bibinfo {volume} {18}},\ \bibinfo {pages} {633} (\bibinfo {year}
		{2002})}\BibitemShut {NoStop}%
	\bibitem [{\citenamefont {Kahlweit}\ and\ \citenamefont
		{Ostner}(1973)}]{kahlweit:estimation}%
	\BibitemOpen
	\bibfield  {author} {\bibinfo {author} {\bibfnamefont {M.}~\bibnamefont
			{Kahlweit}}\ and\ \bibinfo {author} {\bibfnamefont {W.}~\bibnamefont
			{Ostner}},\ }\bibfield  {title} {\bibinfo {title} {An estimation of the
			interfacial tension between the nematic and isotropic states of a liquid
			crystal},\ }\href@noop {} {\bibfield  {journal} {\bibinfo  {journal} {Chem.
				Phys. Lett.}\ }\textbf {\bibinfo {volume} {18}},\ \bibinfo {pages} {589}
		(\bibinfo {year} {1973})}\BibitemShut {NoStop}%
	\bibitem [{\citenamefont {Langevin}\ and\ \citenamefont
		{Bouchiat}(1973)}]{langevin:molecular}%
	\BibitemOpen
	\bibfield  {author} {\bibinfo {author} {\bibfnamefont {D.}~\bibnamefont
			{Langevin}}\ and\ \bibinfo {author} {\bibfnamefont {M.~A.}\ \bibnamefont
			{Bouchiat}},\ }\bibfield  {title} {\bibinfo {title} {Molecular order and
			surface tension for the nematic-isotropic interface of {MBBA}, deduced from
			light reflectivity and light scattering measurements},\ }\href@noop {}
	{\bibfield  {journal} {\bibinfo  {journal} {Mol. Cryst. Liq. Cryst.}\
		}\textbf {\bibinfo {volume} {22}},\ \bibinfo {pages} {317} (\bibinfo {year}
		{1973})}\BibitemShut {NoStop}%
	\bibitem [{\citenamefont {Yokoyama}\ \emph {et~al.}(1983)\citenamefont
		{Yokoyama}, \citenamefont {Kobayashi},\ and\ \citenamefont
		{Kamei}}]{yokoyama:boundary}%
	\BibitemOpen
	\bibfield  {author} {\bibinfo {author} {\bibfnamefont {H.}~\bibnamefont
			{Yokoyama}}, \bibinfo {author} {\bibfnamefont {S.}~\bibnamefont
			{Kobayashi}},\ and\ \bibinfo {author} {\bibfnamefont {H.}~\bibnamefont
			{Kamei}},\ }\bibfield  {title} {\bibinfo {title} {Boundary dependence of the
			formation of new phase at the isotropic-nematic transition},\ }\href@noop {}
	{\bibfield  {journal} {\bibinfo  {journal} {Mol. Cryst. Liq. Cryst.}\
		}\textbf {\bibinfo {volume} {99}},\ \bibinfo {pages} {39} (\bibinfo {year}
		{1983})}\BibitemShut {NoStop}%
	\bibitem [{\citenamefont {Naggiar}(1943)}]{naggiar:phenomenes}%
	\BibitemOpen
	\bibfield  {author} {\bibinfo {author} {\bibfnamefont {V.}~\bibnamefont
			{Naggiar}},\ }\bibfield  {title} {\bibinfo {title} {Ph\'enom\`enes
			d’orientation dans les liquides n\'ematiques. {U}ne nouvelle m\'ethode de
			mesure de la tension superficielle applicable \`a ces liquides},\ }\href@noop
	{} {\bibfield  {journal} {\bibinfo  {journal} {Ann. Phys. (France)}\ }\textbf
		{\bibinfo {volume} {11}},\ \bibinfo {pages} {5} (\bibinfo {year}
		{1943})}\BibitemShut {NoStop}%
	\bibitem [{\citenamefont {Schwartz}\ and\ \citenamefont
		{Moseley}(1947)}]{schwartz:surface}%
	\BibitemOpen
	\bibfield  {author} {\bibinfo {author} {\bibfnamefont {W.~M.}\ \bibnamefont
			{Schwartz}}\ and\ \bibinfo {author} {\bibfnamefont {H.~W.}\ \bibnamefont
			{Moseley}},\ }\bibfield  {title} {\bibinfo {title} {The surface tension of
			liquid crystals},\ }\href@noop {} {\bibfield  {journal} {\bibinfo  {journal}
			{J. Phys. Chem.}\ }\textbf {\bibinfo {volume} {51}},\ \bibinfo {pages} {826}
		(\bibinfo {year} {1947})}\BibitemShut {NoStop}%
	\bibitem [{\citenamefont {Ogilvy}(1990)}]{ogilvy:excursions}%
	\BibitemOpen
	\bibfield  {author} {\bibinfo {author} {\bibfnamefont {C.~S.}\ \bibnamefont
			{Ogilvy}},\ }\href@noop {} {\emph {\bibinfo {title} {Excursions in
				geometry}}}\ (\bibinfo  {publisher} {Taylor and Francis},\ \bibinfo {address}
	{Mineola, NY, USA},\ \bibinfo {year} {1990})\ \bibinfo {note} {(Unabridged
		and corrected republication of the work originally published by Oxford
		University Press, New York)}\BibitemShut {NoStop}%
\end{thebibliography}

%

\end{document}